\documentclass[12pt]{report}
\usepackage{epsfig,graphicx}
\begin{document}
\def\lsim{\mathrel{
\def\arraystretch{.4}
\begin{array}{c}
$$<$$\\
$$\sim$$
\end{array}}}
\def\gsim{\mathrel{
\def\arraystretch{.4}
\begin{array}{c}
>$$\\
$$\sim$$
\end{array}}}

\vspace{0.1 in} 
\
\begin{center} 
{\bf A  Novel Non perturbative Self-consistent and General Approximation 
Method in Quantum Theory$^*$} 
\end{center}

\begin{center}
 
   {\bf Nabaghan Santi$^*$$^*$}
\end{center}
\vspace{.3 in}

\begin{center}
         {\bf Abstract}                  
\end{center}

A  new method of approximation scheme with potential application to a general 
interacting quantum system is presented. The method is non-perturbative, self-
consistent, systematically improvable and uniformly applicable for arbitrary 
strength of interaction. It thus overcomes the various limitations of the 
exsting methods such as the perturbation theory, the variational method, the
WKBJ method and other approximation schemes. The current method has been 
successfully applied to a variety of interacting systems including the 
anharmonic/ double-well oscillators ( with quartic-, sextic- and octic
 couplings ) and the scalar field theory with quartic-coupling in the 
symmetric phase. The method yields important insight in to the structure and
stability of the interacting-vacuum of the theory. The results are in good 
agreement with the exact predictions of supersymmetry where ever applicable.
Possible further applications in the areas of quantum statistics, finite 
temperature field theory, condensed matter physics etc are also discussed.\\

\hrule
\vspace{1.0 in}
*{\bf Ph.D thesis submitted to Sambalpur University, \\ September 2006.}
\vspace{.2 in}

  **{\bf Reader in Physics, B. J. B (Autonomous) College,\\ Bhubaneswar-751014, 
Orissa, India.}\\
{\bf email:  ng\_ santi@yahoo.com.in}

\newpage
{\Large \bf 1. Introduction}\\

 It is now generally believed  that quantum theory (QT) describes$^{1,2}$ the 
observed features of the physical universe at the fundamental level,
both at the microscopic- and the macro-
scales, which span a vast range in distance, typically from
$~\sim~10^{-16}~$ cm  to  light years and
 time scales, from $~10^{-15}~$second to the age of the universe
$~\sim ~10^{10}~$ years. It is truly amazing that phenomena as diverse
as the colour of the sky, the emission of light by heated
bodies, stability of matter, the evolution of stars, super
conductivity and  other properties of matter at extreme conditions of
 temperatures, density and pressure etc, $\it {all}$ yield to
 description$^{2}$ by the QT.\\

However, inspite of the vastness of the range of applicability of QT,
the exact,
$\it{analytic}$ predictions of the theory  based upon the solution of
the underlying  fundamental  dynamical 
equations are rather sparse$^{3}$. It is not difficult to count and 
classify the cases where exact analytic solutions are possible. In
fact, it is  currently believed$^{4,5}$ that the analytic solvability 
of the dynamical equations of quantum theory  is mainly based 
upon the  factorisation property  of the corresponding
quantum-Hamiltonians. 
 In that context, recent applications of super-symmetric 
quantum mechanics (SUSYQM)$^{6}$  have  resulted in considerable
extension$^{6}$ of the range  and class of exactly solvable potentials  in
non-relativistic  QT.\\

In view of the rather limited range of physical phenomena amenable to
exact analytic 
solutions in QT, it has become inevitable to develop various
approximation methods of solution. This aspect was realized soon after 
the discovery of QT. The history of approximation methods is,
therefore, as old as 
the theory itself$^{7}$. Consequently, this has generated a vast amount
of literature devoted to the various approximation schemes (AS).\\

In discussing the merits and limitations of the various AS in QT
certain basic criteria are generally applied. These criteria may, for 
example, include the following:\\(i) {\it  general~ applicability}
~(ii)  { \it simplicity~ of ~implementation}
~(iii) {\it achieved  accuracy} (iv) {\it  systematic~ improvability}~~(v)
~ {\it  rapidity~ of~ convergence~and ~adaptability }~ 
{\it for~numerical computation (when  the obtained results are not in closed 
 form)}etc.  In spite of the vast  amount of accumulated literature  and the
broad range of the existing AS, there is considerable current effort
in searching for further improved AS which would satisfy most of the
criteria listed above. \\

The present thesis records original research aimed at achieving  a
novel method of approximation in QT, which is non-perturbative, 
self-consistent, systematically improvable and simple to implement.
More importantly, it has the potential to be applicable, in principle, 
to any general and arbitrary  interacting system including quantum field theory
 and quantum statistics.\\

In order to further motivate and understand  the scope of the  present
scheme, it is necessary to review
and  survey  the various existing approximation schemes (AS) in QT and
discuss  their relative merits and limitations as against the set of
desirable criteria listed  
above. However, such a task is understandably rather arduous due to
the vast amount of accumulated literature. Moreover, several popular
schemes
of approximation already find mention in texts on QT. We shall,
therefore,  confine ourselves to the $~\it{main}~$ AS's  and shall be
rather brief. In the following sections a survey along the stated
lines is presented for the following main categories of the AS in QT :
                                                                  \\ 
       \\
(1){\it ~Methods~based~upon~Perturbation~theory}
                                                                   \\
\\
(2){\it ~The~Variation~ Method~of~Approximation}
                                                                   \\
\\
(3){\it ~The~Hill-determinant~and related~ methods}                                                                     \\ 
\\
(4){\it  Combination of variation method and perturbation techniques}
\\
\\
(5){\it ~The ~Algebraic ~Method ~of ~Operator~Expansion}
\\
                                                                    \\
(6){\it ~The~JWKB  approximation method}
                                                                      \\  
\\
(7){\it ~The~Method~of~Approximation~ employing~
  the~Canonical~Transformation ~in~QT}
                                                                      \\
\\
(8){\it ~The ~Approximation~ methods~ based ~upon~super~
  symmetric~quantum~mechanics}
                                                                  \\
\\
(9){\it ~The~methods~based~upon~the~ re- summation~ techniques}
                                                                   \\
\\
(10){\it ~The~Approximation~schemes~ based ~upon ~the~path-integral~ 
formulation~of~QT }
\\
                                                                    \\
(11){\it ~The~self-consistent~schemes~of~approximation}
                                                                      \\
\\
(12){\it ~ Other~methods~of~approximation}
\\
\\
In the next {\bf Chapter}, we make a brief survey in dealing with each
scheme of approximation and highlight the main features of each with
special emphasis on its merit and limitation. Each method is also
applied to the case of the quantum an-harmonic oscillator (QAHO) taken 
as the reference standard. This quantum system is chosen for the above 
purpose since it is also the $~\it{traditional~choice}~$ as the
theoretical laboratory for testing various  AS in QT$^{8,9}$. 
Consequently, the QAHO is perhaps among the most widely studied 
quantum  system, see, e.g., refs. [8,9].\\
\\
\\

{\Large \bf 2.  Survey of  Approximation Schemes  in Quantum  Theory }\\

In the following sub-sections we present a brief survey of the
different main schemes of approximation in QT as listed in the
previous {\bf Chapter}.
\\
\\
(1) {\it Methods~based~upon~Perturbation-Theory ~} 
\\
\\
This scheme is among the earliest approximation methods 
recorded in the literature. In the context of classical physics, the
method was applied by Lord-Rayleigh$^{7}$  as early as 1873. In QT the
application of the method is as old as the theory itself, see,
e.g. ref.[7]. Since the 
salient features of the method  are well discussed  in texts (e.g.[3]) 
only brief discussion follows. For simplicity, we  restrict  
to the case of {\it non-degenerate} and {\it discrete} spectrum.\\

The  generic central feature of the perturbation method consists in
addressing to those problems where the system Hamiltonian can be split
in the following fashion : 
\begin{equation}
H~=~H_{s}~+~\lambda H_{I}~,
\end{equation}
 where $~H_{S}~$ corresponds to an exactly {\it solvable} Hamiltonian 
whereas the other part,  $~\lambda H_{I}~$ can be regarded as  a $~\it
{perturbation}$-interaction term  added to the dominant Hamiltonian
represented by  $~H_{s}$,  the implicit assumption being that the
perturbation term is ``small''
 in some-sense ( to be specified below). In  the above equation, the
parameter `$\lambda$'
~generically  corresponds to an expansion-parameter. This may be  either
 naturally present in the interaction term or {\it explicitly}
introduced  by hand to keep a count on the {\it order} of
successive approximation  involving powers of the perturbation-term
$~H_{I}$. In the latter case, it is set to unity, $~\lambda
\rightarrow~1$,~ at the end of the computation to the desired order of 
approximation.\\

In order that the above scheme works, at least the
following  obvious requirements must be met :
                                                              \\
(a)  the system  Hamiltonian {\it permits} a split in the above
manner as in  eqn.(1).
                                                             \\
 (b) the  perturbation-interaction term  {\it $\lambda H_{I}$}~
{\it  remains sub-dominant}  to the solvable unperturbed part,
 $~H_{s}$. In view  that the Hamiltonian is an operator, the above statement
requires  precise quantification. A commonly  accepted criteria, which ensures
 the condition of sub-dominance of the perturbation term is  the 
following$^{3}$:
\begin{equation}
|~<~n_{s}~|~H_{s}~|~n_{s}~>~|~\geq~|~<~n_{s}~|~\lambda
H_{I}~|~n_{s}~>~|~ ,
\end{equation}
where $~|~n_{s}~>~$ is defined by the eigen value equation,  
\begin{equation}
H_{s}|~n_{s}~>~ =~ E^{s}_{n}~|~n_{s}~>
\end{equation}
(c) The method $~\it{yields}~$ meaningful results  if the actual
 $~\it{physical}~$ spectrum can be computed as a power-series, i.e.
\begin{equation}
E_{n}~=~E^{s}_{n}~+~\lambda~ E^{(1)}_{n}~+~\lambda^{2}~E^{(2)}_{n}~+~.....~,
\end{equation} 
and
\begin{equation}
|~n~>~=~|~n_{s}>~+~\lambda ~|~n^{(1)}~>~+~\lambda^{2}~|~n^{(2)}~>~+~.....~,
\end{equation}
\\
(d)~ It ~is~ further ~required that the above ~sequences~ represent~ either 
{\it convergent or~asymptotic} series in order that the eqs.(4)
 and (5)
make sense as an approximation if truncation of the same after a 
{\it finite} number of terms would yield results representative of the 
actual physical situation.
                                                                           \\
\\
(e) An important corollary of the above requirement is, therefore, that the 
  ~{\it rate~ of~convergence} of the series  ( or, the rate of decrease of 
the sub-asymptotic sequence, as the case may be ) are important 
considerations  bearing  upon the {\it practical}  application of the 
method.  
                                                                        \\
\\
The series-development in eqs.(4) and (5) can be  generated by  
iteration starting from the Schr$\ddot{o}$dinger equation for the states 
$~|~ n~ >~$ and $~|~n_{s}~>$. Following standard method$^{3}$, one obtains the
following general expression valid for arbitrary order :
\begin{equation}
|~n~>~=~|~n_{s}~>~+~Q_{n}~(H_{s}-z)^{-1} ( E_{n}~-~\lambda H_{I}~-~z)
~|~n~>~,
\end{equation}
After iteration, eqn.(2.6) yields the following result :
 \begin{equation}
|~n~>~=~\sum_{m}~[~Q_{n}~(H_{s}-z)^{-1} ( E_{n}~-~\lambda H_{I}~-~z)~]^{m}
~|~n_{s}~>~.
\end{equation}
Similarly, for the energy shift one obtains : 
\begin{equation}
E_{n}~=~E^{s}_{n}~+~\sum_{m}~<~n_{s}|~\lambda H_{I}~[~Q_{n}~(H_{s}-z)^{-1} 
( E_{n}~-~\lambda H_{I}~-~z)~]^{m}~|~n_{s}~>~,
\end{equation}
where, $~z~=~$ arbitrary  complex  parameter;  $~Q_{n}~\equiv~ \bf{1}~-~
|~n_{s}~><n_{s}~|$, is a projection  operator with the property of 
projection to states orthogonal to $~|~n_{s}~>$, i.e. 
$<n_{s}|~Q_{n}~|~\psi~> =~0$,  for any arbitrary 
representative state $~|~\psi~>~$. The role of the arbitrary parameter `z' is 
to  generate by  iteration,  the two popular versions of perturbation 
expansion, namely  the Rayleigh-Schr$\ddot{o}$dinger(RS)$^{7}$  and  the 
Brillouin-Wigner (BW)$^{10}$ versions for different choice of z.
Thus, for $~z~=~E^{s}_{n}~$ the RS-series is obtained while for  $~z~=~E_{n}$,
 the BW-series follows. More explicitly, the separate expressions are given 
below for the two schemes :
                                                                  \\
\\
{\underline{\bf RS-Scheme}}
\begin{equation}
|~n~>~=~\sum_{m}~[~Q_{n}~(H_{s}-E^{s}_{n})^{-1} ( E_{n}~-~\lambda H_{I}~-~E^{s}_{n})~]^{m}~|~n_{s}~>~,
\end{equation}
Similarly, for the energy-eigen value one obtains 
\begin{equation}
E_{n}~=~E^{s}_{n}~+~\sum_{m}~<~n_{s}|~\lambda H_{I}~[~Q_{n}~(H_{s}-E^{s}_{n})^
{-1} ( E_{n}~-~\lambda H_{I}~-~E^{s}_{n})~]^{m}~|~n_{s}~>~,
\end{equation}
Similar expressions in the BW scheme are given below.
                                                                     \\
\\
{\underline{\bf BW-Scheme}}
\begin{equation}
|~n~>~=~\sum_{m}~[~Q_{n}~(H_{s}-E_{n})^{-1} (-~\lambda H_{I})~]^{m}
~|~n_{s}~>~
\end{equation}
\begin{equation}
E_{n}~=~E^{s}_{n}~+~\sum_{m}~<~n_{s}|~\lambda H_{I}~[~Q_{n}~(H_{s}-E_{n})^
{-1} ( -~\lambda H_{I})~]^{m}~|~n_{s}~>~.
\end{equation}
\\
\\
It may be remarked here that the above (formal) expressions, which are valid 
to arbitrary order, become useful only when approximated to any given 
{\it finite} order. A comparision of the two schemes reveals that although 
the BW-expressions look simpler, yet it involves the {\it unknown}, exact
energy of the system in the denominator, which thus requires further expansion
up to the desired order in `$\lambda$'. This accounts for the popularity of 
the RS-scheme. On the other hand, however, there are systems for which, the 
BW-scheme is preferred as it converges more rapidly than the RS-scheme.\\

{\it Merits and limitations of perturbation theory}
\\
\\
Decidedly, the main merit of the method is the built-in provision for 
systematic, order-by-order improvement over the zeroth-order (unperturbed) 
result. However, in its na$\ddot {i}$ve formulation as elaborated above 
( which, we denote as na$\ddot {i}$ve perturbation theory (NPT) in the 
following) the 
applicability gets severly restricted due to several reasons. We mention  some
 of these in the following.
\\
\\
The main limitation arises when eqn.(2) fails to
 hold for the system. This may happen due to several reasons. Firstly, this may
 mean that the range of values of `$\lambda$' gets restricted (often, the 
condition: $~\lambda <~1~$)  provides a necessary {\it but not} sufficient 
condition. Secondly, the nature of the problem may be such that the 
inequality (eqn.(2)) fails (  {\it even if $~\lambda<<1~$}). A third reason 
leading to the inapplicability of the NPT  may arise when the desired 
condition, eqn.(2) does not hold for {\it arbitrary}  value  of the 
excitation label `n'.\\

Even when the basic requirement stated in eqn.(2) is satisfied, the 
applicability of the method may be restricted on the grounds of convergence
of the NPT. This may happen, for example, when the physical quantity sought to
 be calculated by perturbation series, is {\it independently} found to be 
{\it non-analytic} at $~\lambda~=~0$. There is a large variety of physical
 phenomena, which fall under this catagory, e.g. tunneling, decay, 
phase-transitions, critical-phenomena, collective and cooperative phenomena 
such as super-conductivity,
super-fluidity etc. It may be pointed-out in this context, that in  most of
 the cases listed above,  the
{\it perturbative} ground state ( i.e., the ground state of the unperturbed
 Hamiltonian, $H_{s}$) is shown to be unstable. In addition to the above 
limitations of the NPT on theoretical grounds, there may be {\it practical} 
difficulties  arising, for example, from computing corrections beyond the 
first-order, which generally involve the  (infinite) sum over the intermediate
states. This  problem has been addressed  by several authors  and it has  been
 found in specific examples that the difficulty can be surmounted  by 
appealing to  special techniques, such as the  Dalgrano-Lewis
method$^{11}$.
\\
\\

Some of the above aspects are exemplified  when one considers the application 
of the NPT to the problem of the QAHO  as discussed below.  
\\
\\
{\it The QAHO problem and the NPT }
                                                                        \\
                                                                         \\
  The Hamiltonian for the QAHO (in one space-dimension) is given by the 
following expression :
\begin{equation} 
H~=~\frac {1}{2} p^{2}~+~\frac {1}{2} gx^{2}~+~\lambda x^{2k}~;~k~=~ 2,~3,~4,.
.....
\end{equation}
In the above equation, $~p~=~-i\frac {\partial}{\partial x}$
~and $~g,~\lambda~$ are real,  positive parameters. (The  units are chosen 
such that $~m~=~1 ~and~\hbar~=~1$).
\\                                                                  
\\
The NPT can be developed by recognizing that for $~\lambda~=~0$, the system 
Hamiltonian reduces to that of the simple harmonic oscillator and is exactly 
solvable. Hence, it is natural to choose the anharmonic term as the 
perturbation-correction  and express the Hamiltonian in the form given by 
equation (1), where $~\lambda H_{I}~\equiv~\lambda x^{2k}$ and $~H_{s}~
\equiv~\frac {1}{2}p^{2}~+~\frac {1}{2}gx^{2}$.   
\\
\\
In order to apply  the NPT in the above formulation, it becomes apparent that
the coupling strength be restricted to small values : $~\lambda <~1~$. 
However, it was discovered that the  NPT-series {\it fails} to converge, even 
if the above restriction is imposed. Specifically, it was demonstrated by the 
pioneering work by Bender and Wu$^{12}$  who studied the large-order 
behaviour
of the NPT-series for the quartic-AHO and came to the conclusion that the 
co-efficient of  $~\lambda ^{n}~$ grows as n! for any value of $~\lambda~$
even  when $~\lambda \rightarrow~0~!~$ It was 
subsequently  established$^{13}$  that the factorial-growth of the
series worsened
further for the cases of  higher-anharmonicity.  Even prior to the
work  under refs.[12,13], it was shown by Lam$^{14}$  that the
 development of
the {\it perturbation theory as a power-series in $~\lambda~$ has  vanishing} 
{\it radius of convergence due to the occurence of an essential singularity at}
$\lambda~=~0$.\\

  Since then, a lot of investigation into  the large-order behaviour of the  
NPT  for the QAHO have been carried out and the results have found mention 
in texts$^{15}$. From these investigations it has become clear that
the divergence of the NPT could be ultimately traced to the failure of the 
essential condition expressed in eqn.(2) {\it due to the eventual dominance}
{\it  of the perturbation-term  $\lambda x^{2k}$}{ \it over the unperturbed }
{\it part for large amplitude of oscillation}.\\

In this context, it has now come to be recognized that the problem of 
divergence of the NPT may not be restricted to the case of the QAHO  alone.
 For example ,in the case of quantum electrodynamics (QED), the possibility of
 failure of the  NPT was conjectured by Dyson$^{16}$  much earlier. This has 
led to the conjecture that the problem of divergence  of the NPT could be 
{\it generic} and the NPT-series may at most be an {\it asymptotic series}. A 
crude estimate of the optimal number, $~n_{0}~$ of terms in the NPT-series 
beyond which, the subsequent terms start diverging could be obtained from the 
criteria :     
\begin{equation} 
\lambda^{n_{0}}R_{n_{0}}~\approx~O(1)~,  
\end{equation}
where, it is assumed that the NPT-series  (for the observable, say, energy E)
 is represented by    
\begin{equation} 
E~=~ \sum_{n}\lambda^{n}R_{n},  
\end{equation}
where, $~\lambda~$ is the coupling strength. Then, assuming factorial growth,
 $~R_{n}~\approx~ n!~$ for large n, and using the Stirling-approximation, one
obtains from eqn.(14), the desired estimate :
\begin{equation}  
n_{0}~\approx~O(\frac {1}{\lambda}).
\end{equation}
It is clear, therefore, that the relevant coupling-strength must be
sufficiently small in order that the NPT-series in eqn.(15) could
make sense even as an asymptotic, divergent series !  We elaborate on
these aspects further in the following while discussing 
{\it summability} methods. ( It may be noted, in the context, that
eqn.(16) may provide the rationale why the QED-perturbation series
can still be trusted as an asymptotic series since the former starts
diverging only after $~\sim ~137$ terms since
$~\lambda_{QED}~\approx~1/137~$). 
\\ 
\\
It may be relevant to note also the following aspects while discussing 
application of the NPT to the QAHO-problem :
\\
\\
(i)  For $~g~=~0~$ (see, eqn.(13)), the an-harmonic term :
 $~\lambda x^{2k}~$ can not be treated as a perturbation  to the
 remaining part of the Hamiltonian,  which is  the kinetic-energy term 
 (KET). This is because of the totally different nature of the
 resultant spectrum- the KET generates a {\it continuous} spectrum
 whereas the addition of the an-harmonic term changes this to a
 discrete one.
\\
\\
(ii)  Arguments based upon scaling  and dimensional  analysis$^{17}$  
can be used  for the  QAHO-problem  to show that the latter is
essentially  a single parametric-problem even though there are two
couplings : $~g~$ and $~\lambda~$ occuring in the Hamiltonian. Thus,
for the {\it quartic}-AHO, the following scaling law holds$^{17}$  :
\begin{equation} 
E~(g,~\lambda)~=~\lambda^{\frac{1}{3}}~E(g\lambda^{-\frac{2}{3}},~1)~,  
\end{equation}     
\begin{equation} 
\psi~(x~;~ g,~\lambda)~=~\psi~(~x~\lambda^{\frac{1}{6}}~; g~\lambda^{-\frac{2}
{3}},~1~)~,  
\end{equation}
where E~=~ energy and $~\psi~=~$ wave function of the system.
\\                                                          
\\
(iii) The above scaling-property together with re-parameterization
techniques can be effectively used$^{18}$ to develop perturbation theory 
for {\it strong coupling} regime, i.e., for $~\lambda>>~1$ since the
effective single-coupling for the problem becomes $~g\lambda^{-2/3}~$
(see, eqns.(17-18). This is useful since the strong-coupling
perturbation theory may be used as a complement to the usual
small-coupling RS-series in studying smooth transition across
$~\lambda~=~1$.
\\   
\\
(iv) It may also be relevant to note that recently a
novel-perturbation theory has been proposed by Turbiner$^{17}$ where the 
Schr$\ddot{o}$dinger-equation is first converted to the
Riccati-equation  in the variable $~y(x)~\equiv~\frac{-d[ln
  \psi(x)]}{dx}$, ( where $~\psi(x)~=~$wave-function) and then 
developing  a perturbative expansion for $~y(x)~$ through a recursive
and iterative procedure. The resultant series for the energy and the
wave function are characteristically different from those obtained in
the usual NPT and are claimed to be rapidly convergent.
\\
\\
We close this sub-section noting that the difficulties encountered
with the NPT as applied to the QAHO-problem have led to several
alternative approaches, as listed earlier which are discussed in
subsequent sections. We next turn our attention to the method based
upon variational-approximation.\\  

( 2 ) {\it  The~Variational-Method~of~Approximation ~(VMA) }
                                                             \\
\\
The VMA is a powerful method  in obtaining the approximate  eigen-values
 and eigen-states of observables in QT. Like perturbation theory (PT), 
 the application of the method has a long history- initial application 
 of the method being ascribed to  Lord Rayleigh$^{19}$  and to 
W.Ritz$^{20}$.  The VMA becomes especially suitable when perturbation theory
 (PT) fails or becomes inapplicable to the considered problem.  VMA is 
 also used to test results based upon PT and for analysis of stability 
 of the system considered.
\\
\\
The essential ingredient of the method is based upon the 
 variational theorem (Theorem-I), which provides an {\it upper bound} for
the ground state energy of the system :
\begin{equation} 
\frac{<~\psi~|~H~|\psi~>}{< ~\psi|\psi~>}~\equiv~E~[\psi]~\geq~E_{0}~,  
\end{equation}
where, $~|\psi >~$ is any {\it arbitrary}, normalisable state;  
{\it H} is the  Hamiltonian, and $~E_{0}~$ is the ground-state energy of the 
system. ( In the above equation, we have used the
standard Dirac-notation for expectation values and also denoted by
$E[\psi]$, the energy- {\it funtional}). The {\it proof} of the
theorem is based upon straight forward application of
eigen-function  expansion method in  QT and can be found in any
standard texts$^{3}$.
\\                                                                       \\
\\
A  second theorem (Theorem-II)  found useful applications, states that the
state $|~\psi~>$, which renders the energy-functional $~E[\psi~]$
 {\it stationary} i.e.  $~\delta E~[\psi]~=~0$, satisfies
the stationary Schr$\ddot{o}$dinger equation : 
$H|~\psi>~=~E|~\psi>$.  Again  the
proof  can be found in standard texts$^{3}$  and is therefore, omitted.
\\
\\
For practical application to obtain a close approximation,  one follws
the Ritz-method$^{20}$, which consists in specifying the {\it trial}
-state $~|~\psi~>~$ as a {\it function} of one or more free-parameters 
: $\{\alpha_{i}\}$ and then {\it minimising}   the energy functional
 $~E[\psi~]$ with respect to these parameters : 
\begin{equation}
\partial{E}[\psi]/\partial\alpha_{i}~=~0~;~~\partial^{2}E[\psi]/\partial
\alpha_{i}^{2}~>~0.
\end{equation}
\\
\\
Thus, the Ritz-method enables one to obtain a least-upper bound (LUB), 
for a given choice of the trial-state, which is then the closest
approximation to the ground-state energy for that choice. Some other
aspects of the VMA are discussed below :
\\
\\
(a) The equality sign in eqn.(19) holds if (and only if) the
trial-state, $|~\psi~>~$ coincides with the true ground state,
$|~\psi_{0}~>$ of the system. (b) An error of $~1^{st}~$ order, i.e.,
$\sim~O(\epsilon)$ in the choice of the trial-state gives rise to an
error of $2^{nd}$ order, i.e., $\sim~O(\epsilon^{2})$ in the computed
value of the energy, which means that the energy is more accurately
determined by the VMA than the wave function and consequently, the
over-all accuracy of the method crucially depends on the ingenuity and 
insight in the choice of the trial-state, consistent with the physical 
boundary conditions for the system. (c) The generalization of the
Rayleigh-Ritz method of the VMA to obtain estimates/approximation for
the higher excited states can be achieved by several methods. Some of
these methods are outlined below : 
\\
\\
(i) A tower of trial states,
$~|~\psi_{\alpha}~>~;~\alpha~=~1,~2,~3......,$ approximating the
higher excited states of the system may be constructed with the
requirement that these are mutually orthogonal to each other, as well
as, orthogonal to the variational-ground-state. The construction can
be achieved by any suitable method of orthogonalisation, such as the
Schmidt-method$^{21}$. Denote by $E_{\alpha}$, the energy-functional for the 
$\alpha$-th state {\it minimized} with respect to its
free-parameters and then arranged as an decreasing sequence, i.e., 
\begin{equation}
E_{\alpha}~\equiv~\Big[~\frac{<|\psi_{\alpha}~|H|~\psi_{\alpha}~>}{<|\psi_
{\alpha}~|~\psi_{\alpha}~>}\Big]_{min}~;~\alpha~=~1,~2,~3,.........
\end{equation}
The resulting sequence, $E~<~E_{1}~<~E_{2}~<~E_{3}~.......,$ then
represents the varitational approximation to the energy of the higher
excited states in the order shown. 
\\
\\
The other standard method to deal with the excited states within VMA
is known as$^{22}$ the ``method of linear variational approximation
(LVA)''. This consists of the choice of the trial-state as a 
{\it linear}
 superposition of a suitably chosen set of eigen-functions
consistent with the boundary conditions and otherwise appropriate for
the system, as follows :
\begin{equation}
|~\psi~>~=~\sum_{n}c_{n}~|~u_{n}~>~;~~\sum_{n}|u_{n}|^{2}~=~1~;~~<~u_{n}~|~n_{m}~>~=~\delta_{nm}.
\end{equation}
The energy functional as given in eqn.(19) computed with this
trial-state leads to the following set of equations after minimization 
with respect to the coefficients $c_{n}$ (see, eqn.(22)) :
\begin{equation}
\sum_{n}c_{n}^{*}~(H_{nm}~-~E\delta_{nm})~=~0~;~~m~=~1,~2,~3,.....
\end{equation}
\\
\\
Condition for existence of non-trivial solutions of the above set of
equation, is given by the vanishing of the secular determinant :
\begin{equation}
det~(~H_{nm}~-~E\delta_{nm}~)~=~0
\end{equation}
\\
For practical purpose, the determinant has to be truncated at some
finite order `$k$'. The solution of eqn.(24) then provides, in
general, $k$-roots for the energy $E$, {\it which} then correspond to
the approximate energies of  the first `$k-excited~ states$'. It is
obvious that the above method becomes efficient in practice when the
truncation-error (due to the chosen finite value of $k$) becomes small 
and further that the spectrum {\it stabilizes} (with increasing value
of $k$). Clearly, therefore, {\it intelligent} choice of the
basis-states, $|~u_{n}~>$ becomes crucial in the above method. 
\\
\\
For certain problems, a variant of the above formalism is often called 
for when it becomes more convenient to work with a  non-orthogonal basis
set. In that case, eqn.(24) is replaced by the following equation :
\begin{equation}
det~(~H_{nm}~-~ES_{nm}~)~=~0,
\end{equation}
where, now $S_{nm}~\equiv~<~u_{n}~|~u_{m}~>$, is the overlap function.
\\
The above basic formulation of VMA can be improved in several ways some
of which, are described in the following.
\\
\\
{\it (a) Improved choice of the trial-states}\\

This obvious generalization that has to be tried first for a given
problem requires the choice of more realistic trial-states, which
often involve large number of variation-parameters. These lead in
general, to results that are more accurate. The main {\it limitation}
of this approach is related to the fact that there is no systematic
and standard method to follow. In the context of the QAHO, several
trial-states have been tried with varying degree of success. A sample
list is given under ref.[23]. 
\\
\\
{\it (b) Obtaining lower bounds on the spectrum}\\

The idea in this approach is to obtain {\it lower bounds} to the
energy eigen-values in addition to the upper bound provided by the
standard VMA. Carrying out the optimization of both the lower- and
upper- bounds then leads more accurately to the actual
eigen-value. Lower bounds are obtained by employing standard
inequalities and other techniques. In the context of the QAHO some
important work, in this direction, are listed in ref.[24].\\

{\it (c) The Gaussian approximation}\\

This method employs the choice of a Gaussian as the trial state for a
general problem. This approach relies upon the general 
{\it  applicability of the theory of small-oscillations to any}
{\it arbitrary  potential/interaction possessing a stable-minimum}. The
variation parameters are usually chosen to be the `width' and the
location of the peak of Gaussian trial-state. For example, in one
space-dimension, the trial wave-function for the ground state could be 
of the generic form, $\psi(x)~=~A~exp[~\frac{-(x-x_{0})^{2}}{\sigma^{2}}]$,
where `$x_{0}$' and `$\sigma$' are the free-parameters and `{\it A}' is a
normalization constant. This method has a long-history mainly because it has
 been `rediscovered' and `refined' time-and-again. It has been applied to a
veriety of problems in QT including quantum field theory. We refer to the work
of Stevenson and collaborators$^{25}$, which also provides guide to the 
earlier literature.\\

{\it (d) The method of minimum enrgy variance (MEV)}\\

In stead of the Hamiltonian, {\it H }itself, upper bounds can be derived, in 
principle, for any {\it arbitrary} function $f(H)$ by the
generalization of the standard Rayleigh-Ritz method. The
generalization when the chosen function is considered as the 
{\it  variance} of the Hamiltonian, has been termed$^{26}$ 
accordingly as ``method of minimum energy variance (MEV)''. The MEV
has been shown$^{26}$ to result in improved estimate of the trial-state
as well as, the energy eigen-value. In particular, it has been
demonstrated that the estimates go beyond the simple Gaussian
approximation$^{26}$ . 
\\
\\
In addition to the above, there are other methods, which combine the
variation method with other approximation methods, e.g., perturbation
theory, super symmetry, the Hill-determinant method etc. in order to
improve the results from the VMA alone. Some of these latter-schemes
are {\it separately} discussed in the following subsections.
\\
\\
( 3 ) {\it  The Hill-determinant and related  methods}\\

In this method, the expectation value of the Hamiltonian of the problem is 
computed in the trial state function, which is expanded in terms of  of a 
suitably chosen set of basis functions (which need not be
orthogonal) with unknown coefficients. The condition for the determination of 
these coefficients translates into the condition of vanishing of the  
 resulting secular determinant (Hill-determinant). This latter equation is 
then solved  numerically  for the approximants  of the Hill-determinant 
obtained  by truncation. More specifically, the trial-state proposed by  the
 authors$^{27}$, who pioneered this approach, is given as :\\

\begin{equation}
\psi(x)~=~exp~( -x^{2}/2)\sum_{n}c_{n}~x^{2n}.
  \end{equation}
Substitution of this trial-state in the Schr$\ddot{o}$dinger equation for the 
quartic-AHO then leads to the following three-term difference equation for the
coefficients, $c_{n}$'s :\\

\begin{equation}
2~(n+1)~(2n+1)~c_{n-1}~+~(E-1-4n)~c_{n}~-~\lambda~c_{n-2}~=~0 ,
  \end{equation}
where, $\lambda$ is the an-harmonic coupling strength and  E = energy of the 
system. \\( Alternatively, due to {\it theorem-II} mentioned earlier, the same
result is obtained after minimisation of the energy functional, $E~[\psi]$ 
with respect to the $c_{n}^{*}$ treating the latter as the  variation 
parameters.)
The condition of existence of solutions of the eqn.(27) then requires that 
the corresponding (infinite-dimensional) Hill-determinant D of the problem
vanishes, D~=~0.(The explicit form of D is given in ref.[27]). The 
authors 
of ref.[27] numerically solve the truncated approximant, $D_{n}~=~0$, by 
noting that these satisfy a three-term recursion relation : 
\\
\begin{equation}
D_{n}~-~(E-1-4n)~D_{n-1}~+~16\lambda n(n-1/2)(n-1)(n-3/2)D_{n-3}~=~0.
\end{equation}
The solution, $D_{n}~=~0$, then leads, after ensuring convergence and 
stability, to n-roots for the energy E, which corresponds to the energy of the
 first n- levels of the even-parity states, when rearranged as an increasing 
sequence. ( The energy levels of states with odd-parity can similarly be 
obtained by substituting, $n\rightarrow n+1/2$).\\

Obviously, therefore, stability, convergence, the computational time and 
effort needed, are the main criteria for the practical implementation of this 
method. The authors of ref.[27]  obtained, by this method the 
eigen-values of the $\lambda~x^{2k}$-anharmonic oscillators separately for
$ k ~=~ 2,~ 3~ and~ 4$. The results compare well with the accurate numerical 
results obtained by other methods.\\

However, several authors$^{28,29}$ have later pointed out the limitation of 
the approach in ref.[27]. One main problem concerns the normalizability of the
 resultant eigen-functions. Another problem is the failure of the method to 
reproduce the so called ``terminated (polynomial)solutions''(later known as 
the ``quasi-exact solutions''$^{30}$). Still another  criticism$^{28}$ of the 
original method$^{27}$ is the failure to reproduce correct  spectral-properties
when applied to a general an-harmonic oscillator having arbitrary couplings 
for the quadratic-, cubic- and quartic an-harmonic terms. Several 
modifications, involving improved choice of the variational trial-state, have 
been suggested$^{28,29}$. For example, the following substitution of the 
exponential convergence factor (see, eqn.(26)), has been proposed :\\
\begin{equation}
exp~(~-x^{2}/2~)~\rightarrow~ exp~(~-\alpha~x^{4}~+~\beta~X^{2}~) ,
\end{equation}  \\

where, the constants $\alpha,\beta$ are either treated as {\it variation} 
parameters or determined otherwise. Even though the above `{\it modified}'
Hill-determinant method removes certain limitations of the original 
approach$^{27}$, there still remain certain controversy$^{31}$ 
regarding  correct reproduction of the eigen-values and `moments'.\\

A related approach is the so-called ``{\it variational sturmain approximation}
''. The concept of the {\it sturmain basis functions} was introduced$^{32,33}$
 as a non-perturbative approximation scheme for solving the stationary 
Schr$\ddot{o}$dinger equation in the context of several potentials considered 
in atomic- and molecular physics problems. This method involves, akin to the 
Hill-determinant method, the simultaneous solution of an infinite number of 
algebraic equation and, therefore, to the vanishing of the secular determinant
as the necessary condition. For numerical solution it is necessary to solve
the finite order approximant to the secular determinant and thereafter ensure
convergence  and stability in respect of chosen size of the latter. Recently,
it has been shown$^{34}$ that the basic sturmain approximation can be 
variationally improved to yield better results. The authors in ref.[34] have 
obtained accurate results for the quartic-AHO and the exponential potential.
For details, ref.[34] can be consulted, which also provides guide to the 
earlier literature.\\    
  
( 4 ) {\it  Combination of Variational and Perturbation techniques }\\

In order to retain the merits but to over come the limitations separately of
 perturbation theory and variation methods, several approaches have been 
proposed, where both the methods have been used in a single formulation. 
Various authors have worked using this combined approach, including the
following: Halliday and Suranyi$^{35}$, W.Caswell$^{36}$,J.Killingbeck$^{37}$,
Hsue and Chern$^{38}$, Feynman and  Kleinert$^{39}$ and  collaborators$^{40}$,
Patnaik$^{41}$ and Rath$^{42}$. The review of each individual  work as 
mentioned above falls beyond the scope of the present thesis. However, it may 
be relevant here to note certain common features of the techniques used by 
these authors and the consequent achievements attained.   \\

In most of the approaches listed above the {\it basic Hamiltonian of the system
}{\it is altered by the addition and subtraction of terms involving certain}
{\it parameters}. These parameters are {\it apriori} arbitrary but fixed later
by the variational  minimisation of energy or by imposing other constraints,
which simplify  computation. {\it Perturbation techniques can then be applied}
{\it with} {\it redefinition of the unperturbed Hamiltonian and the 
perturbation}{\it correction.} The resulting  perturbation series is often 
shown$^{35,36,41,42}$ to be convergent.{\it Thus, the problem of convergence 
of the naive perturbation theory as well as the absence of of a built-in 
mechanism of systematic improvement in the VMA,}  are overcome to a large 
extent in these hybrid method, which may be called {\it variation-perturbation
 method (VPM)}. Apart from the above common underlying feature, each approach 
differs in detail, which may be found in the individual reference cited above.
\\

(5) {\it  The operator Method of Approximation (OMA)}\\

The OMA has been pioneered  by Feranchuk and Komarov$^{43}$. The 
application 
of the OMA  to the QAHO-problem  and to several other problems has been 
described in ref.[44], which also provides a guide to earlier works of 
the authors. The basic idea of this method is to formulate the problem in the 
Fock-space of operators, instead of working with the co-ordinates and momenta.

 The relevant transformation is provided by the following relations :

\begin{equation}
x~=~( a~+~a^{\dagger})/\sqrt{2\omega}~;~ p~=~i~( a^{\dagger}~-~a )/\sqrt{(2/
\omega)}, ~[~a,a^{\dagger~}]~=~1
\end{equation}  \\
where, $a,a^{\dagger}$ are the annihilation- and creation-operators 
respectively  and `$\omega$' is a positive real number having the significance
 of the frequency of the associated simple-harmonic oscillator.
The operators satisfy the standard commutation relation as shown in eqn.(30)
in order to be compatible with the equal-time commutation relation, 
$[x,p]~=~i$.\\

The development of the OMA then consists of the following steps :\\

(i) The Hamiltonian of the system is first re-expressed in terms of the Fock-
space  operators ( by the help of eqn.(30)),\\

(ii) The unperturbed Hamiltonian, $H_{0}$ is chosen by clubbing 
{\it all the  diagonal} terms in {\it H} ( i.e., those involving polynomials of
the number operator $a^{\dagger}a$),\\

(iii) The remaining off-diagonal terms are together chosen as the 
perturbation, V,\\

(iv) A modified perturbation theory is then developed using the above choice
of the unperturbed Hamiltonian and the perturbation term,\\

(v)  Finally, the arbitrary parameter `$\omega$' is fixed by variational 
minimization of the energy. The  procedure is indicated as shown below :\\

\begin{equation}
H( x,p )~\rightarrow~H( a,a^{\dagger})~=~H_{0}( a^{\dagger}a) ~+~V( a,a^{
\dagger})
\end{equation}  \\

To illustrate the method consider the QAHO problem with the Hamiltonian given 
by

\begin{equation}
H( x,p )~=~\frac{1}{2}p^{2}~+~\frac{1}{2}x^{2}~+~\lambda~x^{4}~ ,~\lambda~>~0.
\end{equation}

Then the OMA-unperturbed Hamiltonian $H_{0}$ and the perturbation term V, as
defined in eqn.(31) are given by the following expression$^{44}$ :

\begin{equation}
H_{0}~=~ (1+2N)(\omega^{2}+1)/4\omega + (3\lambda/4\omega^{2})(1+2N+2N^{2})~,
\end{equation}  
\\
\begin{equation}
V~=~-[(\omega^{2}-1)/4\omega]((a^{\dagger})^{2}+a^{2})+(\lambda/4\omega^{2})[2
(a^{\dagger})^{2}(2N+3)+(2N+3)2(a^{2})+(a^{\dagger})^{4}\\+a^{4}]
\end{equation}

In the above equations, N stands for the number operator, $N~=~a^{\dagger}a$.
The energy eigen-value corresponding to $H_{0}$ is easily obtained :

\begin{equation}
E^{(0)}_{n}~=~(n+1/2)[ (\omega^{2}+1)/2\omega]+(3\lambda/8\omega^{2})
(n^{2}+n+1/2)
\end{equation}

Minimization of the energy as given in eqn.(35) above then leads$^{44}$ to 
the following equation, for the frequency `$\omega$':

\begin{equation}
\omega^{3}-\omega-6\lambda(1+2n+2n^{2})/(1+2n)~=~0,
\end{equation}

which determines this parameter as a function of the excitation level `n' and
 the strength of the anharmonic coupling  $\lambda$. Substitution of the 
solution of eqn.(36) in eqn.(35) then leads to the leading-order 
approximation of the energy levels, which are found to be {\it accurate within}
{\it  a few percent} over the entire range of `n'  and $\lambda$. 
Furthermore, this leading order result can be systematically improved order-by
-order by developing the RS-perturbation theory with the perturbation term 
chosen to be V as given in eqn.(34). Besides, this perturbation series has
been  argued$^{44}$  to be {\it convergent}  unlike the na$\ddot{i}$ve 
perturbation theory. Similar results have been obtained$^{44}$ for the cases 
of higher an-harmonicity, double-well oscillators and a variety of other
 systems.\\

The main limitation of the model appears to be the implementation of the basic
{\it ansatz} defining the split given in eqn.(31), particularly for 
non-polynomial interactions, interaction representing analytic functions 
expressible in infinite series, problems in higher dimensions, quantum field 
theory etc. In such cases, {\it additional} assumptions/methods/skills have to
be employed$^{44}$.\\

(6) {\it  The JWKB approximation method}\\

This method is quite old and popular due to its {\it model-independent}
nature- it can be applied to any {\it arbitrary, smooth potential}. The 
formalism has been dealt at length in texts$^{45}$. Therefore, it will not be 
repeated here. We note, however, the main features of the approximation scheme
 as applied to the discrete bound state problems in QT. The relevant central 
formula for the above purpose is the so-called {\it JWKB quantization rule} :

\begin{equation}
\int^{b}_{a}~k(x)~dx~=~( n+\frac{1}{2})~\pi~,~n~=~0,~1,~2,~3......
\end{equation}

where,$~k(x)~=~{2m}/{\hbar^{2}}~( E~-~V(x))~$; a ,b~ are a set of adjacent 
``turning points'' obtained by solving the equation, $k(x)~=~0~$ and other
terms have standard meaning. For those cases where the above integral can be 
evaluated, eqn.(37) can be expressed as :

\begin{equation}
f( E,\lambda,g...)~=~(n+\frac{1}{2})~\pi,
\end{equation}  

where, the l.h.s. of eqn.(38) represents the value of the integral as a 
function of the energy (E) and coupling-strengths$ ~(\lambda,g....)~$ occuring
 in the potential V(x).\\  

The determination of energy E as a function of the other parameters 
$(n,\lambda,g...)$ therefore, requires the {\it inversion } of eqn.(38).
This task is far from simple for most cases of interest since the inversion 
can not be achieved in closed, analytic form except in a few known cases 
( which, also happen to be exactly solvable by standard methods ). However, 
eqn.(38) can be inverted by numerical-methods to the desired order of 
accuracy. The complexity of the procedure increases for the cases, for which
the WKBJ-integral can not be obtained in closed form. Similarly, the cases 
involving multiple ( closely spaced ) turning points$^{46}$ and cases 
requiring higher order of approximation in the WKBJ-series, ( e.g. to deal
with rapidly varying potentials) become increasingly difficult to implement. 
Nevertheless, several useful information /in sight about the energy levels can
be obtained in limiting cases of the small-coupling regime, the strong-
coupling and/or the large- n limits. Application to the case of the QAHO for
studying the large- order behaviour of the na$\ddot{i}$ve perturbation theory
was carried out in ref.[47]. The case of the DWO has been dealt in ref.[48].
The connection  between  multi-instanton solutions  in the path-integral
formalism, the JWKB method and large-order behaviour of perturbation theory 
has been discussed by several authors. We refer to Garg$^{46}$ who discusses 
the issue and lists earlier references.\\

(7) {\it Approximation methods based on quantum-canonical transformation (QCT)}\\

This is a powerful non-perturbative method, which has been successfully 
applied in QT, particularly in many-body systems exhibiting collective- and 
co-operative phenomena, e.g. superconductivity and 
super-fluidity$^{49-51}$.
The method had been primarily expounded by Bogoliubov$^{52}$ and  hence
 more  familiarly  known  as  the ``Bogoliubov-Transformation''. These 
transformations
connect the Hillbert-space of the {\it interacting} system to that of the 
interaction-free case, while {\it preserving the canonical structure of the} 
{ \it basic ( equal-time) commutation rules.} The method is particularly
 useful in variational studies where the ansatz for the interacting vacuum 
state (IVS) generated through QCT can be tested to dynamically establish the 
{\it stable} ground state of the system. The basic formalism of the method  
can be illustrated, for a Bosonic-system with single-degree of freedom, 
through the following example.
 Consider such a system, which is described, in absence of interaction  by the
 free-field annihilation- and creation operators, $ a, a^{\dagger}$. The 
dynamic `field' $~\phi~$ and the canonical momentum~ p~ are parameterized in 
terms of these operators as : $\phi~=~( a~+~a^{\dagger})/\sqrt{2}~$ and 
$~p~=~i~( a^{\dagger}~-~a~)/\sqrt{2}$ and satisfy the standard equal-time 
commutation relation :[$~\phi~,~p~]~=~i$. In presence of interaction, it is 
natural to assume that the interaction- free operators, $~a,a^{\dagger}~$ 
change to the {\it interacting} ones , $~b,b^{\dagger}~$. Similarly, let the 
vacuum states for the corresponding cases be denoted as :$~|~0>~$ and 
$~|vac>~$respectively. By definition, these satisfy : $~a~|~0>=~0$, 
$~b~|vac> =~0$. The canonical- commutation rules  satisfied by both the set of
 operators are identical by physical requirement, i.e.,
\\                 
\begin{equation}
 [~ a,~a^{\dagger}~] ~=[~ b,~b^{\dagger}~] = 1
\end{equation}
\\
The  general Bogoliubov-transformation relating the two sets of operators 
is given by  
\\
\begin{eqnarray}
b~=~a~ \cosh(\alpha) - a^{\dagger}\sinh (\alpha)~,
\nonumber\\
b^{\dagger}~=~a^{\dagger}\cosh(\alpha) - a~  \sinh (\alpha)~,
\end{eqnarray}
\\
where, the parameter $\alpha$ is real and a {\it priori} arbitrary but the 
same can be determined dynamically through variational minimization of energy 
and/or by other physical requirements. The corresponding unitary 
transformation, which relates the two vacua is given by,
\\
\begin{equation}
|~vac~>~ =~ exp~ [~1/2~ \tanh~ (\alpha)~ (a^{\dagger} a^{\dagger}~ -~ a~ a)~]~
 |~0~>~,
\end{equation}
\\
such that the parameter, $\alpha~$ embodies the {\it full} effects of 
interaction : \\$~|vac>~\rightarrow ~|~0>$ as $~\alpha~\rightarrow~0$. This 
vacuum state representing the situation in presence of interaction can be used
 as a trial state for a given Hamiltonian, such as given by eqn.(31) in 
order to study the stability and to compute the ground state energy. The 
energy for the excited-states can then be determined by computing the 
expectation value: $~ <~n~|H(a,a^{\dagger})|~n~>$, where the state $~|~n~>~$ 
is given by the standard formula : $~|~n~>~=~[(b^{\dagger})^{n}/\sqrt{n!}]~
|vac>$.\\

The formalism has been employed in  refs.[53,54] for the QAHO and the DWO 
problems. In the context of $~\lambda \phi^{4}~$ quantum-field theory, QCT 
has been employed in refs.[55,56,57]. QCT also provides important in-sight into  the structure of the interacting vacuum state$^{55,56,58,59}$.\\

The limitations$^{60}$ of the method appear to be those of variational methods
 as discussed earlier.  
\\
\\
(8) {\it  Approximation methods based on super symmetric quantum mechanics}
\\
\\
The method of super symmetric quantum mechanics ( SUSYQM ) has been dealt in 
texts and several review articles. As a representative text$^{6}$, may
 be 
consulted, which provides guidance to earlier literature. As has been noted 
earlier, the property of {\it shape-invariance} in SUSYQM$^{6}$  has
 been 
successfully employed to considerably extend the class of exactly solvable
potentials in QT. The method of SUSYQM can be profitably applied to improve 
and extend the scope of known approximation methods, e.g. perturbation 
theory, variational method and JWKB-approximation scheme. In the context of 
non-relativistic QT, several exact results follow in SUSYQM valid for 
{\it partner potentials}$^{6}$  such as the property of 
{\it iso-spectrality} {\it level degeneracy and positivity}. These exact 
results provide a testing  ground for approximation schemes applied to such
 potentials.\\

The SUSY-improved perturbation theory starts from an initial guess of the 
ground state wave function, which can be based, for example, on a realistic
variational {\it ansatz}. {\it The super-potential} corresponding to this 
trial-ground state wave function can then be constructed by computing the 
logarithmic derivative of the latter. The unperturbed Hamiltonian is then
chosen to be the one obtained from the super potential and the perturbation
correction is taken to be the difference of the original Hamiltonian and the
unperturbed one. The development of the RS-perturbation theory then becomes
straightforword. For illustration of this method, ref.[6] can be consulted.\\

The SUSY- based  JWKB approximation provides the following modification of the
quantization-rule :
\\
 \begin{equation}
 \int^{b}_{a}\sqrt{2m~[ E^{(1)}_{n}~-~W^{2}(x)]}~dx~=~n\pi\hbar~;~n~=~0,~1,~2,
~3,...
\end{equation}    
\\
where, W is the {\it super-potential} and $E^{(1)}_{n}$ is the n-th 
energy-level of the $H_{1}$, which is one of the partner-Hamiltonians.
Similar expression for the energy levels of the other partner-Hamiltonian,
$H_{2}$ is given by,  
\\
\begin{equation}
 \int^{b}_{a}\sqrt{2m~[ E^{(2)}_{n}~-~W^{2}(x)]}~dx~=~(n~+~1)\pi\hbar~;
~n~=~0,~1,~2,~3,...
\end{equation}    
\\
 It may be seen from the above equations that the exact result on the 
iso-spectrality / level-degeneracy of the partner potentials, is respected
in the SUSY-JWKB quantization condition. Moreover, the formulae are valid for
{\it all} values of `n'  rather than for large- n as in the case of the 
conventional formulation. The SUSY-JWKB formula is demonstrated$^{6}$ to
 yield
the {\it exact} results for shape-invariant potentials and results with 
improved accuracy for other known cases.\\

In the context of the QAHO/DWO, SUSYQM- methods have been investigated by 
several authors. A sample list is provided in refs.[61-71].  
\\
\\
(9) {\it  The methods based upon re-summation techniques}
\\
\\
The generic problem of divergence of the na$\ddot{i}$ve perturbation series
(NPS) has led to the various re-summation techniques in order to extract 
some meaningful result out of the former. We will mention briefly the two
 popular re-summation techniques described in the literature, namely (i)
 the {\it Borel-re-summation} and (ii) the {\it Pad$\acute{e}$-approximation} 
in the  following.\\
\\
{\it (i) The Borel re-summation method}
\\
\\
The method of Borel re-summation$^{72}$ is a well-studied 
technique$^{73,74}$, which permits, {\it under suitable conditions}, 
re-summation upto the 
optimal number of terms of an asymptotic power series exhibiting  the growth
at large-order `n' as n!. Consider such a series :
\\
\begin{equation}
F(g)~=~\sum_{n}f_{n}~g^{n}~, 
\end{equation}     
\\
where, $g$ is a real parameter ( `coupling-strength' ) and $f_{n}~\sim~n!$,
for large- n. Then, the {\it associated} Borel re-summed series is defined as
\\
\begin{equation}
B(z)~=~\sum_{n}f_{n}~z^{n}~/n!~, 
\end{equation}     
\\
This latter series is expected to have at least a {\it finite} radius of 
convergence by construction. The orginal series, (44) can be recovered,
at least formally, from eqn.(44) as
\\
\begin{equation}
F(g)~=~(1/g)~\int^{\infty}_{0}exp~(-z/g)B(z)~dz~, 
\end{equation}     
\\
by noting that $~~\int^{\infty}_{0}exp~(-z/g)z^{n}~dz~=~n!g^{n+1}$, provided
that the integral on the r.h.s. of eqn.(46) can be {\it uniquely} 
evaluated. The integral representation, eqn.(46) is known as the 
{\it Borel-transform} and this represents, within the radius of convergence
of the series for B(z), the sum of the asymptotic perturbation series, 
eqn.(44). In order to be useful, the detailed knowledge of the analytic 
structure of B(z) in the complex- z (Borel-)plane is necessary. If the 
singularities of B(z) lie off the positive real axis then the integral,
eqn.(45) can be evaluated by counter-integral after appropriate distortion
of the contour around those singularities.\\

However, if one ( or more) singularity in the Borel-plane lies on the 
real-axis, then the evaluation by contour integral becomes ambiguous; the 
integral representation, eqn.(46) no longer exists and, hence the case is
not Borel re-summable. The other requirement is that the radius of 
convergence of B(z) must be larger than ( or equal to) the magnitude of the 
coupling-strength, `$g$' in order that the re-summation is relevant.\\

There is another situation, which invalidates the re-summation. This occurs
when the terms in the original series {\it all} have equal phase. In that case
 the Borel-transform, B(z) itself grows too fast at infinity. Consequently,
the Borel-integral, eqn.(46) can not be evaluated. There may be other 
situations in addition to those already stated, when the singularity structure
of B(z) prevents Borel re-summation. For the details, ref.[75] may be 
consulted.\\

Finally, it may be stressed that the method fails when the coupling becomes
large,$~|~g~|~\geq~1$ because then the original perturbation series becomes 
invalid  due to the divergence of the {\it individual} terms. {\it Thus, the}
{\it re-summation techniques are necessarily restricted to the small coupling}
 {\it regime}.  \\

Borel  re-summability for the quartic-AHO has been proved in ref.[76]. The 
relation of the singularity in the Borel-plane to the corresponding Euclidean 
classical action was first established by Lipatov$^{77}$. An early study of 
 Borel re-summation method was made by Jaffe$^{78}$. Use of conformal
 mapping
 to enhance the rate of convergence of the Borel-transform has been discussed 
in ref.[79]. Bounds on the optical sum of divergent series, in the context of 
QT, have been obtained in ref.[80] using a novel variational method for
 the 
Borel-transform. Borel-summability of interacting  quantum field theory has 
been reviewed, for example, in ref.[81]. Some modern applications of 
re-summation techniques have been discussed in ref.[82,83]. Singularities 
( `Renormalons') in the coupling-plane due to the effects of renormalization
 of field theories and their significance have been discussed in ref.[84].  
\\
\\
(ii) {\it Pad$\acute{e}$-approximation}
\\
\\
 Re-summation of the original perturbation series, eqn.(44) via 
Pad$\acute{e}~$\\approximation can be expressed as the ratio of two polynomials,
$~ P_{l}(z)~/Q_{m}(z)$ of approximately same degree, as follows :
\\
\begin{equation}
[l/m]_{F}\equiv P_{l}(z)/Q_{m}(z)=(p_{0}+p_{1}z+p_{2}z^{2}+...+p_{l}
z^{l}) /(1+q_{1}z+q_{2}z^{2}+....+q_{m}z^{m})
\end{equation}   
{ \it The  l~+~m~+~1 coefficients,$~ p_{0},~p_{1},~p_{2},.......p_{l} ;~ q_{1},$}
~{ \it $q_{2}....q_{m}$}{ \it are chosen in such a way that the Taylor-series}
 {\it expansion of the r.h.s. of eqn.(46) agrees with the orginal }
{\it perturbation expansion up to the desired order for small $|z|$, i.e.,}   
\\
\begin{equation}
F(z)~-~[~P_{l}(z)/Q_{m}(z)~]~=~O~(~z^{l+m+1}~)~, z~\rightarrow~0 
\end{equation}     
\\
\\
The condition, eqn.(48) requires the solution of $~l+m+1~$ linear equations 
for the coefficients of the polynomials $P_{l}(z)$ and $Q_{m}(z)$. If this
 system
of equations has a solution, the Pad$\acute{e}$-approximant can be expressed 
as a ratio of two determinants which depend upon the coefficients, $f_{0},~
f_{1},~f_{2},......f_{l+m+1}$ of the original perturbation series, eqn.(44).
The generalization to {\it multi-point}  Pad$\acute{e}$-approximation is 
straightforward. If the original asymptotic expansion can be made about a set
 of given points, $~z~=~z_{\alpha}~;~\alpha~=~0,~1,~2,,.... $ then following 
the above procedure, the approximant to the original series can be developed 
for expansion about each point. Details may be found in ref.[85,86] and 
references contained therein.\\

Earlier works on Pad$\acute{e}$ approximation for the AHO are contained in
ref.[87,\\88]. It has been found$^{87,88}$ that single-point 
 Pad$\acute{e}$ 
approximation  works for the quartic- and sextic-AHO cases although the 
convergence is poor for the latter case. However, the same method 
fails$^{87-89}$ for the case of the octic-AHO. More recently,
 it has been shown in ref.[90], that clever use of scaling and transformation 
of the  expansion  variable results in accelerated convergence over a larger 
domain of z, by the use of two-point  Pad$\acute{e}$ approximants for the case
 of AHO's  with quartic-, sextic- and  octic an-harmonicity. It may be noted, 
however, that sophisticated numerical  analysis using symbolic manipulation 
are  required for obtaining results by the above method.
\\
\\
(10) {\it Approximation Schemes Based upon the Path-Integral Formulation of QT}

The Path-Integral (PI) formulation of Quantum Theory$^{91-93}$ has 
traditionally served  as an important basis  to formulate and improve upon 
approximation methods- it may be recalled that the Feynman-diagram technique,
 which is perhaps the most popular method of computation in perturbation 
theory, originated from the PI-approach$^{94}$. In the context of quantum 
mechanics,a very comprehensive account of the PI-formulation is given in 
ref.[9], which also provides extensive reference to earlier work. The 
approximation methods can be formulated either in the real-time 
formulation$^{91,92}$  or as the  imaginary time, Euclidean-formulation$^{9}$.
 The latter is found to be ideally suited for extending the application to 
quantum-statistics.
\\
\\
Various  approximation  schemes  including the Semi-classical  approximation 
methods$^{9,95}$, Stationary-phase approximation$^{9,96}$,  loop-wise 
expansion$^{9}$,  large-order estimate of perturbation  theory$^{9,95}$, 
 variational estimation using \\inequalities$^{9,96}$, variational perturbation
 theory$^{9}$  and  many  more methods$^{9}$ have been profitably  based upon 
the PI-formulation. Application to quantum field theory and statistical 
physics /critical phenomena are extensively treated in the text$^{95}$.\\

In the context of the QAHO-problem, the stationary-phase  approximation was 
used$^{97}$ in the PI-formulation to infer the in-applicability of the 
na$\ddot{i}$ve perturbation theory  due to the existence of an essential
singularity at the origin of the coupling strength plane. A variational method
 was proposed in the Euclidean formalism of the PI in ref.[96] to obtain 
an  approximation to the partition function of the QAHO and thereby, to obtain 
an  estimate of the ground state energy. The method has been improved upon and 
extened in application in ref.[9]  under the name of 
`variational-perturbation theory (VPT)'. Wide applications and convergence 
tests of VPT are described in detail in ref.[9]. In ref.[98], variational 
lower-bound  as well as, upper bound on energy of the QAHO have been obtained 
by the use of inequlities for the partition function. Semi-classical 
approximation of partition functions for one-dimensional potentials using the 
PI-formalism has been described in ref.[99]. The same in higher dimensions is 
dealt in ref.[100]. The multi-instanton  effects for potentials with 
degenerate minima  is  described in  ref.[101]. In the context of scalar field
 theory, ref.[102] describes  non-standard expansion techniques for the 
generating functional in the PI  formulation. Several other applications in 
field theory and statistical  physics can be found in ref.[95].\\
\\
\\
The major limitation of the PI-formulation of approximation method appears 
to be the mathematical complexity involved in evaluating multi-dimensional 
integrals - even the simple, exactly solvable potentials such as the hydrogen 
atom problem, require rather complex techniques$^{9}$ for implementation. In
 many cases, it is therefore, preferable to use the standard operator based 
formalism for implementing the approximation methods.\\
\\
(11) {\it  The self-consistent schemes of approximation}
\\
\\
These schemes are popular in the context of many-body systems and otherwise 
known under the name of ``mean-field approximation'' and the 
``Hartree-approximation'' including various generalizations of the latter, 
 (for example, the ``Hartree-Fock'' method, `` Hartree-Fock-Bogoliubov'' 
method  etc). The {\it basic} concept of the scheme consists in approximating 
the given potential  by an exactly solvable one-body potential. 
 If the original Hamiltonian is {\it H } and the approximating Hamiltonian is 
$H_{0}$, then perhaps the simplest way to ensure self-consistency in the 
approximation  could  be  imposition of  the following constraint :
\\
 
 \begin{equation}
 <~n|H|n~>~=~<~n|H_{0}|n~>~,
\end{equation}
\\           
where, the states $|~n~>$ are the eigen-states of $H_{0}$. In practice, the 
approximating Hamiltonian may be chosen with  {\it a-priori} unknown, 
adjustable parameters, which then get determined through the constraint, 
eqn.(49) and other requirements such as the variational principle and / or 
further simplifying conditions. The non-linear feedback that usually 
characterizes the self-consistency  condition can then be  easily ensured.
 Thus, the simple-looking equation, eqn.(49) possesses the potential to 
include interaction effects non-perturbatively and simultaneously  preserve 
the non-linearity of the original Hamiltonian. The above method of 
implementation often yields results which are reasonably accurate even in the 
leading order of approximation, which can be further improved through standard
recursive procedure as in the case of the Hartree-Scheme.\\

In the context of $~\lambda \phi^{4}~$ quantum field theory and the QAHO 
problem, such a scheme was considered in ref.[103] where, however, the 
expectation value (see,eqn.(49)) was 
restricted to the ground state only, in implementing the self-consistency
requirement. The limitation has been overcome in proposing a ``generalised 
Hartree-method'' in ref.[58]. The `mean-field' approach has also been tried in
 ref.[104].
\\
\\
The main limitations  of the method appear to be the lack of {\it uniqueness}
{\it of the initial choice of the approximating Hamiltonian  and the rate of}
 {\it convergence of the recursive procedure for subsequent improvement}.
 \\
\\
(12) {\it  Other Methods of Approximation}  
\\
\\
In addition to the methods listed above there are several other methods of 
approximation in QT. Since it is not possible to discuss all such methods 
within the limited scope of the present thesis, we list below only a small 
sample  which, we believe, are worthy of mention- the choice is entirely 
personal.
\\
\\
(i) {\it The  $\delta$-expansion method}
\\
\\
This method$^{105}$ is similar in many respects to the approaches of 
the Gaussian\\ approximation$^{25}$, the method of Caswell$^{36}$,
 Killingbeck$^{37}$,
Halliday and Suranyi$^{35}$ and other methods discussed above under the 
category : 
`{\it Variation-perturbation methods}'.
 The basic idea is to split the Hamiltonian  into a `free' part and an 
`interaction-part', both involving one ( or more) {\it additional} parameter, 
say $\Omega$, which is {\it not} present in the original Hamiltonian. In
addition, an expansion parameter, `$\delta$' is multiplied to the interaction
 part for 
the purpose of development of perturbation theory. After calculation to the 
desired order, this parameter is set to unity, $\delta~\rightarrow~1$  at the
 end. Hence, {\it the method is intrinsically non-perturbative although }
{\it calculation  can be done order-by-order in the artificial parameter},
 $\delta$. The result of approximation to any finite order depends upon 
$\Omega$ whereas the {\it sum to all orders must be independent of the same}.
 The sensitivity of the result to this parameter at any given order is 
minimized by application of the `` principle of minimum sensitivity 
(PMS)''$^{106}$.
The convergence of the `perturbation theory' through the 
 $\delta$-{\it expansion method, has been rigorously established for the }
{\it QAHO/QDWO} problem$^{107}$. Application to quantum field theories 
have  been described in ref.[108]. However, the method, in its orginal 
formulation,
fails to reproduce the wave-function at any order due to the variational
nature of the approximation implied by the PMS. This defect has been remedied
in ref.[109] by generalization  of the scheme to allow co-ordinate 
dependence of the $\Omega$ {\it parameter} and using the PMS criteria also for
the  wave-function.
\\
\\
(ii) {\it The non-perturbative renormalization group(NPRG) method}\\

The NPRG formulation uses the Wilson-effective action (WEA)$^{110}$ as 
developed to study the critical phenomena, non-perturbative aspects in 
statistical mechanics and quantum field theory. The method has been applied
to quantum mechanics by Aoki  and collaborators$^{111}$. The basic 
equation 
in the formalism was derived by Wegner and Houghton (WH)$^{112}$. 
However,
the WH-equation, which represents the exact cut-off dependence of the WEA,
can not be analytically solved exactly. Hence some approximation has to be 
made. In the so-called ``local potential approximation (LPA)'', quantum 
mechanical systems, including the AHO and the DWO, have been numerically 
investigated$^{113}$ and shown to yield accurate results. Details can 
be found  in ref.[113] and references to earlier related work contained 
therein.
 \\
\\
(iii) {\it The method of multiple scale perturbation theory} (MSPT)
\\
\\
The above method has been propounded in  ref.[114]. It is especially 
suitable
to the cases where the ordinary perturbation theory diverges due to 
{\it secular terms}~~ (terms that grow rapidly with the co-ordinate or the time
variable) in the chosen perturbation. The AHO/DWO problem is an obvious 
example  where the perturbation $~\lambda x^{4}~$ soon predominates over the
 harmonic term, $~gx^{2}~$ no matter how small $~\lambda~$is. MSPT recognizes 
different characteristic properties of the system, such as the frequency of 
oscillation, different asymptotic behaviour of the wave function etc, at 
different time/length-scales when secular terms are present. In ref.[114],
 the
 MSPT techniques have been applied to solve the Heisenberg-equations of motion
 for the QAHO problem, obtain the energy spectrum and the eigen-states. It 
also provides references to related earlier work.
\\
\\
(iv) {\it The coupled cluster expansion method } (CCEM)
\\
\\
The CCEM is a known technique in many-body theory. In that area, the CCEM has
been widely applied$^{115}$ and extensively reviewed$^{116}$. 
The basic idea 
is due to Hubbard$^{117}$ who proposed that the true interacting vacuum 
state of the system can be generated by an operator, {\it exp(S)} acting 
on the free,
non-interacting vacuum state. In that respect, the above relation can be 
generated through appropriate canonical transformation, as discussed above. 
The method has been extended to quantum mechanics and quantum field theory in
ref.[118]. In ref.[119], the technique has been applied to a 
{\it general} AHO
with quadratic, cubic and quartic  self-interaction such that various cases of
QHO/DWO are covered. The starting point is a parameterization of the 
interacting  vacuum, given by the relation :
\begin{equation}         
|vac'>~=~ exp~(~sa^{\dagger}~+~ta^{\dagger 2}~)|0>~,
\end{equation}
where, s and t are real parameters and other terms have standard meaning as 
explained earlier. The creation/annihilation operators $~b,~b^{\dagger}$ for 
the  interacting vacuum state $~|vac'>~$ can be constructed by standard 
method$^{119}$. Using the above  trial vacuum  state and the tower of 
 higher 
excited states generated there from, the usual variational calculation can be 
done$^{119}$ for the AHO and it yields fairly accurate result as 
discussed  earlier.\\
\\
For the situation of degenerate ground state as happens in case of the DWO, 
the trial vacuum-state can be constructed as a linear superposition of the 
degenerate vacua and variational calculation can be made to calculate the
 energy  split as well as the average energy of a given level. The CCEM method 
consists  in systematically improving the above initial trial state through a
 series of linked cluster operators parameterized by the relation :
\begin{equation}
|vac>~=~exp~(\sum^{\infty}_{k~=~l}S_{k})|vac'>~,
\end{equation}
where,$~S_{k}~=~u_{k}(b^{\dagger})^{k}$. Computation of the energy eigen 
values of quantum systems can be based upon the above trial state and higher 
excited states based upon it after suitable approximation to restrict the 
infinite sum in the argument of the expoential in eqn.(51). For details, 
ref.[119] can be consulted.   
\\
\\
This completes our survey of the various approximation methods in QT.
 \\
\\
In summary, it is revealed from  the above brief yet hopefully representative
survey that considerable progress has been achieved in the development of 
approximation schemes to tackle various problems in quantum theory. At the 
same time it is also brought to focus by the above survey that the `{\it ideal
}' method of approximation, based upon the earlier stated, desirable set of 
requirements of general applicability, simplicity, systematic improvability 
and efficiency of computation etc, remains largely elusive till date. In 
view of the above situation, there still remains a lot of scope to
 incorporate, if possible, the above set of desirable criteria into a 
single-scheme.
\\
\\
The present dissertation is devoted to an attempt at achieving the above 
{\it ambitious} objective, namely, to develop a self-consistent,
 non-perturbative, simple yet efficient scheme of approximation, which is 
systematically improvable and generally applicable, {\it in principle} to an 
arbitrary interacting system in QT including the quantum field theory. The 
 scope of the present dissertation is, however, limited in application  to the
self-interacting Bosonic- systems in quantum theory in which, we include the
 quartic an-harmonic oscillator, quartic 
double-well potential, sextic an-harmonic and double-well oscillator, the 
octic-an-harmonic oscillator and the $\lambda \phi^{4}$  quantum field theory
{\it without} spontaneous symmetry breaking. In the next {\bf Chapter},
 we describe the formulation$^{120}$.\\ 
\\
\\

{\Large \bf 3.  The New General Approximation Scheme\\(NGAS)}\\

{\large \bf 3.1   Formulation of NGAS for self-interacting Systems }\\

Consider a generic Hamilitonian $ H_{\lambda}(\phi,p)$  describing
a  self-interacting quantum system involving the  {\it field}  `$\phi$'
and conjugate momentum `p', given by\\

\begin{equation}
               H_{\lambda}(\phi,p)~ =~ H_{s} (\phi,p) + \lambda~ H_{I}(\phi),
\end{equation}
where, $H_{s}$ is the unperturbed Hamiltonian and $ \lambda H_{I}$ is the
 self- interaction with $\lambda$  as the coupling strength. We use the
 language of field theory identifying quantum mechanics as field theory in
 $(0+1)$dimensions.\\

Many physically important systems are described by the
 above Hamiltonian including the anharmonic oscillators ( AHO ); the 
 double- well oscillators ( DWO ); the  $\lambda\phi^{4}$ quantum field theory
 in the symmetric-phase as well as, in the spontaneously symmetry broken (SSB)
 phase ; {\it  pure} Yang-Mills fields  with the quartic `{\it gluon}'- self 
 interaction etc.\\

As has been discussed earlier, perturbation theory treating $~H_{s}(\phi,p)~$
as the unperturbed Hamiltonian and the entire interaction $~\lambda H_{I}(\phi
)~$ as perturbation, fails to converge even for infinitesimal $~\lambda~$.
 Other non-perturbative methods often suffer from limitations as reviewed in
{ \bf Chapter-2}. With a view to overcome these limitations and difficulties 
the  formulation of NGAS consists of the following steps:\\

{\bf 3.1.1   Choice of the approximating potential :$~V(\phi)~$ }\\

The aim is to replace the orginal interaction $~\lambda H_{I}(\phi)~$ by a
``suitably chosen'' approximating potential$~\lambda V(\phi)~$ such that it
preserves the symmetries of the original interaction but leads to an `` exactly
solvable'' Hamiltonian, i.e. the ``{\it  Effective Hamiltonian} (EH)''
 generated by $V(\phi)$ and defined by

\begin{equation}
            H_{o}(\phi,p)~ =~ H_{s} (\phi,p) + \lambda~ V(\phi),
\end{equation}
is {\sl exactly solvable}, i.e.
\begin{equation}
       H_{0}|n>~=~E_{n}^{(0)}|n>, ~ <m|n>~=~ \delta_{mn},
\end{equation}
where, the spectrum {$~|n>~$} and the eigen-values  $ E_{n}^{(0)}$,
 are known. ( We consider, for simplicity, that the spectrum is discrete
 and non-degenerate ). We  refer to this  requirement, eqn.(54), as the 
``{\it  condition of exact solvability}'' (CES). The next requirement on
$V(\phi)$ is the condition of equal quantum average, as explained below.\\

{\bf 3.1.2   The Principle of Equal Quantum Average }\\

The effective Hamilitonian is constrained to yield the same quantum
 average (QA) as the original, i.e.

\begin{equation}
          <n|H(\phi)|n>~ =~ <n|H_{o}(\phi)|n>,
\end{equation}
which   implies,

\begin{equation}   <n|H_{I}(\phi)|n>~ =~ <n|V(\phi)|n>
\end{equation}
We refer  to this requirement, eqs.( 55-56) as the ``{\sl condition
 of equal quantum average} ( CEQA)''. The next step is to {\it optimize}
 the approximation as described below:\\

{\bf 3.1.3   Variational Optimisation of  $~V(\phi)$}\\

Let $~V(\phi)~$ involve a set of free-parameters: \{ $~\alpha_{i}~$\}.
 Then the requirement of optimisation consists in the variational 
minimisation of $H_{0}$ with respect to the free-parameters  $(\alpha_{i})$
 which characterize$~V(\phi)~$:

\begin{equation}
  \frac {\partial}{\partial{ \alpha_{i}}}  < H_{0} > = 0
\end{equation}
where, the notation is
\begin{equation}
      < \hat{A} >~ \equiv~<n|\hat{ A} |n>
\end{equation}
We refer to this condition, eqn.(57) as the ``{\it condition  of optimality} 
(CO)''.\\

{\bf 3.1.4   The Self-Consistency and other features of the  Method }\\

The steps which are outlined above are the essential ingredients of the
 proposed approximation scheme in the leading order(LO). The following 
several  observations regarding the approximation scheme are in order:\\
                                                                     \\
(i) It is to be noted that in a restricted form, i.e. when the quantum
average in eqn.(57) is restricted to the ground state only, the CEQA, as
expressed in  eqn.(55), corresponds to the Hartree-approximation/ mean
field approximation in quantum field  theory$^{103}$ . In view of this,
 the
 NGAS can be regarded as a {\it ``generalised'' Hartree-approximation 
method} $^{58}$.\\
                                                                      \\
(ii) The {\sl self-consistency} of the procedure is implicit in eqs.( 53-55):
 the states $|n>$ which are obtained as the solution of $H_{0}$ (see,
 eqn.(54) ), are used as input (in eqn.(56)) to determine $V(\phi)$  which,
 in turn, defines $H_{0}$ (eqn.53), thus making the ``feed-back loop''
complete. This can be schematically represented as:\\

~~~~~~~~~~~~~~~~~~~~~~~~~~~ $|n>\Longrightarrow V(\phi) \Longrightarrow H_{0}
\Longrightarrow|n>$.\\
                                                                     \\
(iii) The leading order (LO) approximation in NGAS consists of finding the 
spectrum $~|n>~$ and the energy eigen-values $E_{n}^{0}$. This is easily 
achieved because of the CES (eqn.(54)). {\it It may be emphasized, however,
 that even the LO results capture the dominant contribution of the full 
(nonlinear) interaction through the requirement, eqn.(55),  eventhough one
 always deals with an exactly solvable Hamiltonian}, $H_{0}$. We consider this
 as a key feature of the approximation method.\\

{\bf 3.1.5  The Vanishing Quantum Average of the Effective\\
 Interaction and its Significance }\\

The other important result which follows trivially from eqn.(55) is
 that the modified interaction$~\lambda H'~$ defined by the relation:

\begin{equation}
 \lambda H'\equiv\lambda (H_{I} - V),
\end{equation}
has vanishing QA for arbitrary `$\lambda$' and `$n$', i.e.

\begin{equation}
                  <n|\lambda H'|n> =0.
\end{equation}
In otherwords, {\it only the non-diagonal matrix elements of$~H'~$ are non-
vanishing in the basis which renders $~H_{0}~$ diagonal}.\\

This result is of considerable significance and it naturally suggests a
 scheme of improved perturbation theory (IPT) in which, $H_{0}$ is chosen
as the unperturbed Hamiltonian   and $~\lambda H'~$is considered to be the
perturbation. The  IPT thus developed, can be shown to be convergent, owing
 mainly to eqn.(60) and thus can be used to systematically  improve the  LO
result order-by-order ({\bf Chapter~ 10}). The various steps outlined in 
equations (53-60) define the scheme in the LO.\\

 In the following {\bf Chapters}, we implement and  illustrate the general
 approach  described above to specific quantum systems with self-interaction.\\
\\
\\

{\Large \bf 4.  Application of   NGAS to the case of the Quartic-Anharmonic 
 Oscillator: formulation for the leading order(LO)}\\

{\large \bf 4.1  Application of the NGAS to the case of the Quartic Anharmonic 
Oscillator in Quantum  Theory}\\
\\
\\

In the following subsections we illustrate the NGAS applied to the case of
 the quartic anharmonic oscillator(QAHO) in quantum theory. However, it may
 be worth while to first describe the importance of the QAHO problem due to
 its applications in diverse areas of quantum physics and its significance
 from a  theoretical stand-point as well.
\\
\\
{\bf  4.1.1   The Quartic Anharmonic Oscillator(QAHO) in Quantum
 Theory: its importance and applications}\\

The quantisation of the simple harmonic oscillator is perhaps among the
 most celebrated examples$^{8,9}$ due to its immense
 significance and 
application. In  particular, small-oscillations about the equlibrium- 
position are described by (normal-mode) simple-harmonic oscillators.
 Thus, quantum theory of the simple harmonic oscillator describes small
 oscillations of complex dynamical systems about the equlibrium. This 
property leads, for example, to quantisation of free-field systems, which
 are shown equivalent to an assembly of independent  harmonic oscillators.
 Similarly, most lattice-dynamic studies in the area of condensed matter 
physics employ the harmonic approximation as the leading  description of 
the underlying physics. However, realistic approximation of actual physical
 systems need the incorporation of the anharmonic-interaction  when deviation
 from small-oscillations or inclusion of self-interaction among oscillators
 become necessary on physical ground. This is the primary reason which endows 
the study of anharmonic oscillators with considerable importance and 
significance and explains its application in diverse areas of
 physics$^{8,9}$. Anharmonic oscillators are, therefore, among the most widely 
studied models in quantum theory  in which the quartic-anharmonic oscillator
 ( QAHO ) is  the simplest  system ~exhibiting~ self-interaction. This~ system
 has been extensively investigated leading to a vast amount of 
literature$^{8,9}$. Its importance arises due to physical
 applications in areas including field theory, condensed matter 
physics$^{121}$, statistical  mechanics$^{122}$, non-linear 
systems$^{123}$, classical and quantum 
chaos$^{124}$, inflationary cosmology$^{125}$, lattice dynamics, 
plasma  oscillations etc, to cite only a few cases. Besides, the QAHO
has also served as a {\it theoretical laboratory}~ to study convergence of
 perturbation  theory$^{126}$, development of non-perturbative  
approximation  methods$^{127}$, renormalisation$^{25,113}$,
 vacuum structure$^{128}$ and  stability  analysis$^{129}$ etc.\\

The  Hamiltonian of the system is given by
\begin{equation}
H~=~\frac{1}{2}p^{2}~+~\frac{1}{2}~g~\phi^{2}+
\lambda\phi^{4},
\end{equation}
where $~\lambda~,~ g~$ are real and positive. Note that the canonical momentum
 conjugate to the `field'  $~\phi(t)$ is given by $~ p(t) = d\phi(t)/dt~$.
 In the notation of eqn.(52), the free field Hamiltonian corresponds to
\begin{equation}
H_{s}(\phi) = \frac{1}{2}p^{2}~+~\frac{1}{2}~g~\phi^{2}
\end{equation}
and the interaction is
\begin{equation}
 \lambda H_I(\phi) = \lambda\phi^{4}.
\end{equation}
In order to develop the NGAS  for the  QAHO,  we follow the
steps outlined  in the {\bf Chapter-3}, as described below.\\

{\large \bf 4.2   Formulation of the NGAS for the QAHO }\\

{\bf 4.2.1  Choice of  $V(\phi)$ }\\

The first step is to find a suitable approximating potential $V(\phi)$.
The following {\it ansatz}  is naturally suggested on the grounds of 
simplicity and exact solvability:

\begin{equation}
                 V~(\phi ) = A~\phi^{2} - B~ \phi + C.
\end{equation}
where A,B,C are  parameters to be determined self consistently. It would
 appear from eqn.(64) that the global symmetry of the orginal Hamiltonian,
 eqn.(61), under  $\phi \rightarrow -\phi$, is not respected by the ansatz,
  eqn.(64). However, this is {\it illusory}  since the co-efficient  `B' in
 eqn.(64) is dynamically determined to be propertional to $<\phi>$ ( see eqn.
(87) below),  which  ensures  the  preservation  of the above global symmetry.\\

{\bf 4.2.2   Solution of the eigen value equation of the  Resultant\\
 Effective Hamiltonian(EH) }\\
                                                                \\

To obtain the exact analytic solution of the spectrum of  $~V(\phi)~$ we
 proceed as follows.\\

We define the ``Effective Hamiltonian (EH)''   by the following equation

\begin{equation}
 H_{0}(\phi,p)~=~\frac{1}{2}p^{2}~+~\frac{1}{2}~g~\phi^{2}+\lambda V(\phi)
\end{equation}
The EH defined in eqn.(65) can now be recast in the form of a  diagonalizable
structure as outlined below. Substitution of eqn.(64) in eqn.(65) leads to
 the following equation

\begin{equation}
 H_{0}(\phi,p)~=~\frac{1}{2}p^{2}~+~\frac{1}{2}~g~\phi^{2}+\lambda(A~\phi^{2} -
 B~ \phi + C) 
\end{equation}
The simplification of the eqn.(66) leads into the following Hamiltonian:

\begin{equation}
H_0~=~  \frac{1}{2}~p^{2}~+~\frac{1}{2}~\omega^{2}~( \phi - \sigma )~^{2} + h_
{0},
\end{equation}
where,

\begin{equation}
           \omega^{2}~=~ g + 2\lambda A,
\end{equation}

\begin{equation}
           \sigma~=~\lambda B/\omega^{2},
\end{equation}
and
\begin{equation}
           h_0~=~\lambda C~-~\frac{1}{2}\omega^{2}{\sigma}^{2}.
\end{equation}
It may be atonce recognized  that the EH given by eqn.(67) corresponds
 to a ``shifted'', effective, harmonic oscillator where both the field,
 as well as, the energy are respectively shifted  by `$\sigma$'  and `$h_
{0}$'. Note further that the c-number parameters `$ \omega $'  and `$\sigma$'
 are restricted by  physical requirement,  to satisfy $~\omega > ~0$; $~
\sigma$ ~ =~ real (since,$~ \phi~=~\phi^{\dagger}$).
{\it Diagonalisation} of $H_{0}$ can be  achieved by the standard
method of invoking the creation-and annihilation operators, defined by

\begin{equation}
        \phi(t) = \sigma +( b + b^{\dagger})/ \sqrt{2\omega},
\end{equation}
\begin{equation}
  p(t) =i\sqrt{\omega/2}~ (b^{\dagger} - b),
\end{equation}
along  with the standard equal-time (canonical) commutation relation (ETCR)
 given by

\begin{equation}
[~ b,~b^{\dagger}~] = 1.
\end{equation}
We first introduce the number operator: $~N_{b} \equiv b^{\dagger}b~$ and its 
eigen-states by the equation
\begin{equation} 
N_{b}|n>~ =~ n|n>.
\end{equation}
Because of the {\it hermiticity and positivity of the operator$~N_{b}$};
 the  eigen values are positive-definite and the eigen states are 
orthonormal: 

\begin{equation}
 <m|n>~ =~ \delta_{mn}, 
\end{equation}

\begin{equation}
\sum_{n}|n><n|~=~1.
\end{equation}
Further, as proven in standard text books$^{130}$, by the use
of the  ETCR (eqn. 73), the spectrum of eigen values is restricted to:

\begin{equation}
n~=~0,1,2,...........
\end{equation}

Then, using the~ defining~ parametrisation in  eqns.(71-72)~ one 
immediately  obtains the Hamiltonian in diagonal form: 

\begin{equation}
H_{0}~=~\omega~ (N_{b} + 1/2 ) + h_{0}.
\end{equation}
The energy spectrum is then trivially obtained and given by

\begin{equation}
              E_{n}^{(0)}~=~\omega~ \xi  + h_{0} ,
\end{equation}

where $\xi~=~ ( n + 1/2 )~$  with `n' taking values according to eqn.
(77).\\

At this stage, the following remarks/ observations and results may be
 noted:\\
                                                                   \\
(i) By using the standard properties of the creation-/annihilation 
opertaors, the QA of polynomials of field  `$\phi $' and momentum `p' can 
be easily  evaluated. In particular, we note  the following results  for
 subsequent use (in the following, the vacuum-state is denoted by $|vac>$ 
and is defined by  the property, $ b|vac>~ =~ 0$):

\begin{equation}
  <vac |\phi |vac >~ =~< n |\phi |n >~ \equiv~ < \phi >=~ \sigma~ ,
\end{equation}

\begin{equation}
<~\phi^{2}~>~ =~ \sigma^{2} + (\xi / \omega)~ ;~< p^{2} >~=~\omega \xi,
\end{equation}

\begin{equation}
<~\phi^{3}~>~ =~ \sigma^{3} + 3~\sigma (\xi / \omega),
\end{equation}

\begin{equation}
<~\phi^{4}~>~ =~ \sigma^{4} + 6~\sigma^{2}~(\xi / \omega) + 3~( 1 + 4~\xi^{2} )
/8 \omega^{2} , 
\end{equation}

\begin{eqnarray}
< \phi^{6} >~ =~ \sigma^{6} + 15 \sigma^{4} ( \xi/\omega ) + 45 \sigma^{2}
( 1+ 4\xi^{2})/8\omega^{2}
\nonumber\\
 + (5/8) (\xi/\omega^{3})(5 + 4\xi^{2}).
\end{eqnarray}

The QA of the original Hamiltonian {\it H} defined in eqn.(61) can now be 
evaluated  with the help of eqs.( 80-84) and given by
\begin{equation}
< H > ~=~ \frac{1}{2}< p^{2} > ~+ ~\frac{1}{2}~g~< \phi^{2} > + \lambda< \phi^
{4} >.
\end{equation}

Substituting the  QA values from eqs.(80, 81, 83) and noting that   the 
eqn.(55) holds, eqn.(85) is transformed to:  
\begin{eqnarray}
<n|H|n>~ \equiv~ <n|H_{0}|n> = \omega\xi/2 + ( g + 12\lambda \sigma^{2})(\xi/2
\omega)
\nonumber\\
+(3\lambda /8\omega^{2})(1+4\xi^{2}) + g \sigma^{2}/2 + \lambda \sigma^{4}.
\end{eqnarray}

(ii) Eqn.(80) shows that `$\sigma$' corresponds to the vacuum expectation
value (VEV) of `$\phi$'.  In view of this result  and eqn.(69), the 
coefficient `B' of the linear term in $V (\phi)$ (see eqn.(64)) can be 
re-expressed as :
\begin{equation}
B = ( \omega^{2}/\lambda) < \phi >.
\end{equation}

Eqn.(87) demonstrates that the {\it global symmetry of the original
 Hamiltonian under the transformation, } $\phi ~\rightarrow~ -~ \phi$, {\it is 
preserved by the potential~} $~V (\phi)~$,  which is not otherwise transparent
 in eqn.(64).\\
                                                                            \\
(iii) If one denotes by $~a~~ and~~ a^{\dagger}~$, the corresponding creation- 
and annihilation operators of the `free' theory (i.e. defined by $~H_{s}(\phi,
p)$), then  `$\phi$'  and `p' can also be expressed in terms of these 
operators  analogous to eqs.(71-72) as:
\begin{equation}
\phi(t) = \sigma +( a + a^{\dagger})/ \sqrt{2\omega_{0}},
\end{equation}

\begin{equation}
p(t)~ =~i\sqrt{\omega_{0} /2}~ (a^{\dagger} - a),
\end{equation}

where  $~\omega_{0} \equiv \sqrt g~$. It is important to note here that both
the sets of creation- and annihilation operators  satisfy identical (equal-
time) commutation relation:
\begin{equation}
 [ a, a^{\dagger} ] = 1 = [ b, b^{\dagger}]
\end{equation}

It  may also be mentioned that the VEV of $~\phi~(<\phi>~=~ \sigma)~$ 
remains invariant in the two descriptions. Eqn.(90)  implies that the 
two sets of operators  must be related by a {\it quantum canonical 
transformation} (`` Bogoliubov transformation'')$^{52}$. This
result 
has  crucial implications for the vacuum structure and stability of the 
approximate theory. This is discussed in detail in {\bf Chapter 9}.\\

Returning to the implementation of the NGAS, the next task is to
determine the free parameters, `$\sigma$' and `$\omega$' (or equivalently,
A, B and C occuring in eqn.(64)). This is achieved as follows:\\

{\bf 4.2.3   Determination of the free parameters}\\

The conditions expressed under CEQA and CO given in eqs( 56-57) are 
sufficient to fully determine the free parameters involved in the 
approximation. These requirements translate in this case, to the
following equations:

\begin{equation}
   < \phi^{4} >~ =~  A~< \phi^{2} > - B~ < \phi > + C~ ;
\end{equation}

$~\partial< H_{0} >/\partial\omega = 0~$ and $~\partial< H_{0} >/\partial
\sigma~ =~ 0$ where,$~ <H_{0}>~$ is given by eqn.(86). 
Carrying out the explicit variational minimisation of $<H_{0}>$ with respect 
to `$\omega$' i.e.,$~\partial< H_{0} >/\partial\omega = 0~$ one obtains the 
following equation :

\begin{equation}
\omega^{3}~-~\omega ( 12 \lambda \sigma^{2}~+~g )~-~6 \lambda f(\xi)=~0,
\end{equation}
Again carrying out the explicit variational minimisation of $<H_{0}>$ with 
respect to `$\sigma$' i.e., $\partial< H_{0} >/\partial\sigma~ =~ 0~$,  the
 following equation is obtained:

\begin{equation}
\sigma (   4\lambda \sigma^{2} + g + 12 \lambda\xi/\omega) ~=~0,
\end{equation}
where, in eqn.(92), $~f (\xi )~ \equiv~\xi + 1/4\xi$. Eqs.(92-93) 
are to be solved simultaneously to determine `$\omega$' and `$\sigma$'
 as functions of $\lambda$, g and $~\xi$. In the following, we refer
 to these  eqs.(92-93) as the ``gap equation (GE)'' and `` the
 equation for the ground state (EGS )'' respectively. The constants
 A, B and C appearing in eqn.(64) can  now be  determined as
 follows:\\

To determine the parameter `A'  we use the eqs.(68) and (92). Rearranging
the ``gap-equation" given in eqn(92) we obtain the relation

\begin{equation}
\omega^{2} =  ( g + 12 \lambda \sigma^{2} ) + \frac{6 \lambda f(\xi)}{\omega} 
\end{equation}

Eqn.(94),~ considered together with eqn.(68) lead to a determination of the 
constant `A', given by 

\begin{equation}
A =~6~\sigma^{2}~ +3f(\xi)~/\omega~~~~~~~~~~~~~~~~~~~~~~~~
\end{equation}

To determine the parameter `B' we use the eqs.(69) and (93). Rearranging 
the eqn.(69) leads to the following relation:
\begin{equation}
\sigma ( -1 + \frac{\lambda B}{\sigma\omega^{2}}) = 0
\end{equation}

Comparing  eqs.(93) and (96) we obtain the parameter `B' as
\begin{equation}
B~ =~( 1 + g )~( \sigma \omega^{2}/ \lambda)~ +~4 \omega^{2}
\sigma^{3} + 12\omega\sigma\xi
\end{equation}

To obtain the parameter `C' we use  equation(91):
\begin{equation}
C = <~\phi^{4}~> - A < \phi^{2} > + B <\phi>,~~~~~~
\end{equation}

On substitution of the  values $~ < \phi  >,~<~\phi^{2}~>$, and
$<~\phi^{4}~>~$ from eqs.(80, 81, 83), along with the values of
`A' and `B' determined above, the parameter  `C' is determined. Using
the gap-equation (92) further the parameter C is given by
\begin{eqnarray}
C~=~\frac {3}{8\omega^{2}}(1 + 4\xi^{2}) - \frac {3\xi
  f(\xi)}{\omega^{2}} + \frac {\sigma^{2}\omega}{\lambda}[ \omega^{3}
+ \omega
\nonumber\\
 -6\lambda f(\xi) + 12\lambda \xi - \frac {3\lambda
   f(\xi)}{\omega^{2}}] - \sigma^{4}( 5 + 8\omega^{2})
\end{eqnarray}
On substitution of eqn.(99) in eqn.(70) we obtain the relation
\begin{eqnarray}
h_{0}~=~~\frac {3\lambda}{8\omega^{2}}(1 + 4\xi^{2}) - \frac {3\lambda\xi
  f(\xi)}{\omega^{2}} +  {\sigma^{2}\omega}[ \omega^{3}
+ \frac {\omega}{2}
\nonumber\\
 -6\lambda f(\xi) + 12\lambda \xi - \frac {3\lambda
   f(\xi)}{\omega^{2}}] - \lambda\sigma^{4}( 5 + 8\omega^{2})
\end{eqnarray}
Thus, the parameters defining the approximating potential $~V(\phi)~$ are 
completely determined.\\

Next, using eqn.(86) and noting that $~<H_{0}>~=~<H>~\equiv
E_{n}^{(0)}~$ one obtains the energy spectrum:

\begin{equation}
E_{n}^{(0)} = \omega\xi/2 + ( g + 12\lambda \sigma^{2})(\xi/2
\omega)
+(3\lambda /8\omega^{2})(1+4\xi^{2}) + g \sigma^{2}/2 + \lambda \sigma^{4},
\end{equation}
where `$\omega$' and `$\sigma$' are solutions of eqs.(92-93). The 
different solutions thus obtained are discussed in the following:\\

{\bf  4.2.4   The Leading-Order (LO) Results - Determination of the 
Spectrum of $H_{0}$}\\

Solution of the gap-equation (GE) eqn.(92) and the equation for
ground state (EGS) eqn.(93) constitute the key ingredients in the 
calculation of the energy spectrum. It is convenient to first obtain 
the solution of the EGS, eqn.(93). For the case of the QAHO ($~ g,
\lambda > 0~)~$, eqn.(93) has two  solutions:
\begin{equation}
(i)~~~~~~~~~~~~~~~~~~~~\sigma ~=~0~~~~~~~~~~~~~~~~~~~ 
\end{equation}
and 

\begin{equation}
(ii)~~ ~~~~~~~~~~~~~~~g + 4\lambda \sigma^{2} + 12 \lambda\xi/\omega) ~=~0.~~~~~~~~~
\end{equation}
We now analyse the eqn.(103) which gives

\begin{equation}
\sigma^{2} = -\frac {1}{4\lambda}( g +  12 \lambda\xi/\omega)~ <~ 0
\end{equation}

Since $~g, \lambda~ >~ 0$, eqn.(104) does {\it not} lead to 
acceptable solution, which requires $~\sigma^{2}~>~0~$. Therefore, 
the only physically accepted  solution for QAHO is the solution (i) 
given in eqn.(102), i.e., $~\sigma~=~0$. ( This is also intuitively 
obvious since the single-well shape of the ``classical'' potential can 
{\it  not} get altered to double-well shape by quantum - fluctuations).
 Substitution of eqn.(102) in eqn.(92), then leads to the following 
simplified GE for the QAHO:
\begin{equation}
         \omega^{3} - g\omega - 6 \lambda f(\xi) = 0.
\end{equation}

{\sl It may be emphasized at this point  that this GE ( with g=1) has been
derived by several authors$^{131}$, but starting from widely different 
considerations}. On substitution of $~\sigma~=~0~$ in eqn.(101) we obtain the
 equation

\begin{equation}
 E_{n}^{(0)}~ =~\frac {1}{2}\omega\xi + \frac{1}{2}(\frac {g\xi}{\omega}) +~
\frac{3\lambda}{8\omega^{2}}(1~+~4\xi^{2}).
\end{equation}

Using the GE, eqn.(105) and noting that 
\begin{equation}
f(\xi) \equiv \frac {( 1 + 4\xi^{2})}{4\xi},
\end{equation}

the last term in eqn.(106) can be simplified to be

\begin{equation}
\frac{3\lambda}{8\omega^{2}}(1~+~4\xi^{2}) = (\frac{\xi}{4}) (\omega - \frac{g}
{\omega})
\end{equation}

Substitution of eqn.(108) in eqn.(106) then leads to the following
 simple expression for the energy spectrum of the QAHO in the LO: 
\begin{equation}
E_n^{0} = \frac{\xi}{4} (3\omega + \frac{g}{\omega})
\end{equation}

where `$\omega $' is obtained as a solution of the GE for the QAHO 
given by eqn.(105). It is to be noted that the  ``gap-equation'' 
given by eqn.(105) is in the form of a cubic equation of the type

\begin{equation}
         x^{3} - 3Px - 2Q = 0;~  P, Q > 0
\end{equation}

The real solution of this eqn.(110) is given by,

\begin{equation}
x = Q^{1/3}[ ( 1 + \sqrt {(1 - \frac{P^{3}}{Q^{2}})})^{1/3} + ( 1 - \sqrt {(1 -
 \frac{P^{3}}{Q^{2}})})^{1/3}]
\end{equation}

Then comparing the coefficients of eqs.(110) and (105) the real 
solution is obtained explicitly as

\begin{equation}
\omega~=~(3\lambda f(\xi))^{1/3}[(~1 + \sqrt {(1 - \rho)})^{1/3}~+~(~1 - \sqrt 
{(1 - \rho)})^{1/3}],
\end{equation}

where, $~\rho^{-1}~=~243\lambda^{2}f^{2}(\xi)/g^{3}~$. It may be noted
that, for the case when$~ g = 1~$, the solution for `$~\omega~$' as given
above  has the correct  limiting behaviour, $~\omega~ \longrightarrow~
1$ for $~ \lambda \longrightarrow 0~$ and further that it {\it
exhibits the non - analytic dependence on the coupling $~\lambda~$ at 
the orgin characteristic of the non-perturbative nature of the NGAS}.\\

As has been noted by several authors $^{131}$,~ {\it the  formula, 
eqn.(109) is accurate to within a few percent of the `exact'
result for the full allowed range of $~g,\lambda >~0~$ (both in the `weak 
coupling-' and the `strong coupling' regimes) and for all values of the 
excitaion level $n\geq 0$}. In particular, the accuracy in the strong 
coupling regime can be judged by the following result for the
computation of the ground state energy. The `$exact$' asymptotic
result ( for $g~=~1$) is given by$^{44}$, for 
 $ \lambda\rightarrow\infty$,
 $~E_{0}|_{exact}~=~0.668\lambda ^{1/3}~$ which is to be compared to the
LO-result in NGAS: $  E_{0}|_{NGAS}~=~0.681\lambda^{1/3}$ in the 
same limit.
\\                                                                  
In {\bf Table-1}, the LO results of the present approximation scheme for 
$~g~=~1~$are presented for sample values of `$\lambda$' and `n' along with 
`exact' results ( obtained by numerical methods (Hsue and Chern,
ref.[23])) and  results from earlier computaion$^{44}$ 
( obtained by  different analytical methods) for comparision. As can be seen 
from the comparison, the LO results,  lie within $~ 0.2 - 2 ~\%$ of the
 results from the `{\it  exact}' numerical calculations$^{23}$. In the
 same{~\bf Table-1~}, we also dispaly the improvement of results
 obtained by  inclusion of the first non-trivial correction in the 
improved perturbation theory (IPT), which is discussed in {\bf Chapter-10}.\\
We next apply the method to the case of the quartic -double well oscillator in 
the following {\bf Chapter}.\\

\begin{table}
\vspace {.4in}
 {\bf Table- 1: }
 {\it Leading order(LO) results and perturbation correction in the
 $~2^{nd}~$  order of IPT for the energy levels of the QAHO computed
 for $~g~=~1~$ and sample values of `$\lambda $' and `n' shown along with
 analogous results of ref.[44], compared  with the `exact'(numerical)
results of (Hsue and Chern ref.[23]). The relative percentage  errors 
in the two schemes  are also shown}.
   
   \vspace{0.20in}
   \begin{center}
   \begin{tabular}{c c c c c c c c c}

   \multicolumn{1}{ c }{$\lambda $}&
    \multicolumn{1}{ c } {n}&
   \multicolumn{1}{ c }{$E_{n}^{(0)}$ }&
   \multicolumn{1}{ c }{\quad Exact}&
   \multicolumn{1}{ c }{$E_{n}^{(2)}$}&
   \multicolumn{1}{ c }{Error(\%)}&
   \multicolumn{1}{ c }{$E_{n}^{(2)}$}&
   \multicolumn{1}{ c} {$Error(\%)$}\\
&\quad   &\quad  &\quad ref.[23]      &\quad      &\quad    &\quad 
              ref.[44]&\quad ref.[44]\\
                                                           \\
0.1      &0   &\quad0.5603&\quad 0.5591&\quad0.5591&\quad0.007
              &\quad0.5591&\quad 0.007 \\

         &1   &\quad1.7734&\quad1.7695&\quad1.7694&\quad0.005
              &\quad1.7694&\quad0.005\\
         
         &2   &\quad3.1382&\quad3.1386&\quad3.1391&\quad0.016
              &\quad2.9006&\quad7.580\\
         
         &4   &\quad6.2052&\quad6.2203&\quad6.2239&\quad0.058
              &\quad5.4795&\quad11.96\\
        
        &10   &\quad17.266&\quad17.352&\quad17.374&\quad0.127
               &\quad14.539&\quad16.32\\
        
        &40   &\quad94.843&\quad90.562&\quad95.766&\quad5.75
              &\quad76.152&\quad15.91\\
                                                            \\
1.0      &0   &\quad0.8125&\quad0.8038&\quad0.8032&\quad0.070
              &\quad0.8032&\quad0.07\\
        
         &1   &\quad2.7599&\quad2.7379&\quad2.7367&\quad0.043
              &\quad 2.7367&\quad0.043\\
         
         &2   &\quad5.1724&\quad5.1792&\quad5.1824&\quad0.061
              &\quad4.4440&\quad14.19\\
         
         &4   &\quad10.900&\quad10.964&\quad10.982&\quad0.17
              &\quad8.8890&\quad18.93\\
        
        &10   &\quad32.663&\quad32.933&\quad33.013&\quad0.243
              &\quad25.833&\quad21.56\\
        &40   &\quad192.79&\quad194.60&\quad195.15&\quad0.282
              &\quad149.87&\quad22.99\\

                                                              \\ 

10.0     &0   &\quad1.5313&\quad1.5050&\quad1.5030&\quad0.131
              &\quad1.5030&\quad0.131\\
         
         &1   &\quad5.3821&\quad5.3216&\quad5.3177&\quad0.070
              &\quad5.3177&\quad0.070\\
         
         &2   &\quad10.324&\quad10.347&\quad10.356&\quad0.090
              &\quad8.6131&\quad16.76\\
         
         &4   &\quad22.248&\quad22.409&\quad22.457&\quad0.210
              &\quad17.651&\quad21.23\\
        
        &10   &\quad68.171&\quad68.804&\quad68.996&\quad0.270
              &\quad52.943&\quad23.05\\
        &40   &\quad409.89&\quad413.94&\quad415.18&\quad0.300
              &\quad316.13&\quad23.62\\
                                                            \\

100.0    &0   &\quad3.1924&\quad3.1314&\quad3.1266&\quad0.150
              &\quad3.1266&\quad0.150\\
         
         &1   &\quad11.325&\quad11.187&\quad11.178&\quad0.080
              &\quad11.178&\quad0.080\\
         
         &2   &\quad21.853&\quad21.907&\quad21.927&\quad0.090
              &\quad18.095&\quad17.40\\
         
         &4   &\quad47.349&\quad47.707&\quad47.817&\quad0.230
              &\quad37.314&\quad21.80\\
        
        &10   &\quad145.84&\quad147.23&\quad147.65&\quad0.285
              &\quad112.79&\quad23.40\\
        
        &40   &\quad880.55&\quad889.32&\quad892.03&\quad0.300
              &\quad677.91&\quad23.77\\

\end{tabular}
\end{center}
\end{table}
\newpage

{\Large \bf 5.  Application to the case of the Quartic Double Well
Oscillator (QDWO):\\ Leading Order(LO) Results }\\
\\
\\

{\large \bf 5.1 Importance of the QDWO  and its Applications}\\

The QDWO is also an extensively studied system$^{132}$ because of its 
theoretical importance and practical application. The Hamiltonian of the 
system is given by:

\begin{equation}
H~=~\frac{1}{2}p^{2}~-~\frac{1}{2}~g~\phi^{2}+\lambda\phi^{4};~~ g~,\lambda~> 0
\end{equation}

The crucial~ -ve ~ sign of the  $ \phi^{2}$- term generates  a  quite  
different  physical  situation than the case of the QAHO,  even in the
classical limit. The `classical' potential, $~V_{c} \equiv - \frac{1}{2}g
\phi^{2}~+~\lambda\phi^{4} ~$ exhibits the familiar double-well shape with 
symmetric minima. For $~g~=~1~$, these minima are located at positions 
$~\pm \frac{1}{2\sqrt\lambda}~$ and each with depth
$~\frac{1}{16\lambda}~$. As $~\lambda~$ becomes smaller and smaller the
depth of the two- wells become deeper and deeper. The actual low lying
energy eigen-states of the problem become radically different from the
trial wave function offered by the harmonic basis.
This fact severely handicaps  the convergence of the resultant perturbation
theory. In principle,  the natural solution would be the simultaneous use of 
two harmonic basis centers around the positions of the minima at $~\pm \frac
{1}{2\sqrt\lambda}~$. However,  the implementation of that idea, although
 possible, implies the use of nonorthogonal states which is rather 
cumbersome.\\
 \\

The ~ other~ difficulty is that the theory is {\sl not} defined for $~\lambda 
\rightarrow 0~$, because the ground state does not exist in that limit due to
 the non-existence of a lower limit to $V_{c}$. In that sense, {~\it the SHO
 is not the free-field  limit of the QDWO}. Therefore, the $~ {\sl na\ddot{i}
ve}$ perturbation theory (NPT) is not  applicable as such, to this case. The 
perturbation expansion of the eigen values $~E_{n}( g, \lambda)~$ in~
powers~ of $~\lambda~$ is divergent$^{12,13}$ for all $~\lambda~>~0~$.
 This fact may be understood  qualitatively  by noting  that the 
addition of the term $~\lambda x^{4}~$ turns a completely continuous eigen-
value spectrum of $~p^{2}~-~g~x^{2}~$ into a completely discrete spectrum 
bounded from below. A  nonperturbative treatmant is therefore, necessary. The
WKB method is one such method which is  well suited  especially  for
lower energy eigen values. However, as shown below, the NGAS can  be 
sucessfully applied to the case of the QDWO  in completely analogous 
procedure as in the case of the QAHO, for a considerable larger range
of values of $~ n, g~$ and $~\lambda$.\\

{\large \bf  5.2 Application of NGAS to the QDWO }\\

{\bf 5.2.1 The Gap Equation for the QDWO }\\

 To  develop  the  NGAS for  QDWO  we use the  identical {\it ansatz} for $~V
(\phi)~$ as given in eqn.(64). The parameters A, B, C can also  be 
determined self-consistently as before.  The `` Effective Hamiltonian(EH)''
 for the system is analogously defined as:

\begin{equation}
 H_{0}(\phi,p)~=~\frac{1}{2}p^{2}~-~\frac{1}{2}~g~\phi^{2}+\lambda V(\phi);
 ~~\lambda,~  g~>~0
\end{equation}

Using the {\it ansatz}~ for $~V(\phi)~$ given in eqn.(64) in eqn.(114),
 the EH for QDWO is then recast into the  diagonal structure analogous 
to the case of QAHO and is given by

\begin{equation}
   H_0~=~  \frac{1}{2}~p^{2}~+~\frac{1}{2}~\omega^{2}~( \phi - \sigma )~^{2} + h_{0},
\end{equation}

where, the parameters are defined as before:
   
\begin{equation}
           \omega^{2}~=~ g + 2\lambda A,
\end{equation}

\begin{equation}
           \sigma~=~\lambda B/\omega^{2},
\end{equation}

\begin{equation}
h_0~=~\lambda C~-~\frac{1}{2}\omega^{2}{\sigma}^{2}.
\end{equation}

In this case  also, the EH defined in eqn.(115) is to be interpreted as a 
 ``shifted'' harmonic oscillator. Also in this case, the frequency
 $~\omega > ~0~$; and the vacuum configuration-parameter $~\sigma ~ =~$ real,  
are the physical requirement.\\

 Using eqs.(71,72,73) and proceeding as in the case of the QAHO,
one can  express $~H_{0~}$ into the desired diagonal form given by the
eqn.(78) from which one obtains the energy spectrum

\begin{equation}
              E_{n}^{(0)}~=~\omega~ \xi  + h_{0}~ ,
\end{equation}

where $\xi~=~ ( n + 1/2 )~$ ;$~ n~ =~ 0,1,2,.....$\\
                                                     \\
In order to implement the NGAS, the next task is the determination 
of the free parameters `$\sigma$' and `$\omega$' (or, equivalently, A,
B and C occuring in eqn.(64) for the QDWO.\\

For this purpose, taking the quantum average of original Hamiltonian defined 
in eqn.(113) and using eqs.(80)(81)(83) leads the following equation 

\begin{eqnarray}
<n|H|n>~ \equiv~ <n|H_{0}|n> = \omega\xi/2 -~\frac{1}{2}~g~ (\sigma^{2} + 
(\xi / \omega))~
\nonumber\\
+\lambda(\sigma^{4} + 6~\sigma^{2}~(\xi / \omega) + 3~( 1 + 4~\xi^{2} )
/8 \omega^{2}),
\end{eqnarray}

Carrying out the variational minimisation of EH given in eqn.(120) with 
respect to free parameter `$\omega$' i.e., $~\partial< H_{0} >/\partial\omega
 = 0~$ leads to the following ``gap-equation (GE)'' for the QDWO:

\begin{equation}
\omega^{3} - \omega(12\lambda\sigma^{2}-g) - 6\lambda f(\xi) = 0,
\end{equation}

Similar procedure  with respect to  the free parameter `$\sigma$' i.e.,$~
\partial< H_{0} >/\partial\sigma~ =~ 0~$, leads to the following equation
for the ground state(EGS) configuration:

\begin{equation}
 \sigma(4 \lambda \sigma^ {2} - g  + 12 \lambda \xi / \omega ) =0.
\end{equation}

 It is here to be noted  that these equations differ from the analogous 
equations,~(92-93) for the QAHO by the substitution, $g\rightarrow-g$, as 
expected. \\

It is again convenient to solve the EGS first. It should be  noted however 
that, in contrast to the  case of the QAHO,~ there are now {\it two realizable 
quantum phases of the system } corresponding to the solution of eqn.(122) for
 the ground states. These are characterised by the solutions:

\begin{equation}
 4 \lambda \sigma^ {2} = g - 12 \lambda (\xi / \omega),
\end{equation}

and

\begin{equation}
\sigma = 0.
\end{equation}

respectively. The two solutions lead to different ``quantum phases'' for the 
QDWO, which is discussed below.\\

{\bf 5.2.2  Different Quantum- Phases of the QDWO and the\\ 
Critical Coupling}\\

Following~ standard~ terminology  used in the literature,~ the solution
 given by eqn.(123) leads to the ``  Spontaneously Symmetry Broken
 (SSB)'' phase whereas, the other solution, eqn.(124), corresponds to the
 `` Symmetry-Restored  ( SR )'' phase. It is shown below that the dynamic 
realization of the two ``phases'' is controlled  by the coupling `$\lambda$' 
such that the SSB phase is energetically favoured when $\lambda\leq\lambda_{c}$
, whereas the SR phase is preferred for $\lambda>\lambda_{c}$ where, $\lambda_
{c}$ is a `{\it critical}' coupling. To demonstrate this we consider the GE in 
the respective phases:\\

{\bf 5.2.3  Solution of the Gap Equation and Determination of the Spectrum 
in the Different Phases }\\

(i) {\it The SSB-phase of the QDWO}
                                                 \\                             The GE for SSB-phase is obtained by substitution of eqn.(123) in
eqn.(121) and is  given by
\begin{equation}
\omega_{a}^{3} - 2g\omega_{a} + 6\lambda p(\xi) = 0,
\end{equation}

where, $ p (\xi)\equiv 5 \xi - 1/(4 \xi)$ and we have denoted by $\omega_{a}$,
the frequency in the SSB phase. To get {\it physicallly} accepted solution of
 eqn.(125), it is convenient to adopt the `trial' solution  given by:
\begin{equation}
\omega_{a}~ =~ \rho~[e^{i\theta/3}~+~e^{-i\theta/3}] \equiv 2~\rho~ cos(\theta
/3).
\end{equation}

where `$\rho$' and `$\theta$' are to be determined. Substitution of eqn.(126)
in eqn.(125) leads to a determination of `$\rho$' and $~cos~\theta~$:
\begin{equation}
\rho~ =  ~\sqrt {(2g/3)},
\end{equation}

and

\begin{equation}
 cos~ \theta~=~-(\lambda/ \lambda_{c}),
\end{equation}

with
\begin{equation}
\lambda_{c} \equiv (2g/3)^{3/2}/3p(\xi) \equiv \lambda_{c}( g,\xi)
\end{equation}

From eqn.(128), one gets either:
 
\begin{equation}
 \theta~=~\pi + cos^{-1}(\lambda/ \lambda_{c}),
\end{equation}

or

\begin{equation}
\theta~=~\frac {\pi}{2} +sin^{-1}(\lambda/\lambda_{c}).
\end{equation}

However, on stability grounds it is found that eqn.(131) leads to the 
acceptable solution as the corresponding ground state energy lies lower than
that obtained using the other solution, eqn.(130). Thus the 
{\it physical} solution of eqn.(125) is given by  

\begin{equation}
\omega_{a}  = 2 \sqrt{(\frac {2g}{3})}~ cos~[\frac {\pi}{6} + \frac {1}{3}sin^{-1}
(\frac {\lambda}{\lambda_{c}})]
\end{equation}

An estimate of `$\lambda_{c}$' for the ground state and for $~ g~ =~ 1~$ is
$\lambda_{c}~(g~=~1,\xi~ =~1/2)\\ = 0.0362886$. Clearly, the solution given in 
eqn.(131) is valid {\it only when}, $\lambda\leq\lambda_{c}$.\\

The energy-levels in the LO of the SSB-Phase are computed in analogous manner
 to the QAHO as follows. One notes that  
\begin{eqnarray}
E_{n}^{(0)}|^{QDWO}~ \equiv~ < n |H^{QDWO}|n >~ \equiv ~ < n |H_{0}^{QDWO}|n >~
\nonumber\\
=~ \frac {1}{2}\omega\xi -~ \frac{1}{2}~g~ (\frac {\xi}{\omega})~+~ \frac 
{3\lambda}{8 \omega^{2}}( 1 + 4~\xi^{2} )
\nonumber\\
~ + ~\sigma^{2}~(-\frac {1}{2}g + \frac {6\lambda\xi}{\omega}) +\lambda\sigma^
{4},
\end{eqnarray}

where $~H_{0}^{QDWO}~$ is given by eqn.(114) and the QA values, as given in 
eqs.(80-83) have been substituted. From the defining eqn.  of  the SSB -
Phase i.e., eqn.(123) one can substitute for $\sigma^{2}$ and obtain:
\begin{equation}
E_{n}^{(0)}|_{SSB}^{QDWO} = \frac {1}{2}\omega_{a}\xi + \frac {g\xi}{\omega_{a}}
+ \frac {3\lambda}{8\omega_{a}^{2}}( 1 - 20\xi^{2}) - \frac {g^{2}}{16\lambda}
\label{f22}
\end{equation}
The third-term  in eqn.(134) can be further simplified  by the use of the GE 
in eqn.(125) and given by 
\begin{equation}
\frac {3\lambda}{8\omega_{a}^{2}}( 1 - 20\xi^{2}) = (\frac{\xi}{4}) 
(~\omega_{a} - \frac{2g}{\omega_{a}}).
\label{f23}
\end{equation}
Substitution of eqn.(135) in eqn.(134) then leads to the final expression
for the energy levels given by
\begin{equation}
 E_{n}^{(0)}|_{SSB}^{QDWO} = ~(\frac{\xi}{4}) (3~\omega_{a} + \frac{2g}{\omega_
{a}}~) - (\frac{g^{2}}{16\lambda}).
\end{equation}
                                                                        \\        (ii) {\sl The SR- Phase of the QDWO}\\
                                                             \\                  For $\lambda~>~\lambda_{c}$, the SR-Phase is dynamically favoured. The 
stability analysis of the phase-structure confirms this. The  GE, in
this case, is obtained by substituing eqn.(124) in eqn.(121):

\begin{equation}
\omega_{s}^{3} + g\omega_{s} - 6\lambda f(\xi) = 0,
\end{equation}

where, we have denoted by $\omega_{s}$, the frequency in the {\bf SR}-Phase.
 Note that the above equation simply follows from the GE of the QAHO,~ eqn.
(105),~ by the substitution: $~ g\rightarrow-g$, as expected, due to the 
underlying single well shape. The energy levels in this phase can be 
calculated rather easily. On substitution of eqn.(124) in eqn.(133) one 
 obtains the equation:
\begin{eqnarray}
E_{n}^{(0)}|_{SR}^{QDWO}~ \equiv ~ < n |H_{0}^{QDWO}|n >|_{\sigma=0}~
\nonumber\\
=~ \frac {1}{2}\omega_{s}\xi -~ \frac{1}{2} (\frac
{g~\xi}{\omega_{s}})~+~ 
\frac {3\lambda}{8 \omega_{s}^{2}}( 1 + 4~\xi^{2} )
\end{eqnarray}

Rearranging the GE for SR-Phase given in eqn.(137) leads to the following
equation for the last term in eqn.(138):
\begin{equation}
\frac {3\lambda}{8\omega_{s}^{2}}( 1 + 4\xi^{2}) = (\frac{\xi}{4}) 
(~\omega_{s} + \frac{g}{\omega_{s}}).
\label{f27}
\end{equation}
Using  the result obtained in eqn.(139) in the simplified eqn.(138), the 
energy levels for SR-Phase are given by the following simple expression:
\begin{equation}
E_{n}^{(0)}|_{SR}^{QDWO} = (\frac {\xi}{4} ) (3\omega_{s} - \frac{g}{\omega_{s}}).
\end{equation}

which, again follows from the corresponding formula for the QAHO, eqn.(109),
 by the substitution, $~ g\rightarrow-g~$. In eqn.(140), $~\omega_{s}~$ is the
solution of eqn.(137) which can be obtained in analogous manner and is given
 by
\begin{equation}
\omega_{s}~=~(3\lambda f(\xi))^{1/3}[(~\sqrt{( 1 + \rho)} +1)^{1/3}~-~(~\sqrt {(1 + \rho)} -~1)^{1/3}]
\end{equation}

where, $~\rho^{-1}~=~243\lambda^{2}f^{2}(\xi)/g^{3}~$.
\\                                                      
In {\bf Table-2}, we present the energy-levels of the QDWO  in
 the LO,~ over a wide range of ~`$\lambda$' and ~`n' for $g = 1$. The
 results are compared  with an earlier computation$^{36}$, which 
employs a modified perturbation  theory  and includes  correction  up
 to {\it twenty} orders of perturbation. In the same  {\bf Table-2}, 
we also dispaly the improvement of results  obtained by inclusion of 
the first,  non-trivial correction in the improved  perturbation theory 
(IPT), which is discussed in {\bf Chapter 10}. From the comparison with 
earlier calculation$^{36}$, it is seen that the LO results  are already 
quite accurate.\\
                                                                   
In the next  {\bf Chapter}  we consider the case of the sextic- anharmonic
oscillator and the sextic- double-well oscillator in the NGAS.\\
\\

\begin{table}
\vspace {.4in}
{\bf Table- 2: }
{\it  The  computed energy levels of the quartic - DWO in the
 lowest  order of NGAS  for sample values of `$\lambda$' and 
`n' compared with  the results of ref.[36] which includes 
perturbation  correction up to  twenty orders in a ``modified''
 perturbation theory. Also shown are  the results obtained after
 inclusion of the perturbation  correction$~E_{n}^{(2)}$, at the
 next order in IPT}.

  \vspace{0.4in}
  \begin{center}
  \begin{tabular}{ c c c c c }

  \multicolumn{1}{ c }{$\lambda $}&
  \multicolumn{1}{ c }{\quad\quad$ n$ }&
  \multicolumn{1}{ c }{\quad\quad$E^{(0)}_{n}$}&
  \multicolumn{1}{ c }{\quad\quad$E^{(2)}_{n}$}&
  \multicolumn{1}{ c }{ \quad\quad ref.[36]}\\
                                                                 \\
0.1    &\quad\quad0&\quad\quad\quad0.5496&\quad\quad\quad0.4606&\quad
       \quad0.4702\\                                                
       &\quad\quad1&\quad\quad\quad0.8430&\quad\quad\quad0.7553&\quad
       \quad0.7703\\
       &\quad\quad2&\quad\quad\quad1.5636&\quad\quad\quad1.6547&\quad
       \quad1.6300\\
       &\quad\quad4&\quad\quad\quad3.5805&\quad\quad\quad3.7232&\quad
       \quad3.6802\\
       &\quad\quad10&\quad\quad\quad12.192&\quad\quad\quad12.517&\quad
       \quad12.400\\                                                 
                                                                   \\
1.0   &\quad\quad0&\quad\quad\quad0.5989&\quad\quad\quad0.5752&\quad
       \quad0.5800\\
       &\quad\quad1&\quad\quad\quad2.1250&\quad\quad\quad2.0800&\quad
       \quad2.1800\\
       &\quad\quad2&\quad\quad\quad4.2324&\quad\quad\quad4.2600&\quad
       \quad4.2500\\
       &\quad\quad4&\quad\quad\quad9.4680&\quad\quad\quad9.5950&\quad
       \quad9.5600\\
       &\quad\quad10&\quad\quad\quad30.530&\quad\quad\quad30.650&\quad
       \quad30.420\\                                                  
                                                                     \\
10.0   &\quad\quad0&\quad\quad\quad1.4098&\quad\quad\quad1.3752&\quad
        \quad1.3800\\ 
        &\quad\quad1&\quad\quad\quad5.0650&\quad\quad\quad4.9910&\quad
        \quad5.0900\\ 
        &\quad\quad2&\quad\quad\quad9.8660&\quad\quad\quad9.9050&\quad
        \quad9.8900\\
        &\quad\quad4&\quad\quad\quad21.561&\quad\quad\quad21.791&\quad
       \quad21.700\\ 
        &\quad\quad10&\quad\quad\quad66.950&\quad\quad\quad67.820&\quad
       \quad67.620\\
                                                                     \\
100.0  &\quad\quad0&\quad\quad\quad3.1340&\quad\quad\quad3.0650&\quad
         \quad3.0700\\ 
        &\quad\quad1&\quad\quad\quad11.175&\quad\quad\quad11.024&\quad
        \quad11.002\\
        &\quad\quad2&\quad\quad\quad21.638&\quad\quad\quad21.715&\quad
        \quad21.700\\
        &\quad\quad4&\quad\quad\quad47.023&\quad\quad\quad47.505&\quad
        \quad47.200\\
        &\quad\quad10&\quad\quad\quad145.27&\quad\quad\quad147.10&\quad
        \quad146.70\\

\end{tabular}
\end{center}
\end{table}  
\newpage
{\Large \bf 6.  Application of NGAS to the Sextic-Anharmonic and Double 
Well Oscillator }\\
\\
\\

{\large \bf 6.1  Importance of the  Sextic-Anharmonic
 Oscillator and its Applications}\\

The sextic AHO-system is an example of higher anharmonicity, which
 is also widely investigated$^{12,13,133}$. This system is interesting
 and important in its own right as it finds application in diverse 
areas of physics. As a theoretical laboratory, this system 
together with its double-well counterpart provide perhaps the 
simplest examples, for which supersymmetric quantum mechanics (SUSYQM)
 has definite predictions$^{61-71}$ for energy-levels for specific values of
  the Hamiltonian-parameters. Hence, various models and  approximation 
methods can be tested against the exact analytic results of SUSYQM.
 Besides, owing to the higher anharmonicity, the divergence of the
 naive-perturbation~  theory at higher-orders~ becomes ~ even~ more 
severe$^{133}$, $E_{n}~\sim~\Gamma[n(m-1)]~$ for $~\lambda\phi^{2m}~$ 
type of AHO. This result endows the sextic-AHO with added importance
for testing convergent-approximation methods. It may also be noted
that the ``Wick-ordering'' method of Caswell$^{36}$ is
{\it not} directly applicable to this case since the method 
generates $~\phi^{4}~$ counter-terms which are {\it not}
present in the original (non-ordered)  Hamiltonian. For the above stated 
reasons of practical applicability and theoretical importance, the 
sextic-AHO provides a unique testing ground for the NGAS, which is 
described in the following.\\     
\\
\\

{\large \bf 6.2  Application of NGAS to the Sextic-AHO}\\

{\bf 6.2.1 The Gap Equation for the Sextic-AHO}\\

The Hamiltonian for the system is given by 

\begin{equation}
H~=~\frac{1}{2}p^{2}~+~\frac{1}{2}~g~\phi^{2}+\lambda\phi^{6},
\end{equation}

where,$~\lambda ~,g~>~ 0$. As in other cases, 
\begin{equation}
H_{s}(\phi) = \frac{1}{2}p^{2}~+~\frac{1}{2}~g~\phi^{2},
\end{equation}

corresponds to the free-field Hamiltonian. The interaction-term is, therefore,

\begin{equation}
 \lambda H_I(\phi) = \lambda\phi^{6}.
 \end{equation}

To apply the NGAS, we follow identical steps as in previous cases.
 An identical~ ansatz is assumed for the potential $~V(\phi)~$, which is given 
in eqn.(64). Hence the  Effective Hamiltonian (EH) for this case is 
analogously defined as:

 \begin{equation}
 H_{0}~=~\frac{1}{2}p^{2}~+~\frac{1}{2}~g~\phi^{2}+\lambda~V(\phi)
\end{equation}

Substituting the {\it ansatz} for $~V(\phi)~$  in eqn.(64) in eqn.(145),
the EH can   be recast  into the diagonizable structure, as in earlier cases:
\begin{equation}
H_0~=~  \frac{1}{2}~p^{2}~+~\frac{1}{2}~\omega^{2}~( \phi - \sigma )~^{2} +
h_{0},
\end{equation}

where, analogously
\begin{equation}
\omega^{2}~=~ g + 2\lambda A,
\end{equation}

\begin{equation}
\sigma~=~\lambda B/\omega^{2},
\end{equation}

\begin{equation}
h_0~=~\lambda C~-~\frac{1}{2}\omega^{2}{\sigma}^{2}.
\end{equation}

The interpretation of $~H_{0}~$ in eqn.(146) is also identical- it is the 
Hamiltonian for a ``shifted'' effective harmonic oscillator.
Following steps identical to those in the case of quartic-oscillators, the 
diagonal form for $~H_{0}~$ is obtained as:
\begin{equation}
H_{0}~=~\omega~ (N_{b} + 1/2 ) + h_{0}~ ,
\end{equation}

where again  $~N_{b}~\equiv~b^{\dagger}b~$ with $~b,b^{\dagger}~$ defined as
 before (see eqs.(71-72)). The eigen value equation as given in eqn.(54)
 then leads to  the energy spectrum  as given below:
\begin{equation}
E_{n}^{(0)}~=~\omega~ \xi  + h_{0}~ ,
\end{equation}

where $\xi~=~ ( n + 1/2 )$; $~ n~ =~ 0,1,2,.....$ and $~h_{0}~$ is given by
eqn.(149). For the determination of the parameters `$\omega$' and `$\sigma$',
we use, as before, the variational minimisation conditions: $~\partial< H_{0} >
/\partial\omega = 0~$ and  $~\partial< H_{0} >/\partial\sigma~ =~ 0~$. For 
this purpose  the quantum  average of eqn.(142) evaluated as: 

\begin{equation}
  < H >~= < H_{0} >~= ~\frac{1}{2}< p^{2} >~+~\frac{1}{2}~g~< \phi^{2} >+
  \lambda < \phi^{6} >
\end{equation}

On substitution of the results given in eqs.(80,81,84) in eqn.(152),
the following equation results:

\begin{eqnarray}
  <n|H|n>~ \equiv~ <n|H_{0}|n> = \omega\xi/2~  +~\frac{1}{2}~g~ (\sigma^{2} +
  (\xi / \omega)~
  \nonumber\\
  +\lambda(\sigma^{6} + 15~\sigma^{4}~(\xi / \omega) + \frac {45\sigma^{2}~
( 1 + 4~\xi^{2} )}{8 \omega^{2}} + \frac {5\xi(4\xi^{2} + 5)}{8\omega^{3}}),
~g~> 0
\end{eqnarray}

Carrying out variational minimisation of eqn.(153) with respect to the
frequency parameter `$\omega$' we obtain the relation

\begin{equation}
\omega^{4}~-\omega^{2}(g+30\lambda\sigma^{4})-45\lambda(\sigma^{2}\omega/2\xi)
(1+4\xi^{2}) -(15\lambda/4)(5+4\xi^{2}) = 0,
\end{equation}

Further, carrying out the variational minimisation with respect to `$\sigma$'
leads the following equation: 

\begin{equation}
\sigma~[ g+6\lambda(\sigma^{4}+ \frac {10\xi\sigma^{2}}{\omega} + \frac {15(4\xi^{2}+1)}{8\omega^{2}})~]=0.
\end{equation}

Eqn.(155) which defines `$\sigma$' provides the corresponding ground
 state~ configuration(EGS) of this system.\\

As in earlier cases, it is convenient to first analyse the EGS which 
leads to the following relations:

\begin{equation}
\sigma = 0
\end{equation}

and

\begin{equation}
  g + 6\lambda(\sigma^{4}+10\xi\sigma^{2}/\omega+15(4\xi^{2}+1)/8
\omega^{2})~=~ 0.
\end{equation}

On  inspection of eqn.(157), it is easily seen that {\it no}
 physical solution for $~\sigma^{2}~$ (i.e.$~\sigma^{2}~>~0~$ ) exists 
for $~ g~ >~ 0$, $~\lambda~>~0$. Hence, eqn.(157) is {\it not} 
physically realisable for the sextic  AHO. Therefore, the  
{\it physical} ground state of the sextic AHO is uniquely determined by the
`$\sigma=0$' solution as given by eqn.(156).\\

{\bf 6.2.2  Solution  of the Gap Equation and Determination of the Energy
Spectrum}\\

The GE given in eqn.(154) simplifies  to the following form on 
substitution of the correct EGS expression, i.e. $~\sigma~=~0~$:

\begin{equation}
 \omega^{4}-g\omega^{2} -(15\lambda/4)(5+4\xi^{2}) = 0
 \end{equation}

The parameters A, B, C defining $~V(\phi)~$ can be determined in analogous 
manner as given below. Rearranging the gap eqn.(154) and using the eqn.(147)
`A' is dtermined as

 \begin{equation}
 A=~15~\sigma^{4} +45\sigma^{2}( 1+4\xi^{2})/4\xi\omega +(15/8\omega^{2})
 ( 5+4\xi^{2} );
 \end{equation}

 Similarly, following the same procedure, `B' is calculated using eqn(148) and
 the EGS given in eqn.(155) and given by

 \begin{equation}
 B=~\sigma~[(1+g)(\omega^{2}/\lambda)
 +6\omega^{2}\sigma^{4} +60\sigma^{2}\omega\xi +(45/4)( 1+4\xi^{2})~];
 \end{equation}

To determine `C' we use the equation as given below: 

 \begin{equation}
 C=~<\phi^{6}>-A~<\phi^{2}> +B~<\phi>,
 \end{equation}

together with expressions for $~<\phi^{6}>,~~<\phi^{2}>~and~<\phi>~$ as 
given in eqs.(80-84). On substitution of the physical solution,
 eqn.(156) for the EGS in  eqn.(153) and using the GE as given in 
eqn.(158) one obtains the following simple  expression for the 
energy-levels of the sextic AHO in the LO:

\begin{equation}
E_{n}^{(0)}|_{sextic-AHO}=\frac{\xi}{3}(2\omega+\frac{g}{\omega}); ~ g>0.
\end{equation}

where, ofcourse, `$\omega$' is the solution of the GE given in eqn.(158).
In {\bf Table-3}, sample results for the energy levels of the sextic AHO
computed in LO are presented for $g=1$, over a wide range of
`$\lambda$' and `n' and compared to the results$^{133}$ ( shown
within parenthesis) with percentage deviation from the latter (shown
within square bracket). It can be seen from this tabulation that the
LO- results obtained in the NGAS are quite accurate, compared to those 
obtained by sophisticated numerical calculations$^{133}$. Further 
improvement in accuracy is achieved by application of the IPT as
discussed in {\bf Chapter-10}.

\begin{table}
\vspace {.4in}
{\bf Table- 3: }
{\it Sample results in the lowest order(LO) of  (NGAS) for the 
sextic - AHO over a wide range of `$\lambda$' and `n' compared for $~g~=~1~$
with the results of ref.[133](shown in the parentheses). The relative
percentage  error is shown in square brackets}.
\vspace{0.4in}
\begin{center}
\begin{tabular}{c c c c c c c c }
\multicolumn{1}{ c }{n}&
\multicolumn{1}{ c }{\quad$\lambda=0.2$ }&
\multicolumn{1}{ c }{\quad$2.0$}&
\multicolumn{1}{ c }{\quad$10.0$}&
\multicolumn{1}{ c }{\quad$100.0$}&
\multicolumn{1}{ c }{\quad$400.0$}&
\multicolumn{1}{ c }{\quad$2000.0$}\\
                                                               \\
      0    &1.193&1.676&2.323&3.947&5.521&8.206\\
           &(1.174)  &(1.610) &(2.206) &(3.717) &(5.188) &(7.702)\\
           &[1.611]&[4.079]&[5.313]&[6.188]&[6.415]&[6.544]\\
                                                             \\
      1    &3.966&5.931&8.420&14.52&20.39&30.37\\
           &(3.901)&(5.749)&(8.115)&(13.95)&(19.56)&(29.12)\\
           &[1.681]&[3.165]&[3.762]&[4.148]&[4.244]&[4.298]\\
                                                            \\
      2    &7.420&11.61&16.74&29.16&41.03&61.18\\
           &(7.382)&(11.54)&(16.64)&(28.98)&(40.78)&(60.81)\\
           &[0.523]&[0.612]&[0.6179]&[0.6157]&[0.6145]&[0.6138]\\
                                                                \\
      4    &16.15&26.48&38.73&68.01&95.90&143.2\\
           &(16.30)&(26.83)&(39.29)&(69.05)&(97.38)&(145.4)\\
           &[0.9170]&[1.302]&[1.426]&[1.499]&[1.517]&[1.527]\\
                                                               \\
      6    &26.88&45.08&66.36&117.0&165.1&246.5\\
           &(27.29)&(45.94)&(67.70)&(119.4)&(168.5)&(251.7)\\
           &[1.50]&[1.870]&[1.98]&[2.043]&[2.058]&[2.067]\\
                                                                \\
     10    &53.24&91.17&135.0&238.7&337.1&503.8\\
           &(54.31)&(93.26)&(138.2)&(244.5)&(345.3)&(516.1)\\
           &[1.967]&[2.245]&[2.323]&[2.367]&[2.377]&[2.383]\\
                                                               \\
     14    &85.01&147.0&218.3&386.6&546.2&816.3\\
           &(86.78)&(150.4)&(223.4)&(395.7)&(559.1)&(835.6)\\
           &[2.047]&[2.230]&[2.279]&[2.306]&[2.313]&[2.316]\\                                                              \\
     17    &111.9&194.4&289.0&512.1&723.7&1082.0\\
           &(114.0)&(198.3)&(294.9)&(522.7)&(738.6)&(1104.0)\\
           &[1.868]&[1.974]&[2.001]&[2.016]&[2.020]&[2.022]\\

\end{tabular}
\end{center}
\end{table}
\newpage
{\large \bf 6.3  The  Sextic-Double Well Oscillator (Sextic-DWO) }\\
\\
\\
{\bf 6.3.1  Importance of Sextic-Double Well Oscillator}\\

The sextic  double well oscillator has been the subject of much
investigation and discussion for a
pretty long time$^{134}$, both from the analytical and the 
numerical points of view. This is because it has important applications in 
quantum field theory and molecular physics$^{135}$ and in other branches
also. By using Hill-determinant method, Biswas et al$^{27}$ have 
calculated 
the eigen values of the oscillators of the type $~\lambda x^{2n}~$. Banerjee
 and Bhattacharjee have 
obtained energy eigenvalues for the potential of the form $~x^{2} + \lambda 
x^{4}~$ using scaled Hill determinant method$^{136}$.  However, 
it has been
pointed out that$^{137}$ this method has limited domain of applicability
 for the sextic oscillator potential of the type$~V(x)~=~\mu x^{2}~ -
~\lambda x^{4} ~+~ \eta x^{6};~ \eta ~>~0~$. There have been some  
improvements$^{138}$ of the procedure for removal$^{139,140}$ of 
dificulties in the Hill-determinant approach$^{141}$. The potential 
cited 
above  is particularly of great interest in scalar field theory$^{142}$ 
and also in the calculations of the vibrational spectra of 
molecules$^{143}$.
 It has been discussed by authors of ref.[144] that the potential 
admits exact analytic solutions for the ground state under certain conditions.
It has also been shown that$^{145}$ the potential is ``quasi exactly
solvable (QES)'' i.e., exact solutions can be obtained if the coupling 
constants  satisfy some constraints. Moreover, supersymmetric quantum 
mechanics (SUSYQM) can be applied$^{71}$ to compute the eigen values
of the double-well potential $~x^{6}~-~3x^{2}~$ and the spectrum can
be related to that of the ``partner'' potential: $~x^{6}~+~3x^{2}~$,
~which is the anharmonic counter-part  . We apply our scheme of
~approximation in the next section, to calculate the energy eigen
values for the sextic-DWO as a method of~ confirmation of our scheme, as 
well as, to test the various exact results following from QES and SUSYQM.
                                                                     \\
                                                                      \\
{\bf 6.3.2  Application of NGAS to the Sextic-DWO}\\

The Hamiltonian for such a system is given by the following expression:

\begin{equation}
H~=~\frac{1}{2}p^{2}~-~\frac{1}{2}~g~\phi^{2}+\lambda\phi^{6},
\end{equation}

where,$~\lambda ~>~ 0$ and  $g~>~0~$. To  apply NGAS, we follow
identical steps as in previous case of the QDWO. An identical ansatz
is suggested for the potential $V(\phi)$ which is given in eqn.(64). Hence the
( Effective Hamiltonian (EH)) for this case can be written
as:

\begin{equation}
H_{0}~=~\frac{1}{2}p^{2}~-~\frac{1}{2}~g~\phi^{2}+\lambda~V(\phi)
\end{equation}

On Substitution the value for $V(\phi)$ given in eqn.(64) in eqn.(164) the 
EH is now  expressed into the diagonizable structure:

\begin{equation}
H_0~=~  \frac{1}{2}~p^{2}~+~\frac{1}{2}~\omega^{2}~( \phi - \sigma )~^2 + h_{0},
\end{equation}

where,

\begin{equation}
\omega^{2}~=~ -g + 2\lambda A,
\end{equation}

\begin{equation}
\sigma~=~\lambda B/\omega^{2},
\end{equation}

\begin{equation}
           h_0~=~\lambda C~-~\frac{1}{2}\omega^{2}{\sigma}^{2}.
\end{equation}

The parameters A,B,C can be determined  through~ identical procedure as 
in earlier cases. In particular, we get back relations given by
eqs.(159-161) with the substitution: $g~\longleftrightarrow~-g$.
Again the eqn.(165) can be regarded  as  the hamiltonian for the
``shifted'' effective harmonic oscillator for sextic- DWO, which is
identical with  the equations developed earlier for the case of QAHO
eqn.(67), QDWO(eqn.(115) and  sextic-AHO (eqn.(146)). This, therefore, 
confirms  the genraral applicability of the method formulated  in
{\bf Chapter 3}. As usual `$~\omega~$' is identified as the frequency 
of the ``shifted'' harmonic oscillator which has the restriction of  satisfying
the physical requirement $~\omega~ > ~0~$. The parameter `$~\sigma~$' accounts 
for the field-shift of the ``shifted'' harmonic sextic-DWO which must
be real, since the field is hermitian, $~\phi~ =~\phi^{\dagger}$. Parameters 
 `$~\omega~$' and `$~\sigma~$' are  determined by following the
 analogous procedure ( see below ). Finally the  EH  defined in 
eqn.(165) is transformed  by standard method into  diagonalised  form as:

\begin{equation}
H_{0}~=~\omega~ (N_{b} + 1/2 ) + h_{0}~ ,
\end{equation}

where the creation- and annihilation  operators and  ETCR  given in
eqs.(71-73) have been used. The  energy spectrum for sextic-DWO
is then trivially obtained as:

\begin{equation}
E_{n}^{(0)}~=~\omega~ \xi  + h_{0}~ ,
\end{equation}

where $\xi~=~ ( n + 1/2 )~$ ;$~ n~ =~ 0,1,2,.....$ and $~h_{0}~$ is given by
eqn.(168).
                                                                             \\
                                                                             \\
{\bf (i) The Gap-Equation  and  the different Quantum- Phases of the Sextic-
QDWO}\\
                                                                           \\
                                                                            \\
For the determination of the parameter `$~\omega~$' and  `$~\sigma~$'
we use the  standard  variational-minimisation conditions  $~\partial<
H_{0} >/\partial\omega~ = ~0~$ and $~\partial< H_{0} >/\partial\sigma~
=~ 0~$. On taking the quantum average of eqn.(163) and  taking the
advantage of eqn.(55) we obtain $~<H_{0}>~$ as given below:

\begin{eqnarray}
<n|H|n>~ \equiv~ <n|H_{0}|n> = \omega\xi/2 ~-~\frac{1}{2}~g~ (\sigma^{2} +
(\xi / \omega))~
\nonumber\\
+\lambda(\sigma^{6} + 15~\sigma^{4}~(\xi / \omega) + 45~\sigma^{2}( 1 + 4~\xi^{2} )
/8 \omega^{2} + \frac {5\xi(4\xi^{2} + 5)}{8\omega^{3}}),  
\end{eqnarray}

From  the variational minimisation of eqn.(171) with respect to
`$~\omega~$', we obtain the relation

\begin{equation}
\omega^{4}~-\omega^{2}(-g+30\lambda\sigma^{4})-45\lambda(\sigma^{2}\omega/2\xi)
(1+4\xi^{2}) -(15\lambda/4)(5+4\xi^{2}) = 0,
\end{equation}

Eqn.(172) corresponds to ``gap-equation'' GE for the case of sextic-DWO.
Similarly, from the variational minimisation of eqn.(171) with respect to 
`$~\sigma~$' leads to the following equation:

\begin{equation}
\sigma~[ -g+6\lambda(\sigma^{4}+10\xi\sigma^{2}/\omega+15(4\xi^{2}+1)/8\omega^{2})~]=0.
\end{equation}

Eqn.(173) has the significance of defining the  ground state (EGS).
This~ equation leads, as~ in~ the~ case of the~ QDWO, to ~the~ SSB-phase and
the SR-phase, corresponding  to the~ `$\sigma^{2} \neq 0$'~ and~ the~
`$\sigma=0$' solutions~ respectively. {\it It is found on
}{\it computation, however, that the SR- phase is
 energetically favoured for all values of}{\it $~\lambda~$ since the  energy 
 levels for this phase, always lie below  the  corresponding  ones for the SSB-
phase}. This result has important consequence in the context of  supersymmetry, as discussed subsequently.
                                                                       \\
                                                                        \\
{\bf (ii) Solution of  the Gap-Equation and determination of the energy 
spectrum of the SR-phase}\\                                                                                                    
                                                                   \\
 The GE, in the case of sextic-QDWO for the SR-phase, is simply  obtained
 by substituting `$\sigma=0$' in the ``gap-equation'' eqn.(172)
which  corresponds to the GE for the AHO (eqn.(158)) by the
replacement, $~g\rightarrow~-g~$: 

\begin{equation}
\omega_{a}^{4} + g\omega_{a}^{2} - (\frac{15\lambda}{4})(5+4\xi^{2}) = 0,
\end{equation}

where again, we have distinguished the frequency of the DWO, by the
 subscript `a'. To determine the energy levels one substitutes
`$~\sigma~ =~0$' in the eqn.(171) for $~<H_{0}>~$ one obtains:
\begin{equation}
E_{n}^{(0)}|_{sextic-DWO}~ \equiv~ < H_{0} >|_{\sigma = 0} =
\frac{\omega_{a}\xi}{2} ~-~\frac{1}{2}(\frac {g\xi}{\omega_{a}})
~+~\frac {5\lambda\xi}{8\omega_{a}^{3}}( 5 + 4\xi^{2} ).
\end{equation}

Next, using the GE, eqn.(174), the last term in eqn.(175) can be
simplified  as:
\begin{equation}
\frac {5\lambda\xi}{8\omega_{a}^{3}}( 5 + 4\xi^{2} ) = (\frac
{\xi}{6})(\omega_{a} + \frac{g}{\omega_{a}}).  
\end{equation}
Substitution of the above equation in eqn(175) then leads to energy
spectrum
given below:

\begin{equation}
 E_{n}^{(0)}|_{sextic-DWO}=\frac{\xi}{3}(2\omega_{a} - \frac{g}{\omega_{a}}),
~ g > 0.
\end{equation}

In {\bf Table-4}, sample results for the energy levels of the sextic DWO
computed in the  LO are presented  and
 are compared to the results  of ref.[71]. It can be
 seen from this tabulation that the LO- results obtained in NGAS are quite
accurate, compared to those obtained by sophisticated numerical calculations
of ref.[71]. On comparision it is seen that there is good agreement for
 all values of `n', expect for the ground state. This is further discussed
in the next {\bf Chapter}. Further improvement in accuracy is achieved by 
application of IPT as discussed in {\bf Chapter-10}.\\

In the   {\bf Chapter-7}, we consider the implication of supper symmetric 
quantum  mechanics (SUSYQM) for the case of the sextic oscillators and 
comparision with  the results of NGAS.\\
\\
\\

\begin{table}
\vspace{.4in}
{\bf Table- 4: }
 {\it Sample results  for the energy levels of the  sextic- AHO and DWO in the
 LO of NGAS (for $\beta = 1$) displaying the approximate validity of ISPP
 relation(see eqn.(191)). Also shown for comparision, are the corresponding
 results of ref.[71] obtained by numerical methods based upon SUSY}. 
( Note that the ISPP relations are preserved for arbitray value of `$\beta$' 
due to the `scaling' property given by eqn.(191) of the text ).

\vspace{.4in}
\begin{center}
\begin{tabular}{ c c c c c }
\multicolumn{1}{ c }{\quad\quad$n$ }&
\multicolumn{1}{ c }{\quad\quad\quad$E_{n}^{(AHO)}$}&
\multicolumn{1}{ c }{\quad\quad\quad$E_{n+1}^{(DWO)}$}&
\multicolumn{1}{ c }{\quad\quad\quad$E_{n}^{ (AHO)}$}&
\multicolumn{1}{ c }{\quad\quad\quad$E_{n+1}^{(DWO)}$}\\

      &\quad           &\quad          &\quad ref.[71]&\quad ref.[71]
\\
                                                          \\
  0   &\quad1.95608&\quad2.38721&\quad1.93548&\quad1.93548\\
  
  1   &\quad6.37732&\quad6.24897&\quad6.29849&\quad6.29849\\
  
  2   &\quad11.7352&\quad11.3668&\quad11.6810&\quad11.6810\\
  
  3   &\quad17.9931&\quad17.4785&\quad18.0426&\quad18.0426\\
  
  4   &\quad25.0597&\quad24.4375&\quad25.2546&\quad25.2546\\
  
  5   &\quad32.8581&\quad32.1484&\quad33.2261&\quad33.2261\\
  
  6   &\quad41.3276&\quad40.5427&\quad41.8910&\quad41.8910\\
  
  7   &\quad50.4197&\quad49.5679&\quad51.1979&\quad51.1979\\
  
  8   &\quad60.0950&\quad59.1822&\quad61.1053&\quad61.1053\\
  
  9   &\quad70.3204&\quad69.3513&\quad71.5790&\quad71.5790 \\
  
 10   &\quad81.0680&\quad80.0462&\quad82.5899&\quad82.5899 \\
 
 11   &\quad92.3136&\quad91.2421&\quad94.1129&\quad94.1129 \\
 
 12   &\quad104.036&\quad102.917&\quad106.126&\quad106.126 \\
 
 13   &\quad116.217&\quad115.053&\quad118.611&\quad118.611  \\
 
 14   &\quad128.839&\quad127.632&\quad131.549&\quad131.549  \\
 
 15   &\quad141.889&\quad140.640&\quad144.927&\quad144.927  \\
 
 16   &\quad155.351&\quad154.062&\quad158.728&\quad158.728 \\
 
 17   &\quad169.214&\quad167.887&\quad172.942&\quad172.942  \\
 
 18   &\quad183.467&\quad182.102&\quad187.557&\quad187.557  \\
 
 19   &\quad198.099&\quad196.698&\quad202.561&\quad202.561  \\
 
\end{tabular}
\end{center}
\end{table}
\newpage

{\Large\bf 7.  Comparision of the Results of NGAS with the Exact
 Predictions of Super Symmetry  for the Sextic Oscillator}\\
\\

{\large\bf 7.1   Super Symmetric Quantum Mechanics (SUSYQM) in Brief }\\

Supersymmetry (SUSY) is a symmetry beween Fermions and Bosons$^{6}$ which
  was invoked  by physicists  to obtain a unified 
description of all basic interaction of nature although there has been no 
experimental evidence of SUSY being realised in nature. Nevertheless, in the
 last fifteen years, the ideas of supersymmetry have stimulated new approaches 
 and applications to the~ different~ branches of physics including
 atomic, molecular, statistical and condensed  matter physics as well
 as non-relativistic quantum mechanics$^{6}$.\\

 In this section, we will be concerned with the application of SUSY in the
area of non-relativistic quantum mechanics - a field now known as super
symmetric quantum mechanics (SUSYQM). The investigations in this area have 
rapidly proliferated in the past two-decades which have established~ SUSYQM~ 
 as an~ important and~ ineresting area of reaserch. The application of SUSY to 
quantum mehanics has yielded several new and significant results. In particular, the class of potentials yielding exact analytic solution of the non-
relativistic ~Schr$\ddot{o}$dinger's~ equation have been classified and
 understood readily on the basis of SUSY. Moeover, the results and concept 
of ``partner-potentials'' yielding iso-spctral hamiltonians are among the 
significant predictions of SUSYQM. For reviews of the subject, the 
ref.[6] may be consulted. \\

Our immediate interest in the subject of SUSYQM  arises  in this thesis from
 the fact that the latter has exact-predictions of iso-spectrality of the 
 sextic-AHO and the sextic-DWO which form partner hamiltonians for specific
 choices of the parameters defining the potentials. To this topic, we turn to
in the next section.\\

{\large\bf 7.2  Super Symmetry and Iso-spectrality of Partner Potentials}\\

One of the simplest non-trivial applications of SUSYQM$^{12,13}$  is
 made for the case of the sextic-oscillators (AHO and DWO). Consider the 
 ``super potential'' ref.[6]:

\begin{equation}
W{(\phi)}=\beta \phi^{3},
\end{equation}

This is the simplest anharmonic, supersymmetric and ``parity even'' potential
 with no known analytic solutions. This is perhaps the simplest choice beyond 
the linear solvable case. Acording to SUSY quantum mechanics$^{6}$ the 
`` partner-potentials'' are given by:

\begin{equation}
V^{(-)} \equiv \frac{1}{2} (W^{2} -  W') 
\end{equation}

\begin{equation}
V^{(+)} \equiv \frac{1}{2} (W^{2} +  W') 
\end{equation}

By using  eqn.(178) in eqs.(179) and (180) the ``partner-potentials'' are
generated as given below:

\begin{equation}
V^{(-)} = \frac{1}{2}(\beta^{2}\phi^{6} -  3\beta\phi^{2})
\end{equation}

\begin{equation}
V^{(+)}  = \frac{1}{2}(\beta^{2}\phi^{6} +  3\beta\phi^{2})
\end{equation}

Both the ``partner-potentials'' correspond to parity- even, confining 
potentials. One of these, $~V^{(-)}~$ corresponds to the double-well. The 
ground state
energy of the ``partner-potential'' $V^{(-)}$ occurs at E = 0. In this
case there is no trace of the double well structure as the zero-point
energy is just equal to the well depth.\\

The potentials for the sextic DWO and the  AHO which are used in the 
Hamiltonian given  by the eqn.(163) and eqn.(142) respectively can be written 
as:   

\begin{equation}
V_{1,2}~ =~\lambda\phi^{6} \mp \frac {1}{2} g\phi^{2}
\label{70,7.6}
\end{equation}
 where $~\lambda~>~0~$ and $~g~>~0~$. On the other hand, the  Hamiltonians
 corresponding to the  ``partner-potentials''are:

\begin{equation}
H^{(-)} = \frac{1}{2}p^{2} +  V^{(-)}
\end{equation}
 
\begin{equation}
H^{(+)} = \frac{1}{2}p^{2} +  V^{(+)}
\end{equation}

In standard notation, the  `exact' results of SUSY for the above Hamiltonians
 as given in eqs.(184-185) can be summarized as follows$^{6}$:\\

\begin{equation}
(i)~~~~~~~~~~~~~~~~~~~~~~~~E_{n+1}^{(-)} = E_{n}^{(+)},~~~~~~~~~~~~~~~~~~~~~~~~~~
\end{equation}

\begin{equation}
(ii)~~~~~~~~~~~~~~~~~~~~~~~~E_{0}^{(-)} = 0,~~~~~~~~~~~~~~~~~~~~~~~~~~~~~~~
\end{equation}

\begin{equation}
(iii)~~~~~~~~~~~~~~~~~~~~~~~\psi_{0}^{(-)}(\phi) = A~ exp(-\int^{\phi} W(y)dy),~
\end{equation}

where, n=0,1,2,.....; the ground state wave function for $H^{(-)}$ is denoted
by  $\psi_{0}^{(-)}(\phi)$ and `A' denotes its normalisation. The property 
given by eqn.(186) is referred to [6] as 
{\it ``Iso-spectrality'' of Partner Potentials} (ISPP). Eqn.(187) is a 
rigorous result of exact (unbroken) super 
symmetry, while  eqn.(188) is the prediction for the ground state wave 
function of $H^{(-)}$.\\

  Application of eqs.(181,182,184-188) to the case of the sextic AHO and DWO
  characterised by potentials in eqn.(183) becomes  at  once  obvious, when the
 following specific values for `$\lambda$'  and  `$g$' are  chosen:
\begin{equation}
 \lambda~=~ \frac{1}{2}\beta^{2},~  g = 3\beta;~  \beta ~ >~0.
\end{equation}

For the above choice of `$\lambda$' and `$g$', eqn.(186) can then be  rewritten 
as:

\begin{equation}
 E_{n+1}^{(DWO)}(\lambda,g ) = E_{n}^{(AHO)}(\lambda,g)
\end{equation}\\

{\large\bf 7.3  Comparision of the  Energy Levels obtained in NGAS (LO) with 
the  Exact Results from SUSY}\\

{\it Surprisingly, it is found that the relation given in eqn.(186)(or
 eqn.(190)) is obeyed to a very good accuracy by the LO results from NGAS
 for all allowed values of $~\lambda > 0!~$} In {\bf Table-5}, we demonstrate
the (approximate) validity of the ISPP relation  of SUSY in NGAS, by
comparing the energy level of the DWO for the excitation label `(n+1)'
with that of the AHO for the label `n'. {\it The agreement is seen to
be impressive, particularly at large values of `n', considering that
only the LO-results are used}. It may be observed, in this context, that the
formulae for energy levels of sextic oscillators in NGAS, given by eqs.(162) 
and (177), obey the following  interesting {\sl ``scaling'' law}:

\begin{equation}
 E_{n}^{(0)}(\beta ) = \sqrt\beta E_{n}^{(0)}(1)
\end{equation}

{\it This scaling property guarantees the validity of the ISPP relation,
eqn.(190) for arbitrary values of $~\beta~$, once the relation is established 
for any particular given value of the latter}.\\

{\large\bf 7.4  Positivity Property of the Energy Eigenvalues of the 
Sextic-DWO predicted by  SUSYQM and NGAS }\\

The other observation is regarding the {\sl ``positivity''} property of the
energy- eigen values of the sextic-DWO predicted by SUSY through the eqs.
(186-187), which is otherwise {\sl not} obvious owing to the double-well 
structure of the potential (at least, this is not the case for the SSB-phase of
 the QDWO!). Interestingly, {\sl the positivity of the energy levels of the 
sextic-DWO, as predicted by SUSY, is also dynamically realized in }NGAS. This 
is because of the fact that the SSB-phase is ruled out on grounds of stability
(see, remarks following  eqn.(173)).\\

As a final confirmation of consistency with the exact results from SUSY on
 energy levels of above systems, it is necessary to establish not only the ISPP
-relation, but also the {\it absolute magnitudes} of the former. In 
{\bf Table-4(page-66)}, we  compare the results of ref.[71], on the 
energy levels of
 sextic-oscillators obtained by sophisticated numerical methods, with those 
based upon the simple formulae, eqs.(162,177) in LO of NGAS for $\lambda =
0.5$ and $~ n\leq 20~$. It can be seen from this comparison that there is good
agreement for all values of `n', {\it except for the ground state-energy of the
DWO}. In the latter case there is some deviation from the exact result of SUSY,
 given by eqn.(146). It may be plausible that the discrepancy could be due to
 the departure from the predicted {\it exact} ground state wave function as 
given by eqn.(147) from the wave function of the DWO in LO of NGAS. This 
aspect is further investigated below:\\

{\large\bf 7.5   The DWO-Ground State Wave Function in NGAS and
 SUSYQM}\\

Having demonstrated the approximate validity  in NGAS, of the ISPP relations
and positivity property  of energy levels predicted by SUSY, it remains to
 compare the respective ground state wave functions. The exact result from
SUSY is given by  eqn.(188). For the case of sextic DWO, this result can be 
made  more specific as given below. Using eqn.(178) in eqn.(188) the 
ground state wave function is written as:
\begin{equation}
\psi_{0}^{(-)}~ =~ A~ exp(- \beta\phi^{4}/4)
\end{equation}

The normalisation constant `A' can be calculated easily and is given by
\begin{equation}
A~=~(8\beta)^{1/8}/\sqrt{\Gamma (1/4)},
\end{equation}
where $\Gamma(z)$ is the Euler Gamma-function. Hence, the SUSY prediction for 
the ground state  wave function is given by  
\begin{equation}
\psi_{0}^{(DWO)}(\phi,\beta)|_{SUSY} = (8\beta)^{1/8}  exp(-\beta\phi^{4}
/4)/\sqrt\Gamma(1/4)
\end{equation}

Using the value of  $~\Gamma(1/4)~$, eqn.(7.17) can be approximately 
represented as:  
\begin{equation}
\psi_{0}^{(DWO)}(\phi,\beta)|_{SUSY} = (0.68108)\beta^{1/8}  exp(-\beta\phi^{4}
/4).
\end{equation}
On the other hand, the ground state wave function in LO-NGAS corresponds to
 that of {\it an effective simple harmonic oscillator} with variable frequency
determined by the corresponding~ {\it gap-equation}. For the case of the
sextic- DWO, the NGAS result is given by
\begin{equation}
\psi_{0}^{(DWO)}(\phi,\beta)|_{NGAS} = (\frac{\omega_{a}}{\pi} )^{1/4}e^
{-\omega_{a}\phi^{2}/2} 
\end{equation}

where `$\omega_{a}$' is the solution for the ground state of
``gap-equation'' given in eqn.(174) i.e., 
\begin{equation}
\omega_{a}^{4} + g\omega_{a}^{2} - (\frac{45}{2})\lambda = 0
\end{equation}

The physical, acceptable solution of the above ``gap-equation'' is given by
\begin{equation}
\omega = [(\sqrt{(g^{2} + 90\lambda)} - g)]^{1/2}
\end{equation}

On substitution of $~\lambda ~=~\frac{1}{2}\beta^{2};~g~=~3\beta~$, the  
eqn.(198) leads the following result

\begin{equation}
(\frac{\omega(\beta)}{\pi})^{1/4} = (\frac{1.4747}{\pi})^{1/4}\beta^{1/8} = 0.828\beta^{1/8} 
\end{equation}

Using eqn.(7.22) in eqn.(7.19) we have the following equation

\begin{eqnarray}
 \psi_{0}^{(DWO)}(\phi,\beta)|_{LO-NGAS} = (\omega_{a}(\beta)/\pi)^\frac{1}{4}
exp(-\omega_{a}(\beta)\phi^{2}/2)
\nonumber\\
 = 0.828\beta^{1/8}exp(-\omega_{a}\phi^{2}/2).
\end{eqnarray}

where `$\omega_{a}$' satisfies the gap-equation, eqn.(197); `$\beta$' is
defined by eqn.(189) and $~\xi~ =~ 1/2~$, corresponding to the ground state
of the DWO. Comparision of the coefficients of the exponential terms in eqs.
(194) and (200)  interestingly reveals that not only the
``$\beta^{1/8}$'' factor is common but also that the coefficients:
0.68108 and 0.828 are comparable. We compare 
the two results in {\bf Figure  1} for $\beta = 100$. The quality of
 the approximation can be judged from this figure. It is plausible that the
inclusion of higher order corrections to the ground state wave function in
IPT of NGAS {(\bf Chapter-10)} may further improve the agreement.\\

To summarize the results of this {\bf Chapter}, it is shown that NGAS respects 
and
preserves the exact results of SUSYQM with good accuracy. It would be 
interesting to extend the comparison$^{120}$ to the system of 
self-interactin  oscillators described in SUSYQM by the super potentials:
\\$~W_{\pm}~ \equiv ~\beta\phi^{3}~{\pm}~ \gamma \phi~$, which generate a 
family of sextic-, quartic- and quadratic- AHO/DWO for different values of the
 parameters `$\beta$' and `$\gamma$'. This is, however, beyond the scope of 
the present thesis.\\
\\
\\

\begin{figure}
\centerline{\epsfysize=8.5truecm\epsfbox{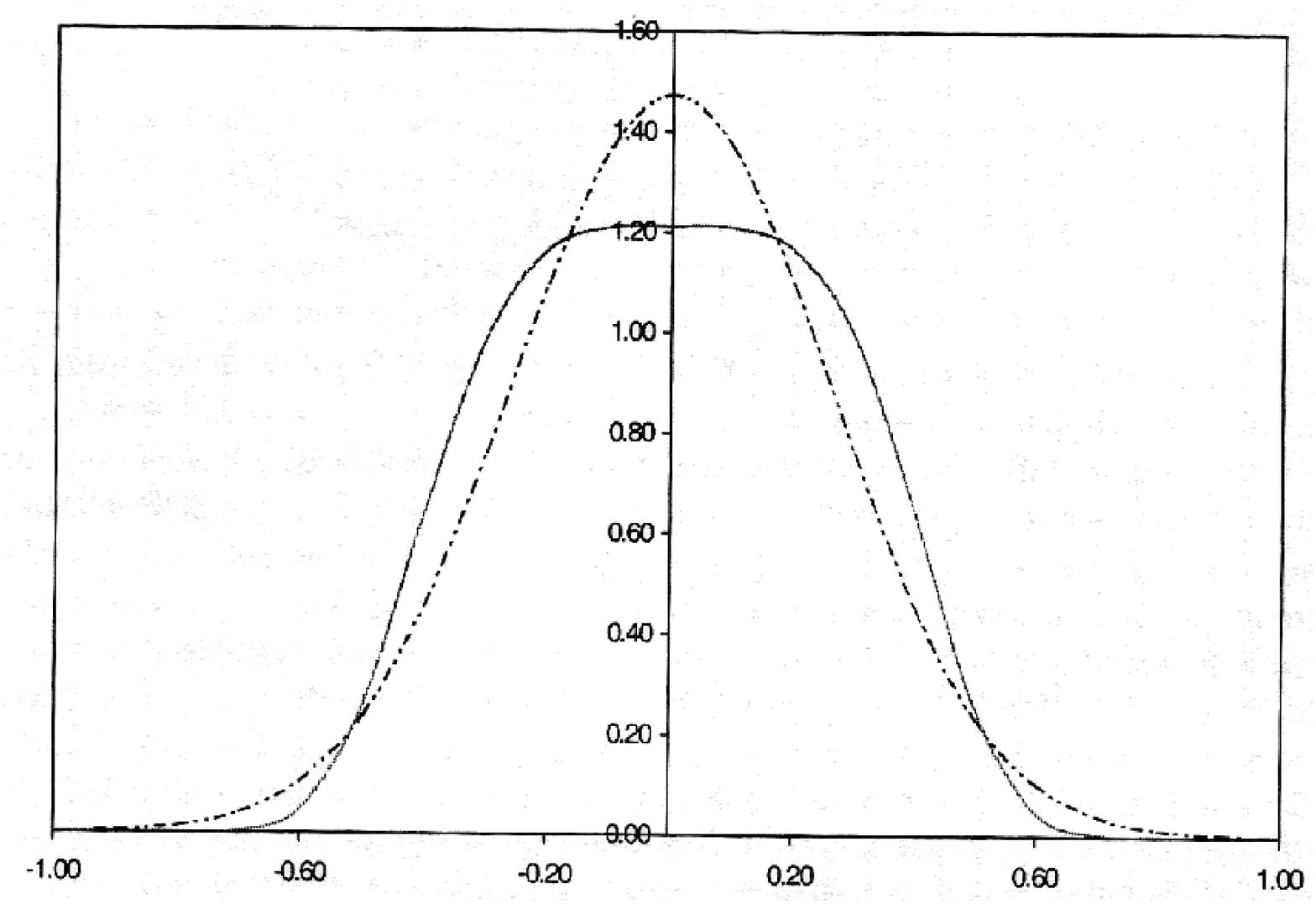}}
\caption~{Comparision of the ground state wave function of the
  sextic-DWO 
predicted by SUSY
( curve with sharper peak ) with that obtained in LO of NGAS for $\beta ~=~100$, see eqs.(195) and (200) of text. }
\label{1}
\end{figure}

\newpage
{\Large \bf 8.  The NGAS for the Octic-Anharmonic\\ Oscillator 
( Leading Order(LO) Results )}\\
\\
\\
{\large \bf 8.1  The Octic-Anharmonic Oscillator:\\ Introduction }\\

The ~ octic-anharmonic oscillator, like its quartic- and~ sextic-
 counterparts,~finds~ applications in modelling molecular physics,
 lattice-vibrations in
 solids~ and in~ quantum chemistry. Because~ of the~ higher~
 anharmonicity, the system also ~provides the theoretical laboratory
 for more stringent-tests for non-perturbative approximation schemes
 in quantum theory since the divergence of the na$\ddot{i}$ve 
( Rayleigh-Schr$\ddot{o}$dinger) perturbation theory becomes still more
severe$^{12,13,14}$  in this case.\\

The system is considered here to test the generality of the application of NGAS
to the case of still higher anharmonicity and hence to test the reliability of 
the said scheme. The detailed application is described below.\\

To demonstrate further the generality and  uniformity of the approximation
(NGAS), we apply the method to the case of the next higher anharmonicity, i.e.,
the octic-anharmonic oscillator, described by the following Hamiltonian:\\

{\large \bf 8.2  Application of NGAS to the Octic-AHO}\\

{\bf 8.2.1  The derivation of the Gap-equation for the Octic-AHO}\\

As before, we start from the Hamiltonian describing the system:
\begin{equation}
H~=~\frac{1}{2} p^{2}+\frac{1}{2}g\phi^{2}+\lambda\phi^{8}; ~ g,~\lambda >0.
\end{equation}

The free field Hamiltonian corresponds to $~H_{s}~=~\frac{1}{2} 
p^{2}+\frac{1}{2}g\phi^{2}$ and the interaction term is given by $\lambda H_{I}
(\phi)~ = ~\lambda\phi^{8}$. To develop the NGAS for the octic AHO we follow
the identical ansatz for $~V(\phi)~$ as given in eqn.(62):

\begin{equation}
V(\phi) = A\phi^{2} - B\phi + C
\end{equation}

The parameters A,B,C can be  determined self-consistently as before. For this
 purpose, consider the ``Effective Hamiltonian (EH)'' in this case given by, 

\begin{equation}
H_{0}(\phi,p)~=~\frac{1}{2}p^{2}+\frac{1}{2}g\phi^{2} + \lambda V(\phi)
\end{equation}

As in the case of quartic- and sextic- anharmonicity, eqn.(203) is transformed
into the following diagonal structure by using the  eqn.(202) and given by:

\begin{equation}
H_{0}(\phi,p)~=~\frac{1}{2}p^{2}+\frac{1}{2}\omega^{2}(\phi - \sigma)^{2} + h_{0} 
\end{equation}

where again,

\begin{equation}
\omega^{2}~=~ g + 2\lambda A,
\end{equation}

\begin{equation}
           \sigma~=~\lambda B/\omega^{2},
\end{equation}

\begin{equation}
 h_0~=~\lambda C~-~\frac{1}{2}\omega^{2}{\sigma}^{2}.
\end{equation}

Eqn.(204) corresponds to ``shifted'' effective Harmonic oscillator. The field 
`$\phi$' is shifted by `$\sigma$' and the energy is shifted by `$h_{0}$'. As 
in other cases here also the physical requirement is that $~\omega > ~0~$ ; 
$~\sigma ~=~$real.\\

Using the creation- and annihilation operators defined by the eqs.(67) and 
(68) along with the equal-time commutation relation defined by the eqn.(69)
the EH, $~H_{0}~$ given in eqn.(204) is expressed into the desired diagonal
form

\begin{equation}
              H_{0}~=~\omega~ (N_{b} + 1/2 ) + h_{0}~ .
\end{equation}

To obtain eqn.(208), we have introduced the number operator  $~N_{b} \equiv 
b^{\dagger}b~$ and its eigen-states by, $~N_{b}|n>~ =~ n|n>$, $~<m|n>~ =~ 
\delta_{mn}$ as usual. This leads to the  energy spectrum  as in the earlier 
cases:
\begin{equation}
              E_{n}^{(0)}~=~\omega~ \xi  + h_{0}~ ,
\end{equation}

where $\xi~=~ ( n + 1/2 )~$ ;$~ n~ =~ 0,1,2,.....$\\

It is next required to determine the frequency `$\omega$' and  `$h_{0}$' 
 defined by the eqn.(207). For this purpose, we note that the quantum average 
of the eqn.(201) is given by:
\begin{equation}
< H >~=~<H_{0}>~=~\frac{1}{2}<  p^{2} >+\frac{1}{2}g< \phi^{2} > + \lambda < \phi^{8} >
\end{equation}

where eqn.(55) has been used and  $~ <~ \phi~ >,~ <~ \phi^{2}~ >,~ < ~ p^{2}~
 >~ $ have been calculated  using the standard properties  of creation-
/annihilation operators which are given in eqs.(80) and (81). The  QA of 
the last term in eqn.(210) can be obtained  by standard  procedure and is 
given by:   

\begin{eqnarray}   
< \phi^{8} > = \sigma^{8} + 28\sigma^{6}(\frac {\xi}{\omega}) + \frac{105}
{4\omega^{2}} \sigma^{4}(4\xi^{2} + 1) + \frac {35}{2}(\frac {\sigma^{2}\xi}{\omega^{3}})(4\xi^{2} + 5) +
\nonumber\\
 \frac {35}{128}(\frac {1}{\omega^{4}})(16\xi^{4} + 56
\xi^{2} + 9) 
\end{eqnarray}

Substituting the QA- values in eqn.(210) we obtain the equation
\begin{eqnarray}
< H_{0} >~=~ \frac{1}{2}\omega \xi +\frac{g}{2}[  \sigma^{2} + (\xi / \omega)] +
\lambda[\sigma^{8} + 28\sigma^{6}(\frac {\xi}{\omega}) + \frac{105} {4\omega^{2}} \sigma^{4}(4\xi^{2} + 1) +
\nonumber\\
\frac {35}{2}(\frac {\sigma^{2}\xi}{\omega^{3}})(4\xi^{2} + 5) + \frac {35}{128}(\frac {1}{\omega^{4}})(16\xi^{4} + 56\xi^{2} + 9)]~~~
\end{eqnarray}

Applying the variational minimisation condition  $~\partial< H_{0} >/\partial
\omega = 0~$ leads to the following equation
\begin{equation}
\omega^{5}-\omega^{3}(g+56\lambda\sigma^{6})-105\omega^{2}(\lambda\sigma^{4}/
\xi)(4\xi^{2}+1)-105\omega\lambda\sigma^{2}(4\xi^{2}+5)-35\lambda h(\xi)=0,
\end{equation}

where $ h(\xi)~=~\xi^{3}+(7\xi/2)+(9/16\xi)$. Eqn.(213) is referred as the
``gap-equation'' for the case of octic AHO. \\

Again carrying out minimisation with respect to $\sigma$ i.e.,$~\partial< H_{0} >/\partial\sigma~ =~ 0$ we have the following equation 

\begin{equation}
\sigma~[ g+\lambda(8\sigma^{6}+168\sigma^{4}(\xi/\omega)+105\sigma^{2}(4\xi^{2}+1)/\omega^{2}+35\xi(4\xi^{2}+5)/\omega^{3})~] = 0.
\end{equation}

Eqn.(214) corresponds to equation to ground state (EGS) for this case. The 
solution of eqn.(214) for $~\sigma~$ leads the two following equations:
\begin{equation}
(i)~~~~~~~~~~~~~~~~~~~~~~~~\sigma~ =~ 0.~~~~~~~~~~~~~~~~~~~~~~~~~~~~~~~~~~~~~~~~~~~~~~~~~~~~~~~~~~~~~~~~~~~~~~~~~~~~~~~~~~~~~~~
\end{equation}

\begin{equation}
(ii)~~[ g+\lambda(8\sigma^{6}+168\sigma^{4}(\xi/\omega)+105\sigma^{2}(4\xi^{2}+1)/\omega^{2}+35\xi(4\xi^{2}+5)/\omega^{3})~]~ =~ 0.\\
\end{equation}

Obviously, eqn.(216) has  {\it no} physically acceptable solution  for 
$~\lambda,~g,~\omega~>~0$.
Hence the `physical' solution of the EGS, eqn.(214), is at $\sigma~=~0$.
Substitution of this value  in eqn.(213) leads to the simplified GE, given by:
 
\begin{equation}
\omega^{5}-g\omega^{3}-35\lambda h(\xi)=0
\end{equation}

It is then straight forward to evaluate A, B, C appearing in the  approximating
potential, eqn.(202) in terms of `$\omega$' and  `$\sigma$'. These are given by
the following expressions:
\begin{equation}
A~ =~ 28\sigma^{6} +\frac{105\sigma^{4}( 1 + 4\xi^{2})}{2\xi\omega} + (\frac {105
\sigma^{2}}{2\omega^{2}})(5 + 4\xi^{2}) + \frac{35 h(\xi)}{2\omega^{3}},~~~~~~~~~~~~~~~~~~~ \\
\end{equation}

\begin{equation}
B~ =~ \sigma [(1+g)\frac{\omega^{2}}{\lambda}
+8\omega^{2}\sigma^{6} +168\sigma^{4}\xi\omega +105\sigma^{2}(1 + 4\xi^{2}) + 
(\frac{35\xi}{\omega})( 5 + 4\xi^{2})], \\
\end{equation}

\begin{equation}
C~ =~ <\phi^{8}>~-~A~<\phi^{2}> ~+~B~<\phi>.~~~~~~~~~~~~~~~~~~~~~~~~~~~~~~~~~~~~~~~~\\
\end{equation}\\

{\bf 8.2.2  Solution of the Gap Equation and Determination of the Energy
Spectrum}\\

The solution of eqn.(217) determinines the frequency `$\omega$' of the 
``shifted'' harmonic oscillator. To obtain the  energy levels one substitutes
$~\sigma~=~0$ and `$\omega$' as the solution of eqn.(217). This leads, after 
 some simplification, to the following simple formula: 
 \begin{equation}
 E_{n}^{(0)}|_{octic-AHO}=(\frac{\xi}{8})(5\omega+\frac{3g}{\omega}),~ g>0.
\end{equation}

where `$\omega$' is obtained by solving  eqn.(217) numerically.\\

In {~\bf Table-5~},we compare this LO- result in NGAS with earlier 
computations$^{133}$ over a wide range of values of `$\lambda$' and 
`n'. 
It can be seen from this comparision that the results obtained in the LO of
 NGAS are already quite accurate over the full range of the parameters, which 
demonstrates the generality of the method and uniformity of the approximation 
with increasing anharmonicity.\\

We next turn our attention to the physics of the {\it effective} vacuum state:
$~|vac>~$, obtained as an approximation to the true vacuum of the theory.\\
\\
\\

\begin{table}
\vspace{.4in}

{\bf Table- 5: }
{\it Sample results  for the octic- AHO in the LO NGAS  compared with
 results of earlier calculations from ref.[133] (shown in parentheses),
 over a wide range of `$\lambda$' and `n'}.
\vspace{.1in}
 \begin{center}
 \begin{tabular}{c c c c c c }
 \multicolumn{1}{ c }{$n $}&
 \multicolumn{1}{ c }{\quad$\lambda=0.1$ }&
 \multicolumn{1}{ c }{\quad$1.0$}&
 \multicolumn{1}{ c }{\quad$5.0$}&
 \multicolumn{1}{ c }{\quad$50.0$}&
 \multicolumn{1}{ c }{\quad$200.0$}\\

0        &1.3005&1.7794&2.3290&3.5565&4.6425\\
         &(1.2410)&(1.6413)&(2.1145)&(3.1886)&(4.1461)\\
                                                       \\
1        &4.4717&6.3946&8.5167&13.172&17.259\\
         &(4.2754)&(5.9996)&(7.9296)&(12.1950)&(15.9519)\\
                                                         \\
2        &8.6264&12.717&17.126&26.698&35.062\\
         &(8.4530)&(12.421)&(16.711)&(26.033)&(34.183)\\
                                                          \\
4        &19.763&30.026&40.863&64.165&84.444\\
         &(19.9930)&(30.4605)&(41.495)&(65.202)&(85.8251)\\
                                                             \\
6        &34.217&52.669&72.044&113.48&149.47\\
         &(35.0560)&(54.1403)&(74.0830)&(116.7629)&(153.83)\\
                                                              \\
8        &51.570&80.013&109.65&172.99&227.97\\
         &(53.146)&(82.6496)&(113.3486)&(178.9215)&(235.82)\\
                                                                 \\
9        &61.239&95.255&130.64&206.23&271.81\\
         &(63.225)&(98.5529)&(135.26)&(213.6157)&(281.5864)\\
                                                                 \\
10       &71.532&111.49&153.01&242.64&318.52\\
         &(73.954)&(115.49)&(158.5991)&(250.5751)&(330.3433)\\
                                                                 \\
11       &82.424&128.68&176.69&279.14&368.06\\
         &(85.308)&(133.42)&(183.3103)&(289.71)&(381.97)\\
                                                                \\
12       &93.893&146.79&201.65&318.67&420.14\\
         &(97.2636)&(152.31)&(209.3443)&(330.9440)&(436.3695)\\
                                                                \\
13       &105.92&165.79&227.84&360.14&474.85\\
         &(109.7967)&(172.11)&(236.6436)&(374.1834)&(493.4143)\\
                                                                 \\
14       &118.49&185.65&255.21&403.50&532.06\\
         &(122.89)&(192.81)&(265.1732)&(419.3737)&(553.0335)\\
                                                                                \end{tabular}
\end{center}
\end{table}
\newpage
{\Large \bf 9.  The  Property, Structure, Stability and the Significance
of the `` Effective'' Vacuum in NGAS}\\
\\
\\
{\large \bf 9.1  The Bogoliubov Transformation relating the Free-Field Vacuum
 to the Effective Vacuum }\\

The study of the properties  and the structure of the vacuum of interacting
quantum systems are of considerable importance$^{146}$. In the present
 scheme,
the vacuum state,$~|vac>~$ of the effective Hamiltonian $H_{0}$, approximates
the vacuum of the true interacting theory in the leading order. To study its
 properties and structure in comparision to the ``free''- field vacuum,$~|0>~$,
 it is useful to start with eqs.(71-73, 88-90). In view of eqn.(90), 
the creation- and annihilation operators of the `free-theory' and the 
approximated theory with self-interaction, are related by quantum-canonical 
transformation (``Bogoliubov-Transformation'')$^{52}$, given by:

\begin{equation}
b~=~a~ \cosh(\alpha) - a^{\dagger}\sinh (\alpha)
\end{equation}

\begin{equation}
b^{\dagger}~=~a^{\dagger}\cosh(\alpha) - a~  \sinh (\alpha)
\end{equation}

The two vacua are then related by the following equations:
\begin{equation}
|vac> = exp [(1/2)~ \tanh (\alpha)~ (a^{\dagger} a^{\dagger} - a a)] |0>~\equiv
U(\alpha;a,a^{\dagger})|0>
\end{equation}

The derivation of eqn.(224) follows from eqn.(222) by using the defining 
property of the vacuum, $b|vac> = 0~$ and the representation of the 
annihilation operator given by, $a = \delta/\delta a^{\dagger}$. The parameter
 `$\alpha$' occuring in the above equations, can be simply related to 
`$\omega$' by using  the eqs.(71,88,222) as  given below:\\
                                                                 \\  
\\
From eqs.(71) and (88) one can obtain,

\begin{equation}
b~ +~b^{\dagger}=~\sqrt\frac {\omega}{\omega_{0}}~(a~+~a^{\dagger}) 
\end{equation}

Again from eqs.(222) and (223) we have the relation 

\begin{equation}
b~ +~b^{\dagger}~=~(a~+~a^{\dagger})e^{-\alpha} 
\end{equation}

By  solving  eqs.(225) and (226) we get the relation between `$\alpha$'
and `$\omega$' and is given by

\begin{equation}
e^{\alpha} ~=~\sqrt\frac {\omega_{0}}{\omega}
\end{equation}

Then,

\begin{equation}
\alpha =   (1/2) ~ ln (\omega_{0}/ \omega),~ \omega_{0} = \sqrt g
\end{equation}

It is useful to have the transformation inverse to eqs.(222-223). This is given
by

\begin{equation}
 a~=~b~ \cosh(\alpha) + b^{\dagger}\sinh (\alpha)
\end{equation}

\begin{equation}
a^{\dagger}~=~b^{\dagger}\cosh(\alpha) + b~  \sinh (\alpha)
\end{equation}

The following significant physical results  follow from the above equations,
eqs.(222-225).\\

{\large \bf 9.2   Structure of the effective vacuum}\\

(i) A non-trivial structure (``dressing'') of the ``effective'' vacuum(EV) of
the theory emerges from the equations. The situation could be analogous to
the case of the ground state of the super-fluid$^{147}$ and the hard 
sphere  Bose -gas$^{49}$. The structure is characterized by the 
non-vanshing  number density of the free particles in the EV and its critical 
dependence on the  strength of the interaction, which is shown below: 
                                                                      \\

The number density of the free particles in the effective vacuum (EV)
is defined by the relation

\begin{equation}
n_0 ~\equiv ~< vac|a^{\dagger} a| vac >~ 
\end{equation}

From eqs.(229) and (230) and using the defining property of the vacuum,\\ 
$~b|vac>~=~0~= ~<vac|b^{\dagger}~$ one obtains the relation

 \begin{equation}
< vac|a^{\dagger} a| vac >~  =~ \sinh^{2} (\alpha) 
\end{equation}

Hence by using eqn.(227) we have the following equation

\begin{equation}
n_0 ~\equiv ~< vac|a^{\dagger} a| vac >~ =~ \sinh^{2} (\alpha) = \frac{1}{4}
(\frac{\omega}{\omega_{0}} + \frac{\omega_{0}}{ \omega} - 2 )
\end{equation}
Considering the case of the quartic- AHO, it can then be shown using the `gap-
equation' that $~ n_0\sim\lambda^{1/3}~ $ for $~ \lambda~>>~1~$. In the limit
 of vanishing  interaction, one recovers the expected behaviour,
 $~n_0\rightarrow 0~$ for $~\lambda\rightarrow 0 ~$ .
                                                                          \\
                                                                           \\
(ii) Secondly, eqs.(222-224,228) imply an entirely new physical interpretation
 of the parameter, `$~\omega~$' which determines (through eqs.(228) and (235))
 the ``vacuum structure function'' `$\alpha$' in the sense that $\alpha \neq 0$
~(i.e.,$~\omega \neq \omega_{0}~$) signifies the non-trivial structure of the
~ EV ~in presence of interaction.\\

{\large \bf 9.3    Stability of the Effective Vacuum}\\

In the remaining part of this {\bf Chapter}, we investigate the stability
properties of the EV.\\
                                                                    
{\it  Instability of the Perturbative(``free-field'') Vacuum}
                                                                        \\
                                                                         \\
It is shown  below that the {\it perturbative} vacuum, $~|0>~$ {\it becomes
unstable compared to the } effective vacuum {\it $~|vac>~$ for all values of
 the coupling strenth} `$\lambda$'. For this demonstration we consider, for
 reasons of simplicity, the case of the QAHO. The standard method for studying
 the stability properties is to consider the ``{\it  effective potential 
(EP)}''. {\it The EP, for any given choice of a vacuum state, is 
defined$^{148}$  to be the expectation value of the Hamiltonian in the 
chosen vacuum-state  and expressed as a function of the VEV of the 
``field''  `$\phi$'}. For the 
case of the QAHO, this is obtained in LO of NGAS, from eqn.(86) by choosing,
 $~g~=~1, n~=~0~$ and `$\omega$' constrained to satisfy eqn.(92). The 
resulting expression is as follows:

\begin{equation}
V_{eff}^{NGAS}(\sigma) = \frac{\omega}{4} + \frac{(1+12\lambda \sigma^{2})}{4
\omega} + \frac{3\lambda}{4\omega^{2}} + V_{c},
\end{equation}
where,

\begin{equation}
V_{c}  = \frac{1}{2}\sigma^{2} + \lambda \sigma^{4},
\end{equation}

is the ``{\sl classical potential}'' and `$\omega$' satisfies  eqn.(92).
                                                                        \\
                                                                         \\
An analogous expression for the corresponding EP based upon the 
 {\it perturbative} (free-field) vacuum is obtained by the substitution, 
$~\omega\rightarrow 1~$ in eqn.(234) above ( this~ follows by comparing, 
eqs.(71) and (88)) and is given by

\begin{equation}
V_{eff}^{Pert}(\sigma) = \frac{1}{2} + 3\lambda (\sigma^{2} + \frac{1}{4})+
 V_{c}
\end{equation}
The ground state energy is defined to be the~ {\it global}~ minimum  of
the effective
potential and corresponds to $~\sigma~ =~ 0~$ in either case. We thus obtain
 the respective ground state energies from eqn.(234) given by the 
following equations :

\begin{equation}
E_{0} = \frac{\omega}{4} + \frac{1}{4 \omega} +
\frac{3\lambda}{4\omega^{2}} =  (1/8)(3\omega + 1/\omega)
\end{equation}

The above relation is obtained by using the GE for the ground state:
$~\omega^{3} - \omega - 6\lambda = 0~$. However,$~E_{0}^{Pert}~$ is obtained 
from eqn.(236), by using $~\sigma~=~0~$ and given by

\begin{equation}
E_{0}^{Pert} = 1/2 + 3\lambda /4.
\end{equation}

(Note that eqn.(237) is also contained in eqn.(109) for the special case 
considered here (i.e.$~ g=1,~n~=~0~$)). Recalling that the GE for the
ground state is given by:$~\omega^{3} - \omega - 6\lambda = 0~$, it
is straight forward to establish that:

\begin{equation}
E_{0}-E_{0}^{Pert}~ <~ 0, ~for~ all~ values~ of~ \lambda.
\end{equation}

which is shown  below.
                                                                \\ 
                                                                 \\    
By using eqs.(237) and (238) we have the relation

 \begin{equation}
 \Delta E(\lambda) = E_{0} - E_{0}^{Pert} = \frac{1}{4} (\omega +
 \frac{1}{\omega}) + 
\frac{3\lambda}{4}(\frac{1}{\omega^{2}} -1 ) - 1/2 .
\end{equation}

Here it is to be noted that as $~\lambda \rightarrow 0~$;$~\omega
\rightarrow~1$ then  $~\Delta E(\lambda)\rightarrow~ 0$ as it should!
Now let us study how $~\omega (\lambda)$ and  $~\Delta E(\lambda)~$
behave as $~\lambda \simeq~0_{+}$.
Recalling the gap equation for the ground state we have the equation :

\begin{equation}
\omega^{3} - \omega - 6\lambda = 0
\end{equation} 

By substituting $~\omega ~=~1~+~\xi(\lambda)~ ;~|\xi(\lambda)|~<<~1~$
in  eqn.(241) then $~\xi \simeq~ 3\lambda~ +~ O(\lambda^{2})~$ as
$\lambda \rightarrow~0~$. Hence in the $\lim_{\lambda \rightarrow~0}$,
$~\omega(\lambda)\simeq ~1~+~3\lambda$. So it is seen that `$\omega$'
is monotonically increasing with `$\lambda$' as `$\lambda$' increases!
 Substituting this value of `$\omega$' in the eqn.(240) and after some 
simplification we get the relation

  \begin{equation}
\lim_{\lambda \rightarrow~0_{+}}~ \Delta E(\lambda) =  E_{0} - E_{0}^{Pert} = 
 - ( \frac {9\lambda^{2}}{2})~  < 0
\end{equation} 

From this relation it is evident that  $~\Delta E(\lambda)~$ is
monotonically decreasing  with increasing `$\lambda$' (for $~\lambda~>~0~$) !
This proves the instability of the perturbative (free-field) vacuum of the
QAHO. It may be noted that, although we have established the above result
for the QAHO, the same can be rigorously demonstrated in all other cases of
anharmonicity considered here.\\

{\large \bf 9.4  Significance of the Effective Vacuum state}\\

In the context of quantum-mechanics, the results presented in the previous 
sections: 9.1 - 9.3 are significant in many respects which are listed below:
                                                                          \\ 
                                                                         \\
(a) In the  first-place, the results contained in eqs.(231-233) 
demonstrate that the effective vacuum state is endowed with non-trivial
  particle-content and structure, which is directly attributable to 
interaction since the  particle-content vanishes  as $~\lambda \rightarrow~0~$
. This result is  analogous to the ``dressing'' of the ``physical''-vacuum and
  is made more  transparent when one considers the $~\lambda\phi^{4}~$-field 
theory discussed  subsequently in {\bf Chapter-11}. 
                                                                      \\
                                                                       \\
(b) The other significant result is the equivalence of the NGAS-vacuum with 
that obtained through quantum canonical/ Bogoliubov-transformation as depicted 
in ~ eqs.(222-230). ~ It also demonstrates that the effective-vacuum (EV)
 state is {\it  not}  obtainable in perturbation-theory due to non-
analytic dependence of the ``vacuum-structure function (VSF)'' `$\alpha$' on 
the coupling-strength `$\lambda$'( see,eqn.(228) and the gap-equation, e.g.,
eqn.(241)).                                                           
                                                                       \\  
                                                                        \\ 
(c) The third  significant  observation is contained in the conclusive 
demonstration of the instability of the ``perturbative''/ free-field vacuum 
 in respect of the  EV, as contained in eqn.( 242). In this context it may 
be further significant that the convergence of the perturbation theory about 
the free-field vacuum discussed in the next {\it Chapter} may have a direct
 bearing with the stability properties of the theory. This is discussed in 
greater-detail in the next {\bf Chapter}.\\ 
\\
\\

{\Large \bf 10.  Improved Perturbation Theory(IPT) in NGAS}\\

{\large \bf 10.1   Considerations of the  Convergence of the `{naive}'
Perturbation Theory and  other Variants }\\

One of the main motivations for proposing the NGAS as described earlier, is the
 possibility of construction of an improved perturbation theory (IPT) which
could be  convergent for {\it all} allowed values of of the coupling
strenth, `g' and `$\lambda$'. This expectation is based upon the result, 
eqn.(59), which is reproduced below:

\begin{equation}
<n|\lambda H'|n> = 0.
\end{equation}

Since $~H~ =~ H_{0}~ +~ \lambda H'~$, eqn.(243) naturally suggests that the 
IPT be constructed  by choosing $H_{0}$ as the unperturbed Hamiltonian and 
 $\lambda H'$ as the perturbation. The convergence of the resulting IPT 
{\it is intuitively} suggested since the basic condition of 
convergence$^{149}$ is satisfied by ensuring that the magnitude of the
 perturbation remains always sub-dominant to the unperturbed contribution:

\begin{equation}
|<n|\lambda H'|n>| \equiv  0 << |<n| H_{0}|n>|
\end{equation}

{\it The important point to note is that eqn.(244) holds for arbitrary values 
of  `g', `$\lambda$' and `n'}. In this context, it may be noted that the 
analogous requirement, which is the necessary condition for convergence of 
perturbation expansion, does not hold good in the case of$~ {\sl na\ddot{i}ve}
~$ perturbation theory (NPT), where the entire  self-interaction, $\lambda 
H_{I}(\phi)$ is chosen as the perturbation  to the `{\it free}' Hamiltonian,
 $~H_{s}(p,\phi)~$(see eqn.(52)). Consequently, the divergence$^{12,13,87}$
 of the NPT is anticipated as it could be traced$^{14}$ to the eventual 
dominance of the perturbation-contribution over the unperturbed one for any 
value of $\lambda >0$, no matter however small. This was explicitly 
demonstrated in ref.[12,13,87].\\

In the next section, we formulate the improvement of the LO results
through the ``( improved ) perturbation theory'' (IPT) developed using
$~\lambda H'~$ as the perturbation while treating $~H_{0}~$ as the
unperturbed  Hamiltonian.\\

{\large \bf 10.2   Improvement of the LO Results through the  IPT in NGAS}\\

The unique feature of NGAS summarized in eqn.(244) leads to the systematic
further (order-by-order) improvement of the LO results (which are already
accurate to within a few percent of the {\sl exact} result). We demonstrate
below, the improvement in accuracy, by inclusion of the higher-order 
contribution in IPT for the case of the QAHO and  the QDWO.\\

In the Rayleigh-Schr$\ddot{o}$dinger (RS) development of the perturbation 
series, the  perturbative correction  to the energy levels is given by the 
standard expansion:

\begin{equation}
E_{n} = E_{n}^{(0)} + \Delta E_{n}^{(1)} + \Delta E_{n}^{(2)} + \Delta E_{n}^{
(3)} + .........
\end{equation}

where, the LO-contribution, $E_{n}^{(0)}$ has already been defined,( see,
 eqs.(54-55)). It is important to note that  {\sl the first order 
contribution } $\Delta E_{n}^{(1)}$ {\it  vanishes due to eqn.(243)}:

\begin{equation}
 \Delta E_{n}^{(1)} =  <n|\lambda H'|n> = 0,
\end{equation}
                                                                       \\
({\it In the above sense, the IPT can be regarded  as optimal and this result,
 eqn.(246), distinguishes the IPT from many other variants  of perturbation 
theory$^{150}$ used earlier, for the problem}). Using eqn.(246), the 
next higher order (HO) contributions are given by the following expressions:

\begin{equation}
 \Delta E_{n}^{(2)} =  \sum_{m\neq n}|(\lambda H_{nm}')|^{2}/\Delta_{nm},
\end{equation}

\begin{equation}
\Delta E_{n}^{(3)} = \sum_{m\neq n,k\neq n}\frac{(\lambda H_{nm}')(\lambda H_{mk}') 
(\lambda H_{kn}')}{\Delta_{nm}\Delta_{nk}}
\end{equation}
                                                                        \\
Similar expressions for still higher-order corrections can be obtained  by
standard$^{149}$  methods. In the above equations, we have used the 
following notations:   $(\lambda H')_{mn} \equiv  <m|\lambda H'|n>$ and 
$\Delta_{nm} \equiv(E_{m}^{(0)}-E_{n}^{(0)})$. The above formulae can be 
applied to the case of the QAHO as discussed below.\\ 

{\large \bf 10.3   IPT- Applied to Quartic-Oscillator (AHO/DWO)}\\

The Hamiltonian for the systems  given in eqn.(61) is
\begin{equation}
~~~~~~~~~~~~~~~~~~~~~~~~~~~~~~H~=~\frac{1}{2}p^{2}~+~\frac{1}{2}~g~\phi^{2}+\lambda\phi^{4},~~~~~~~~~~~~~~~~~~~~~~~~~~~~~~~~~
\end{equation}

The free field Hamiltonian is given by
\begin{equation}
H_{s}(\phi) = \frac{1}{2}p^{2}~+~\frac{1}{2}~g~\phi^{2}
\end{equation}

and the self-interaction term is given by
\begin{equation}
 \lambda H_I(\phi) = \lambda\phi^{4}.
\end{equation}

With the new ansatz $~V(\phi)~$ the effective Hamiltonian is given by
\begin{equation}
 H_{0}(\phi,p)~=~\frac{1}{2}p^{2}~+~\frac{1}{2}~g~\phi^{2}+\lambda V(\phi)
\end{equation}

where $~V(\phi)~$ is given by the equation

\begin{equation}
                 V~(\phi ) = A~\phi^{2} - B~ \phi + C.
\end{equation}

The parameters A,B,C are determined and given by the ens.(95), (97), (99)
 with the choice $~\sigma~=~0~$(for the case of the QAHO). The
 modified interaction $~\lambda H'~=~H~-~H_{0}~
 =~(\lambda\phi^{4}~-~V(\phi))~$ have  vanishing  quantum  average 
for arbitrary `$\lambda~$',~`$g$' and `n' as has  been noted. Hence, the
first order contribution vanishes as  given in eqn.(246). To
calculate the higher order perturbative contribution, the matrix
elements for the case of the QAHO, are given by

\begin{equation}
 (\lambda H')_{mn} = <m|\lambda \phi^{4}|n> - (3\lambda/\omega)f(\xi)<m|\phi^{2}|n>,~ m\neq n
\end{equation}

where,

\begin{eqnarray}
<m|\phi^{2}|n> = (1/2\omega)(\delta_{m,n+2}\sqrt{(n+1)(n+2)} +
\nonumber\\
 \delta_{m,n-2}\sqrt {n(n-1)} ).
\end{eqnarray}

and

\begin{eqnarray}
<m|\phi^{4}|n> =  (1/4\omega^{2})(\delta_{m,n+4}\sqrt {(n+1)(n+2)(n+3)(n+4)} +
\nonumber\\
\delta_{m,n-4}\sqrt { n(n-1)(n-2)(n-3)} +
\nonumber\\
 2(\delta_{m,n+2}(2n+3)\sqrt {(n+1)(n+2)}) +
\nonumber\\
 2(\delta_{m,n-2} (2n-1)\sqrt {n(n-1)}) )
\end{eqnarray}

(The  above matrix elements are easily calculated by using  eqn.(71) and 
 introducing the  basis  states $~|vac>~$, and  $~|n>~;~n~=~1,2,3..$ with the
 defining  property of the effective vacuum  $~b|vac>~=~0~=~<vac|b^{\dagger}$, 
and further relations:  $~|n>~=~\frac{(b^{\dagger})^{n}}{\sqrt n!}~|vac>~$; 
$~\hat{N}~=~b^{\dagger}b~$;  $~\hat{N}|n>~=~n|n>~$; $ ~b^{\dagger}|n>~=~\sqrt
 {n+1}|n>~$ and  $~b|n>~=~\sqrt{n}|n-1>~$).\\

We present in {\bf Table-1},(see page -46) the results for the energy levels 
of the QAHO, with the inclusion of the second-order perturbation correction.
 In the same {\bf Table} we also compare our results with available `exact'
 numerical results  and results of calculation in second order perturbation 
theory of ref.[44], which is based upon the operator methods. It 
may be seen from this  {\bf Table} that the accuracy is considerably improved 
 by inclusion of the perturbation correction and further that the convergence
 of the IPT is found superior (order-by-order) to that in ref.[44]. 
Similar results for the QDWO  after inclusion of the second order perturbative 
correction in IPT, is  presented in  {\bf Table-2} (at page-54). In this 
{\bf Table}, we also compare our results with those obtained by inclusion of 
{\it  twenty orders in the `modified' perturbation theory} of ref.[36].
 Again  uniform improvement in accuracy is seen. It is also seen from the  
{\bf Table} that 
the results of the present analysis which includes only the first non-trivial 
perturbative correction, compares well with the results of ref.[36] 
obtained  by sophisticated  numerical methods. In the context of the above 
results, the  following  observations may be relevant:\\

{\large \bf 10.4   Remarks on Convergence  and other aspects of IPT in NGAS}\\

(a) Corrections up to the fourth-order  in IPT have been computed  for the QAHO
 and the QDWO although we have reported only the second order correction in the
 {\bf Tables-1,2}. {\it It is seen that these higher order corrections remain
uniformly small compared to the LO results over the full range of `$\lambda$'
and `n' and decrease fast with  the order of correction, which is 
consistent with the expectations from a rapidly converging sequence}.
                                                                       \\
                                                                        \\  
\\
(b)  As has been demonstrated  in the previous section, the `perturbative'
ground state (i.e., corresponding to the free-field Hamiltonian  $~H_{s}
(\phi,p)~$, see eqn.(52)) becomes unstable compared to the ground state  of
the `effective' Hamiltonian, $~H_{0}~$. Thus the stability of the theory, and
the convergence of the IPT- both appear to critically depend on the choice of
$~H_{0}~$. It may perhaps be plausible, therefore, to conjecture that  the
convergence of the perturbation theory may be intimately connected with the 
choice of a {\it stable} vacuum resulting from a proper effective Hamiltonian 
chosen as the unperturbed part.
                                                                      \\
                                                                       \\
(c) When compared with the results of some other variants of perturbation
theories$^{150}$ applied to the above systems of anharmonic and double
 well oscillators, the IPT appears  to provide better convergence, when 
compared at each order.
                                                                         \\
                                                                          \\
(d) For the case of the QDWO, the results of IPT, as well as those from the
other variants of perturbation theories, show poor convergence near the
transition point: $~\lambda\sim\lambda_{c}(\xi)~$ as intuitively expected
$^{151}$. However, since $~\lambda_{c}~$ is small  
$~(\lambda_{c}(\xi)\leq 0.0362886)~$, this limitation does not affect most 
applictions of practical interest. In particular, the strong coupling regime,
 $~\lambda~ >>1$  is excellently described by the IPT.
                                                                    \\
                                                                     \\
(e) Although we have provided only plausible argument  in support of the 
convergence of the IPT, a formal proof can be attempted following the methods 
available in the literature$^{152}$ .

In the following {\bf Chapter}, we extend the method to 
$~\lambda\phi^{4}$ - quantum field theory in ( 3 + 1 ) dimensions.\\

{\Large \bf 11.  Application of NGAS to $~\lambda\phi^{4}~$- Field\\
Theory in ( 3 + 1 ) dimensions}\\
\\
\\

{\large \bf 11.1   Formulation of the  NGAS for   $~\lambda\phi^{4}$ quantum 
field theory} \\

We consider, in this {\bf chapter}, the application of the NGAS to
$~\lambda\phi^{4}~$ theory in (3+1) dimensions. In view of the
successful application of the NGAS to the AHO's  and DWO's as
described  in earlier {\bf chapters}, it is but natural to extend  the
formalism to   $~\lambda\phi^{4}~$ quantum field-theory and test the
consequences of the scheme.\\

It may be worth while pointing out here that the  $~\lambda\phi^{4}~$- field
theory  in physical dimensions ( and in lower dimensions ) is an
important physical system which finds crucial applications in diverse
areas of physics, e.g.,~the standard model of particle-physics$^{84}$,
cosmology$^{153}$, condensed matter physics$^{121}$,
 phase-transitions and
critical phenomena$^{75}$ etc. Besides, this theory provides the
simplest theoretical laboratory for testing the various approximation
schemes in quantum field-theory(QFT). Hence it becomes imperative
to test the current approximation scheme NGAS by applying the same to  
 $~\lambda\phi^{4}~$- QFT.\\

In this thesis, we consider the theory in the massive, symmetric-phase 
described by the Lagrangian:
\begin{equation}
{\cal L }= \frac{1}{2}(\partial_{\mu}\phi)(\partial^{\mu}\phi) - \frac {1}{2} m^{2}
\phi^{2} - \lambda \phi^{4}~,
\end{equation}

where $~m^{2}~ >~ 0~$. The Hamiltonian density derived from the above 
Lagrangian is given by

\begin{equation}
{\cal H}~ =~ \frac {1}{2}(m^{2}\phi^{2} + \phi _{t}^{2}+\phi _{\alpha}^{2}) +\lambda\phi^{4}
\end{equation}

where  we have defined: $~\phi _{t}\equiv ~\partial \phi
(\vec{x},t)/\partial t~$ and $~\phi _{\alpha}\equiv
\partial \phi(\vec{x},t)/\partial x^{\alpha}~$.\\ 
\\

To formulate the NGAS for the above theory we follow analogous steps
as in the cases of AHO/DWO considered earlier and choose an
approximating  potential (AP) denoted by $~V(\phi)~$ such that ideally
the defining constraints as given by eqs.(52-57) are
satisfied. However, unlike the case in quantum mechanics, the PEQA
involving multiparticle-states, is hard to implement in QFT. Hence, on 
grounds of simplicity, we relax the condition, eqn.(56) by
restricting the QA  to be evaluated  in the `` few-particle" states
only. To be more specific, we proceed as follows :\\  

{\bf 11.1.1    Choice of $V(\phi)$}\\
                                                          
Using the most general ansatz quadratic in the fields  $\phi$, we
parametrise  $V(\phi)$ as given below : 
\begin{equation}
V(\phi) = A \phi^{2} - B \phi + C 
\end{equation}
As before, the `` effective Hamiltonian'' (EP) $~{\cal H}_{0}~$ is defined as 
: 

\begin{equation}
{\cal H}_{0}~\equiv~ \frac {1}{2}(m^{2}\phi^{2} + \phi
_{t}^{2}+\phi _{\alpha}^{2}) +\lambda V(\phi)
\end{equation}

Substitution of eqn.(259)~ in ~eqn.(260) leads, after some
simplification, to the following expression :

\begin{equation}
{\cal H}_{0}~ = ~\frac{1}{2}M^{2}\xi^{2} + \frac{1}{2}\xi_{\alpha}^{2} +
\end{equation}

where,  

\begin{equation}
\xi(\vec{x},t) \equiv \phi(\vec{x},t) - \sigma~,
\end{equation}

\begin{equation}
 \sigma \equiv \frac{\lambda B}{M^{2}}~,
\end{equation}

\begin{equation}
M^{2} \equiv m^{2} + 2\lambda A~,
\end{equation}
and 
\begin{equation}
h_{0} \equiv~\lambda C - \frac{1}{2} M^{2}\sigma^{2}.
\end{equation}
In addition, $~\xi_{t} \equiv \partial\xi/\partial t~ ;~ \xi_{\alpha}
\equiv \partial\xi/\partial x^{\alpha}$ etc.
Eqn.(261) is atonce identified to be the Hamiltonian density of the
hermitian scalar-field $~ \xi(\vec{x},t)~$. This is
$~\it {not}~$ surprising since the~ AP~ was $~\it{chosen}~$ 
accordingly. However, the important point to emphasize is that the AP
is proposed to incorporate the effects of self-interaction even though 
the EH corresponds to that of an exactly solvable system. To
demonstrate this aspect it is first necessary to obtain the spectrum
of $~H_{0}~$ which is done as follows.\\

The diagonalisation of the EH given by eqn.(261) is straightforward
by using the Fourier expansion in terms of creation- and annihilation 
operators :
\begin{equation}
\xi(\vec{x},t)~ =~ \phi (\vec{x},t) - \sigma~ =~ \int \frac {d^{3}\vec{k}}{\Omega_{k}(M)}[b(\vec{k})e^{-ikx} + b^{\dagger}(\vec{k}) e^{ikx}]~,
\end{equation}

 where

\begin{equation}
\Omega _{k}(M)~ \equiv~ 2 (2\pi)^{3}\sqrt{|\underline{k}^{2}| +
M^{2}}~\equiv~ 2 ~(2\pi)^{3}~\omega_{k}(M)~,
\end{equation}
 and  $~kx \equiv k^{0}t-\vec{k}.\vec{x}$, as usual.
The operators$~b(\vec{k}),b^{\dagger}(\vec{k})$ satisfy the standard 
(equal-time) commutation relations (ETCR)~:

\begin{equation}
[b(\vec{k}),b^{\dagger}(\vec{q})]~ =~ \Omega _{k}(M)\delta ^{3}(\vec{k}-\vec{q}
),
\end{equation}
which is a consequence of the ETCR between the `field' $~\phi(\vec{x},t)~$ and
its canonical conjugate momentum~: $~\pi(\vec {x},t)~ \equiv~ \partial
{\cal L}/\partial\dot {\phi}~$, given by : 
\begin{equation}
[~\phi(\vec{x},t),~ \pi(\vec {y},t)~]~ =~i~\delta^{3}~(\vec{x} -\vec{y}).
\end{equation}
                                                                     \\
{\bf 11.1.2   Diagonalisation  of the Effective Hamiltonian}\\

The energy of the system described by $~H_{0}~$ is obtained by standard methods
and given by : 
\begin{equation}
H_{0}~\equiv~\int d^{3}\vec{x}~{\cal H}_{0}(\vec{x},t)~=~\frac{1}{2}\int \frac {d^{3}\vec{k}}{\Omega_{k}(M)}[b(\vec{k})~ b^{\dagger}(\vec{k}) + b^{\dagger}(\vec{k})~b(\vec{k})]
+\int d^{3}\vec{x}~h_{0}~~~~~~~~
\end{equation}
The  spectrum of the states  are analogously obtained and denoted by :
$~|~vac~>,\\
|~\vec{p}~>,~ |~\vec{p_{1}},~\vec{p_{2}}~>,.....$etc where the
effective vacuum state $~|~vac~>~$ is defined by 
\begin{equation}
b(\vec{k})~|vac>~ =~ 0~,
\end{equation}
and the multi particle-states are generated  by multiple application of the 
creation-operator $~b^{\dagger}(\vec{p})~$ on $~|~vac~>~$:

\begin{equation}
b^{\dagger}(\vec{p})~|~vac~>~ =~ |~\vec{p}~>
\end{equation}

\begin{equation}
\frac {b^{\dagger}(\vec{p_{1}})~b^{\dagger}(\vec{p_{2}})}{\sqrt{2!}}~ =~ |~\vec{p_{1}},~\vec{p_{2}}~>,~etc
\end{equation}
Let us note that, $~H_{0}~|~vac~>~ =~
E_{0}~|~vac~>~;~H_{0}~|~\vec{p}~>~=~E_{1}
~(\vec{p})~|~\vec{p}~>,....etc.~$ where $~E_{0},~E_{1}~$etc correspond
to the energy of the corresponding states.\\

The next step is the implementation of the PEQA. 
\\                                                                  

{\bf 11.1.3     PEQA~ for $\lambda\phi^{4}$ QFT in NGAS}\\

This requirement translates  to the following constraints : 
\begin{equation}
<vac|~\phi^{4}~|vac>~=~<vac|~V(\phi)~|vac>~, 
\end{equation}

\begin{equation}
<\vec{p}~|~\phi^{4}~|~\vec{p}>~=~~<\vec{p}~|~V(\phi)~|~\vec{p}>~,
\end{equation}
and similarly for multi-particle states. The implementation of eqs.(274,
275) require the evaluation of the QA of monomials of the field $~\phi(\vec
{x},t)~$ such as~: \\$~<vac|~\phi^{n}(\vec{x},t)~|vac>~\equiv~<\phi^{n}(\vec{x},t)
~>,~<\vec{p}|~\phi^{n}(\vec{x},t)~|\vec{p}>~$ etc. We first turn to evaluation 
of $~<\phi^{n}(\vec{x},t)>~$. This is readly done using $~\it{translational-
invariance}~$ of the vacuum-state~:$~|vac>~$ and the ETCR,~ as given by 
eqn.(268). Some  useful  results thus obtained are  given  below~:  

\begin{equation}
<\phi(\vec{x},t)>~=~\sigma~, 
\end{equation}

\begin{equation}
<\phi^{2}(\vec{x},t)> ~=~\sigma^{2}~+~ \int \frac {d^{3}\vec{k}}{\Omega_{k}(M)} 
~\equiv~ \sigma^{2}~+~I_{0}~,
\end{equation}

\begin{equation}
<\phi^{4}(\vec{x},t)> ~=~\sigma^{4}~+~6\sigma^{2}I_{0}~+~3I_{0}^{2}~,~etc.
\end{equation}
Similarly,

\begin{equation}
<\phi_{\alpha}^{2}(\vec{x},t)> ~=~ \int \frac {d^{3}\vec{k}}{\Omega_{k}(M)}|\vec{k}|^{2}, 
\end{equation}

\begin{equation}
<\phi_{t}^{2}(\vec{x},t)> ~=~ \int \frac {d^{3}\vec{k}}{\Omega_{k}(M)}~\omega_{k}^{2}(M) 
\end{equation}
\\                                                                   
{\bf 11.1.4    Determination  of   the AP}\\

Using the above results in eqn.(274) one determines the AP uniquely, which
is given by  

\begin{equation}
V(\phi) ~=~6(\sigma^{2}~+~I_{0})\phi^{2} ~-~ 2\sigma^{3}\phi ~-~3(\sigma^{2}~+~I_{0})^{2}.
\end{equation}
In other words, the coefficients A, B, C defining the AP (see, eqn.(259)), are
uniquely determined~:

\begin{equation}
A ~=~6(\sigma^{2}~+~I_{0})
\end{equation}

\begin{equation}
B ~=~ 2\sigma^{3}
\end{equation}

\begin{equation}
C ~=~ ~-~3(\sigma^{2}~+~I_{0})^{2}.
\end{equation}
It is further verified by explicit calculation that the above choice of the AP
guarantees not only the equality of the QA over the vacuum state; i.e. 
$<V(\phi)> = <\phi^{4}>$ ( eqn. (274)),~ but also that for the one-particle 
states as well, i.e. $~<\vec{p}~| V(\phi)|~\vec{p}>~=~<\vec{p}~| \phi^{4}|
\vec{p}>$,~ as required by eqn.(275).\\

It is important to emphasize at this point that eqns.(282-284) when 
considered together with eqns.(263-265) form a complete set of $\it{
self-consistency}$ $\it{conditions}$ which uniquely specify the physical 
consequences of the theory $\it{in the } $ $\it{leading-order(LO)}$,
such as the spectrum, renormalisation, stability properties and the structure
of the effective  vacuum. These physical consequences of the theory are 
discussed in the following sections.
                                                                   \\ 
                                                                    \\
{\large \bf 11.2   The ``Effective  Potential (EP)'' and Renormalisation in 
LO}\\

Before we discuss the ( non-perturbative) renormalisation programme in LO,
it is useful to first investigate the consequences of eqs.(262-265) 
considered together with eqs.(282-284). Substitution of eqn.(282) in
eqn.(264) leads to the following equation :

\begin{equation}
M^{2}(\lambda,\sigma) \equiv  m^{2} + 12\lambda\sigma^{2} + 12\lambda I_{0}(M^{2})~,
\end{equation}

i.e.,
\begin{equation}
M^{2}(\lambda,\sigma) \equiv  m^{2} + 12\lambda\sigma^{2} + 6\lambda \int\frac {d^{3}(\vec{k})}{(2\pi)
^{3}\sqrt{|\vec{k}|^{2}+M^{2}}}~,
\end{equation}

Eqn.(286) can be interpreted as the generation of the `mass-gap'
(i.e. shift in the bare-mass) due to interaction. In analogy with the 
terminology used 
earlier, we refer eqs.(285-286) as the ``gap-equation(GE)'' of the theory.
 This equation plays crucial-role in the subsequent discussions.\\
                                                                   \\

Similarly,~ consideration of eqn.(283) together with eqn.(263)
leads to the  `` equation for the ground-state (EGS)''

\begin{equation}
\sigma~ [~\sigma^{2}~ -~ \frac{M^{2}}{2\lambda}~]~ =~0 
\end{equation}
                                                                    \\
As in  case of the  QAHO and QDWO's considered in earlier {\bf chapters}, it
 is  convenient to first obtain the solution of the EGS,  eqn.(287). The 
$~\it{two}~$ solutions of eqn.(287) are given by~:
 
\begin{equation}
\sigma~=~0~ ,
\end{equation} 
and 
\begin{equation}
\sigma^{2}~=~\frac{M^{2}}{2\lambda} 
\end{equation}
                                                                \\
We show below that eqn.(289) is $~\it{not}~$ an acceptable solution on
physical grounds. To~ establish this result substitute eqn.(289) in eqn.
(285) which leads to~:
\begin{equation}
5M^{2}~=~-( m^{2} + 12\lambda I_{0}(M^{2}) )
\end{equation}   
Since $~M^{2}~\geq~0~$ is the $~\it{physical}~$ requirement for definition 
of the theory ( see, e.g. eqn.( 267 ) and eqn.( 270 )), equation (290)
can $~\it{not}~$ lead to acceptable solution $~\it{unless}~$
\begin{equation}
\lambda~<~0~ i.e.~ \lambda~ \equiv~-g~,~g~>~0~;
\end{equation}
and
\begin{equation}
12~gI_{0}(M^{2})~-~m^{2}~\geq~0
\end{equation}
( Note that $~\lambda~<~0~$ can $~\it{not}~$ apriori be ruled out since 
bare-parameters in the Lagrangian are $~\it{unobservable}$,~see later).
However, now the defining equation, eqn.(289) can then be rewritten as :  

\begin{equation}
\sigma^{2}~=~-~\frac{M^{2}}{2g}~<~0~, 
\end{equation}
which is $~\it{not}~$ acceptable on grounds of the hermiticity i.e. $~\phi^
{\dagger}~=~\phi!~$ Thus eqn. (289) is ruled out as a solution and the
unique physical solution of the ground state corresponds to eqn. (288). 
Having thus fixed the ground-state  configuration, the implementation of the 
renormalisation  programme  can be done by defining the effective-potential
 (EP). The latter is defined as :
\begin{equation}
U_{0}(\sigma)~\equiv~<vac|~{\cal H}_{0}~|vac>~, 
\end{equation} 
such that 
\begin{equation}
 \sigma~\equiv~<vac|~\phi(\vec{x},t)~|vac>.
\end{equation}

It may be noted that the l.h.s. of eqn. (294) is $~\it{defined}~$ to be a
function of $~\sigma~$ alone. This means that any other parameter occuring in
$~{\cal H}_{0}~$ is to be variationally fixed by minimisation of $~<{\cal
  H}_{0}>$. The procedure is made explicit below by working out the current 
example, with $~{\cal H}_{0}~$ defined in eqn.(260). To this end, we first 
calculate $<~{\cal H}_{0}~>~$ by using the eqn.(274) which {\it guarantees}
 the following equation: 
\begin{eqnarray}
<~{\cal H}_{0}~>~\equiv~ <~{\cal H}~>
~=~\frac{1}{2}m^{2}<\phi^{2}>~+~\frac{1}{2}<\phi_{t}^{2}> 
\nonumber\\
~+~\frac{1}{2}<\phi_{\alpha}^{2}>~+~\lambda<~\phi^{4}~>.
\end{eqnarray} 
This works out to be : 
\begin{eqnarray}
~~<~{\cal H}_{0}~>~=~ \frac{1}{2}m^{2}(\sigma^{2}~+~I_{0})~+~ \frac{1}{2} \int\frac
{d^{3}(\vec{k})}{\Omega_{k}(M)}~(~\omega^{2}_{k}(M^{2})~+~|\vec{k}|^{2}~)
\nonumber\\
 ~+~ \lambda~(~\sigma^{4}~+6\sigma^{2}I_{0}~+~3I_{0}^{2}~)~.          
\end{eqnarray} 
This can be rewritten as, ( using
$~\omega^{2}_{k}(M^{2})~\equiv~|\vec{k}|^{2}~+~M^{2}$)
\begin{eqnarray}
<~{\cal H}_{0}~>~=~ I_{1}~-~\frac{1}{2}I_{0}~(~m^{2}~+~12\lambda\sigma^{2}~+~
12\lambda I_{0}~)~~~~
\nonumber\\
~+~\frac{1}{2}m^{2}~(~\sigma^{2}~+~I_{0}~) 
~+~ \lambda~(~\sigma^{4}~+~6\sigma^{2}I_{0}~+~3I_{0}^{2}~)~,         
\end{eqnarray} 
where, we have defined~:
\begin{equation}
I_{n}(x)~ \equiv~ \int~\frac {d^{3}\vec{k}}{\Omega_{k}(x)}~[~\omega_{k}^{2}(x)~]^{n},~~ n~ = ~0,~\pm ~1,~ \pm ~2,....
\end{equation}

These integrals were first introduced  by Stevenson$^{154}$. Eqn.(298)
 can be  simplified further  by using the
``gap-equation'' as given by eqn.(285). One then obtains :

\begin{equation}
<~{\cal H}_{0}~>~=~ I_{1}(M)~-~3\lambda I_{0}^{2}(M)~+~\frac{1}{2}~m^{2}\sigma^{2}~+~\lambda ~\sigma^{4}
\end{equation}   

We thus derive the LO-effective potential of NGAS as given by 

\begin{equation}
U_{0}(\sigma)~=~\frac{1}{2}~m^{2}\sigma^{2}~+~\lambda ~\sigma^{4}~+~
I_{1}(M)~-~3\lambda I_{0}^{2}(M),
\end{equation}   

where, {\it it is implicitly  understood that  the ``gap-equation'',
  eqn.(285)  is  to  be {\underline first}}  {\it solved to obtain $~ M^{2}~$}{\it as a  function of `$\sigma$'}.\\

One can next carry out the renormalisation programme ( in the LO) by
noting that$^{154}$ :
                                                                 \\
                                                                  \\
(i) the vacuum-configuration corresponds to the absolute (global)
minimum of $~U_{0}(\sigma)~$, i.e. by solving :

\begin{equation}
\frac {dU_{0}}{d\sigma}|_{\sigma_{0}}~=~0~;~\frac{d^{2}U_{0}}
{d\sigma^{2}}|_{\sigma_{0}}~>~0
\end{equation}     

(ii)~the $~\it{renormalised}~$ mass in LO is given by :
 
\begin{equation}
m_{R}^{2}~\equiv~\frac{d^{2}U_{0}}{d\sigma^{2}}|_{\sigma~=~\sigma_{0}}~,
\end{equation}     

(iii) ~the LO-renormalised coupling strength is likewise defined to be: 

\begin{equation}
\lambda_{R}~\equiv~\frac{1}{4!}\frac{d^{4}U_{0}}{d\sigma^{4}}|_{\sigma~=~
\sigma_{0}}~,
\end{equation}     

where $~\sigma_{0}~$ corresponds to the vacuum-configuration as
defined by eqn.(302). It is directly verified by minimisation of
$~U_{0}(\sigma)~$ ( see, eqn.(302)), that the {\it global}
minimum  of the former occurs at $~\sigma_{0}~=~0~$, which is
consistent with eqn.(288) as it should be. Next, evaluating
eqn.(303) at $~\sigma_{0}~=~0~$ one gets the renormalised mass :

\begin{equation}
m_{R}^{2}~=~m^{2}~+~12\lambda {I_{0}}(m^{2}_{R})~\equiv~M^{2}(~\lambda,~
\sigma^{2}~=~0)          ,
\end{equation}

Similarly,~ after a straight forward calculation, one obtains ref.[154] the
renormalised coupling as given by: 

\begin{equation}
\lambda _{R}~ = ~\lambda [\frac{1-12\lambda I_{-1}(m_{R})}{1 + 6\lambda I_{-1}
(m_{R})}],
\end{equation}  

where again $~I_{-1}(x)~$ is defined as per the general definition
given in eqn.(299).\\

 At this stage several remarks / observations are in order :
                                                                 \\
                                                                  \\
(a)~~The results contained in eqs. (285,286 ) and eqs. (301-306)
were first derived by Stevenson$^{154}$ in the context of the
``Gaussian effective potential (GEP)'' for the symmetric
$~\lambda\phi^{4}~$ theory and obtained by variational calculation
using a Gaussian-trial wave-function.\\

The reproduction of the results of the GEP in ref.[154] in the LO of NGAS 
demonstrates that  the GEP (see, eqn.(301)) is contained in the NGAS
as the leading order approximation.
                                                                    \\
                                                                     \\
(b)~~The development of the IPT (~see,~{\bf chapter-10}~) has therefore, the
 potential to go {\it beyond  the  Gaussian  approximation} in  a  systematic
 way, order  by  order. ( The application of IPT to $~\lambda \phi^{4}~$ theory
 falls  however, beyond the scope of the present dissertation.)
                                                                       \\
                                                                        \\
(c)~~In view of the equivalence  with the GEP in the leading order, 
$~\it{all}~$ the results obtained in the former approximation, are reproduced
 in the LO of the NGAS. In particular, the demonstration of 
$~\it{non-triviality}~$ of the symmetric $~\lambda\phi^{4}~$ theory in the 
GEP,~ being entirely based  upon the consequences of renormalisation as 
contained in eqs.(305) and (306), is also reproduced in the LO of NGAS. We 
discuss in the following sub-section, some of the results  concerning the 
stability  and non-triviality of the theory.   
                                                                        \\
                                                                         \\
(d)  It must be emphasized, however, that the current scheme, NGAS is based 
upon $~\it{entirely~ different~ starting ~assumptions ~}$ and~ is much $~\it
{more~ general}~$ than the GEP, which is obtained solely due to the
choice of the  AP as given in eqn.(259) and that too, in the leading order.\\

We next discuss some of the consequences  of the above non-perturbative 
 renormalisation  scheme obtained in the LO of NGAS  leading to the $~\it
{stability}~$ and $~\it{non-triviality}~$ of the theory.
\\
\\
\\
{\large \bf  11.3    Stability and  non-triviality of $\lambda \phi^{4}$-theory in the LO of NGAS}\\

For the above purpose, it is convenient to start with  eqs.(285) and (286)
, which involve the {\it divergent} integral due to the momentum-
integration and which,  therefore, need a suitable  method  of subtraction.\\

Using the {\it subtraction-procedure~ devised ~by ~Stevenson}$^{154}$,
 eqn.(306) can be inverted to express $~\lambda~$  in terms of the 
{\it  observable} parameters, $~\lambda_{R}~$  and $~m_{R}~$. This leads to 
{\it two} solutions for $~\lambda~$ of which, the physical one is given
by :  

\begin{equation}
\lambda = [-1/6I_{-1}(m_{R})][1+1/2[\lambda_{R}I_{-1}(m_{R}) + .......]
\end{equation}

( The {\it other} solution is $~\lambda~ = ~-(1/2)\lambda_{R}~ +~ 0(1/I_
{-1}(m_{R})~$. This solution can be shown to lead to  {\it instability},
since the minimum of the EP corresponding to this solution lies (infinitely)
higher than the minimum corresponding to eqn.(307)).\\

It may be noted  at this point that eqn.(307) implies a viable, stable 
$~\lambda \phi^{4}~$ theory~ when ~the ~(unobservable)~ bare-coupling~ becomes
{\it negative~ but~ infinitesimal}. This version of the theory has 
 therefore, been designated in ~ ref.[155]~ as  {\it  precarious}
  $~\lambda\phi^{4}$ {\it  theory}.\\

Substituting for $~\lambda~$ as given by eqn.(307), one can solve for the 
bare-mass, by inverting  eqn.(305), after carrying out the  subtraction 
as per the Stevenson-prescription. This leads to the following expression 
for the bare-mass : 

\begin{equation}
m^{2} = {m_{R}}^{2} + 2I_{0}(m_{R})/I_{-1}(m_{R}) + (sub-leading~ terms)
\end{equation}

With the aid of eqs.(307) and (308),~ the effective potential, as given by 
eqn.(301) can be recast in manifestly  renormalised form involving the 
$~\it{observable}~$ parameters: $~\lambda_{R}~$ and $~m_{R}~$ only. The 
resulting expression is given by :   

\begin{equation}
U_{0}(\sigma)~ =~ U_{min} + \frac {1}{4}t~{m_{R}}^{2}\sigma ^{2} - ({m_{R}}^{4}/128\pi
^{2})(t-1)^{2}-({m_{R}}^{4}/64\pi^{2})(t-1)\eta,
\end{equation}

where 

\begin{equation}
t~=~M^{2}(\sigma)/{m_{R}}^{2}~;~~ \eta \equiv -4\pi^{2}/\lambda_{R}~,
\end{equation}

and

\begin{equation}
U_{min} = I_{1}(m_{R}) - 3\lambda I_{0}^{2}(m_{R}).
\end{equation}

Similarly, the renormalised  version of the ``gap-equation'' is given
by :

\begin{equation}
(1-\eta)(t-1) - (16\pi^{2}/{m_{R}}^{2})\sigma^{2}~ =~ t~ln~t.
\end{equation}

It must be pointed out that, one has to $~\it{first}~$ solve the gap-
equation, eqn.(312) to obtain  $ ~t~\equiv~t(\sigma)~$, which is then to be 
substituted  in eqn.(309) to infer the $~\sigma$-dependence of $~U_{0}
(\sigma)~$.\\

It may be noted that the gap-equation, eqn.(312) is a transcendental 
equation and its solution exists only when,

\begin{equation}
\sigma ^{2}~ \leq ~\sigma_{min}^{2}~\equiv~({m_{R}}^{2}/16\pi^{2})[e^{-\eta}~ +~\eta - 1] 
\end{equation}

The domain of validity of the effective  potential (EP) is thus restricted by
the range of $~\sigma^{2}~$ determined by eqn.(313) for any value of $\eta$.
In particular, in the regime of large-coupling ($~\eta ~\rightarrow~0~$~), the 
domain of the EP shrinks with $~\eta~$ since $~\sigma^{2} _{min}~\rightarrow 0$
in this limit. This is the situation, therefore, of small oscillations about 
$~\sigma~\simeq~0~$ and it corresponds to the pathological situation when 
$~|\lambda_{R}|~\rightarrow~\infty$. On the other hand, the small coupling 
regime  ($~\lambda_{R}~\rightarrow~0$), which corresponds to $~\eta>>~1$, the 
domain of EP increases with $~\eta~$. One can thus summarise that the LO-EP of
the symmetric $~\lambda \phi ^{4}~$ theory is reasonable  and well behaved 
unless the renormalised  coupling is very large. One can thus conclude that 
 a non-trivial and stable theory results in the LO of NGAS provided the 
$~\it{physical}~$ coupling $~\lambda_{R}~$ is $~\it{not}~$ unusually 
large. We further comment upon the issue of stability of the  perturbative 
vacuum of the theory in the following $~\it{subsection}$.\\

To study the stability issue, it is necessary to compute the effective 
potential based upon the $~\it{perturbative}~$ vacuum (i.e. the vacuum 
of the free-field theory ). This is easily achieved by letting $~M~\rightarrow~
m~$ in all formulae ( see, eqn.(264)). Thus starting from eqn.(297) and 
letting   $~M~\rightarrow~m~$ one obtains, after simplification the following
 expression :

\begin{equation}
<{\cal H}_{0}>_{P}~ =~ \frac {1}{2}m^{2}\sigma^{2}~ +~ \lambda \sigma ^{4}~+ ~
\bar{I}_{1}~ +~ 6\lambda\sigma ^{2}\bar{I}_{0}~ +~ 3\lambda {\bar{I}_{0}}^{2}
\end{equation}

In the above,$~<{\cal H}_{0}>_{P}~$  denotes $~<0|{\cal H}_{0}|0>~$ i.e. VEV of $~{\cal H}_{0}~$ in
the $~\it{perturbative}~$ vacuum state ; and $~\bar{I}_{n}~\equiv~I_{n}(m^{2})
$. By definition, the effective-potential based upon the perturbative vacuum 
denoted by $~U^{P}(\sigma)~$ is identified with  $~<{\cal H}_{0}>_{P}~$. Hence one obtains :

\begin{equation}
U^{P}(\sigma)~ =~ \frac {1}{2}m^{2}\sigma^{2}~ +~ \lambda \sigma ^{4}~+ ~
\bar{I}_{1}~ +~ 6\lambda\sigma ^{2}\bar{I}_{0}~ +~ 3\lambda {\bar{I}_{0}}^{2}
\end{equation}

The renormalised-parameters following from the eqn.(315) are like-wise 
computed and denoted by

\begin{equation}
\bar{m}_{R}^{2}~ =~ d^{2}U^{P}/d\sigma^{2}|_{\sigma^{2} ~=~ 0}~;
\end{equation}

\begin{equation}
 \bar{\lambda}_{R}~ =~ (1/4!)d^{4}U^{P}/d\sigma^{4}|_{\sigma^{2}~=~0}
\end{equation}

where $~\sigma^{2}~=~0~$ is again the location of the global-minimum of 
$~U^{P}(\sigma)~$, as can be readily verified. In this context, it must be 
emphasized  that $~\it{the~ integrals~ : ~\bar{I}_{n}~ occuring~ in }$
$~\it{eqn.(315) ~are ~independent~ of~ ~\sigma~}$. Next, computing the 
derivatives of $~U^{P}(\sigma)~$ at the minimum, one obtains the following 
expressions for the renormalised  parameters based upon the perturbative 
vacuum: 

\begin{equation}
\bar{m}_{R}^{2}~ =~ m^{2}~ +~ 12\lambda \bar{I}_{0}
\end{equation}

\begin{equation}
\bar{\lambda}_{R}~ = ~\lambda
\end{equation}

The requirement of the $~\it{finiteness ~of ~\bar{m}_{R}~ and ~\bar
{\lambda}_{R} ~then ~ demands~ that~ \lambda~}$ ${\it must~ be~ -ve (ref.
to~ eqn.(318))}$, for otherwise $\bar{m}_{R}^{2}$ would be infinitely large
 since $\bar{I_{0}}$ is divergent and the bare ( unobservable )
 mass,$m^{2}>0$. However, this
 would lead to $\it{instability}$ since the effective-potential $U^{P}
(\sigma)$ will $\it{not}$ have a lower-bound! This is made manifest by
explicitly writing the EP in terms of the renormalised parameters: 

\begin{equation}
U^{P}(\sigma) = \frac {1}{2}\bar{m}_{R}^{2}\sigma^{2} +\bar{\lambda }_{R} \sigma ^{4} + 3\bar{\lambda} _{R} {\bar{I}_{0}}^{2} + \bar{I}_{1}
\end{equation}

To prevent instability of the theory when renormalised about the perturbative
 vacuum it, therefore, becomes inescapable that 

\begin{equation}
~\bar{\lambda} _{R}~ =~ \lambda~ =~0~,
\end{equation}

which is nothing but the $~\it{triviality~ scenario}$~!\\

A few remarks/observations are in order, in view of the above results: 
                                                                    \\
                                                                     
(i)   We believe that the result in ( eqn.(321)) constitutes perhaps, the
most direct demonstration of triviality of symmetric $~\lambda \phi^{4}~$ 
theory in physical dimensions.  
                                                                      \\
                                                                       
( ii )~At the same time, the result, eqn.(321) also demonstrates that the
conclusion of $\it{triviality of the  theory is an artefact of  the
 na\ddot{i}ve ~  perturbation ~theory}$ $\it{ built ~and ~renormalised~ 
 around~ the~ free-field~vacuum }~$. As demonstrated earlier, the theory 
renormalised about the NGAS-vacuum leads to a perfectly acceptable, stable and
 non-trivial $~\lambda \phi^{4}$- theory ( see, eqs.(309-313) and discussions
 following ).     
                                                                    \\
                                                                     
( iii )~It may be further pointed out that the ground state of the trivial
theory is still unstable as compared to that in the LO of NGAS, i.e.
\begin{equation}
U_{min}~<<~U_{min}^{P},
\end{equation}
which is readily established by referring to eqn.(320) ( with
$~\sigma~=~0~$) and eqn.(311).\\

This completes our results and discussions regarding the stability and the 
triviality of   $~\lambda \phi^{4}$- theory in the context of NGAS in the 
LO.\\

In the next subsection, we discuss the structure of the interacting vacuum
in analogy with the results obtained in the case of the AHO/DWO's considered
earlier.\\

{\large \bf  11.4    Properties of the Interacting Vacuum State (IVS) in
 NGAS}\\

The actual/physical vacuum state in presence of interaction is approximated
in the LO of NGAS by the state: $~|vac>~$ which is the lowest energy state 
of $~H_{0}~$. In analogy with the results obtained for the AHO/DWO's in 
{\bf chapter-9}, the structure and properties of this state can be inferred 
from studying  the quantum-canonical transformation ( Bogoliubov-Valatin
transformation, ref.[52]) connecting the interacting vacuum state (IVS)
with the free-field vacuum ( FFV ) state.\\

For this purpose it is convenient to start from the Fourier-decomposition of 
the field $~\phi(\vec{x},t)~$ in terms of the $~\it{free-field}~$ creation-
and annihilation operators analogous to eqn.(266):

\begin{equation}
 \phi (\vec{x},t) ~=~ \sigma~ +~ \int \frac {d^{3}\vec{k}}{\Omega_{k}(m)}[a
(\vec{k})e^{-ikx} + a^{\dagger}(\vec{k}) e^{ikx}],
\end{equation}

where, now

\begin{equation}
\Omega _{k}(m)~ \equiv~ 2 ~(2\pi)^{3}~\omega_{k}(m)~;~
{k}^{0}~=~\omega_{k}(m)~\equiv~ \sqrt{|\vec{k}|^{2} ~+~m^{2}},
\end{equation}

corresponding to the propagation of the free-field quanta satisfying
the mass-shell condition : $~{k^{0}}^{2}~-~|\vec{k}|^{2}~=~m^{2}~$. The
free-field  operators satisfy the standard commutation relations :

\begin{equation}
[a(\vec{k}),~a^{\dagger}(\vec{q})]~ =~ \Omega _{k}(m)~\delta ^{3}(\vec{k}-
\vec{q}).
\end{equation}

Comparision of eqs.(268) and (325) implies that the modified
operators :

\begin{eqnarray}
B(\vec{k}) ~\equiv~\frac{b(\vec{k})}{\sqrt{\Omega_{k}(M)}}
\nonumber\\
A(\vec{k}) ~\equiv~\frac{a(\vec{k})}{\sqrt{\Omega_{k}(m)}}
\end{eqnarray}

satisfy identical commutation relations :

\begin{equation}
[B(\vec{k}),B^{\dagger}(\vec{q})]~ =~ \delta ^{3}(\vec{k}-\vec{q})~=~
[A(\vec{k}),~A^{\dagger}(\vec{q}))].
\end{equation}\\

It  follows,  therefore, that the two sets  of operators must be
connected through   Bogoliubov-transformation ref.[52]  given by :

\begin{eqnarray}
B(\vec{k})~~ =~ cosh(\alpha_{ k})A(\vec{k}) ~-~ sinh(\alpha_{k})
A^{\dagger}(-\vec{k})
\nonumber\\
B^{\dagger}(\vec{k}) =~ cosh(\alpha_{k} )A^{\dagger}(\vec{k})~ -
~sinh(\alpha_{k})A(-\vec{k}) ,
\end{eqnarray}

whereas the inverse transformation is given by :

\begin{eqnarray}
A(\vec{k})~ ~=~~ cosh(\alpha_{ k})B(\vec{k}) ~+~ sinh(\alpha_{k})
B^{\dagger}(-\vec{k})
\nonumber\\
A^{\dagger}(\vec{k})~ =~ cosh(\alpha_{k} )B^{\dagger}(\vec{k})~ +
~sinh(\alpha_{k})B(-\vec{k}).
\end{eqnarray}   

In the above, $~\alpha_{k}~=~f(|\vec{k}|)~$ , is apriori an arbitrary real 
function of $~|\vec{k}|~$, i.e.

\begin{equation}
\alpha_{-\vec{k}}~=~\alpha_{\vec{k}}~=~\alpha_{\vec{k}}^{*} .
\end{equation}\\

However, eqs.( 268 ), (323) and (328) considered together
further imply that 

\begin{equation}
exp~( 2\alpha_{k})
~=~\frac{\omega_{k}(M)}{\omega_{k}(m)}~=~\frac{\sqrt{|\underline{k}|^{2} 
    + M^{2}}}{\sqrt{|\underline{k}|^{2} + m^{2}}}
\end{equation}\\

To show this, consider eqs.(268) and (323) at $~t~ =~ 0~$,
which can be written as:

\begin{eqnarray}
\phi (\vec{x},t) ~=~ \sigma~ +~ \int \frac {d^{3}\vec{k}}{\sqrt{\Omega_{k}(m)}}[~A(\vec{k})~ +~ A^{\dagger}(-\vec{k})~] e^{i\vec{k}~.~\vec{x}}~
\nonumber\\
 ~=~ \sigma~ +~ \int \frac {d^{3}\vec{k}}{\sqrt{\Omega_{k}(M)}}[~B(\vec{k})~ +~ B^{\dagger}(-\vec{k})~] e^{i\vec{k}~.~\vec{x}}~,
\end{eqnarray}

which implies  that :

\begin{equation}
\{~B(\vec{k})~ +~ B^{\dagger}(-\vec{k})~\}~=~\sqrt{\frac{\Omega_{k}(M)}{\Omega_{k}(m)}}~\{~A(\vec{k})~ +~ A^{\dagger}(-\vec{k})~\}
\end{equation}

However, from eqn.(328) it follows that 

\begin{equation}
\{B(\vec{k})~ +~ B^{\dagger}(-\vec{k})\}~=~exp~(\alpha_{k})\{~A(\vec{k})~ +~ A^{\dagger}(-\vec{k})~\}
\end{equation} 

thus  leading to the desired  result, eqn.(331).\\

To obtain  the information regarding the  particle-content and other
features of the IVS it is instructive to first compute the
number-density of the free-field-quanta  residing in the IVS. To this
end let us note that the free-field-number operator is given by the
standard expression :

\begin{eqnarray}
N~\equiv~\int {\frac  {d^{3}\vec{k}}{\Omega_{k}(m)}a^{\dagger}(\vec{k})a(\vec
{k})}
\nonumber\\
~=~\int { d^{3}\vec{k}}~A^{\dagger}(\vec{k})~A(\vec{k})
\end{eqnarray}  

Hence the desired number density of the free-field quanta in the IVS
is given by

\begin{equation}
n(\vec{k})~=~<~vac~|~\frac {A^{\dagger}(\vec{k})A(\vec{k})}{v}~|~vac > ~,
\end{equation}  

where $~v ~\equiv~$ spatial-volume of quantisation $~\equiv~\int d^{3}\vec{x}$.
Using eqs.(329), eqn.(336) is easily evaluated :

\begin{equation}
n(\vec{k})~=~\frac {sinh^{2}(\alpha^{vac}(\vec{k}))}{(2\pi)^{3}}
\end{equation} 

where $~\alpha^{vac}(\vec{k})~$ is given by eqn.(331) evaluated
for $~M~\equiv~ m_{R}~$=~ free-particle-mass renormalised~ about the
IVS, $~| vac >$.~( It may be recalled that $~M(\sigma~=~0)~=~m_{R}~$
and $~\sigma~=~0~$ define the IVS ). This leads finally to the
expression :

\begin{equation}
n(\vec{k})~=~(\frac{1}{32\pi^{3}}) ~[~
\frac{\omega_{k}(m)}{\omega_{k}(m_{R})}~+~\frac{\omega_{k}(m_{R})}{\omega_{k}
(m)}~-~2~]~.
\end{equation}    
                                                                     \\
To extract further meaningful content from eqn.(338), we note that
the bare-mass is divergent : $~(\frac {m}{m_{R}})~\sim~0(\frac
{\Lambda}{\sqrt{ln~\Lambda}})$ where $~\Lambda~$ =  momentum cut-off
( see, eqn.(309)). Since, according to the standard prescription of
the renormalisation procedure , the cut-off must be removed (
i.e. $~\Lambda \rightarrow~\infty~$)$~\it{prior}~$  to the
calculation of any physical quantity of the theory , one obtains : 

\begin{equation}
\lim_{\Lambda \rightarrow \infty}~( \frac
{n(\vec{k})}{n(\vec0)})~\equiv~\rho (\vec{k})~=~( 1~+~\frac {|\vec{k}|^{2}}{m_{R}^{2}})^{-\frac{1}{2}}~,
\end{equation}       

where $~n(\vec{0})~=~n(\vec{k})|_{max}~=~( \frac{1}{32\pi^{3}})( \frac 
{m}{m_{R}})~$, is the maximum value of $~n(\vec{k })~$, occuring at $~\vec{k}~=~0$. \\

{\it Equation (339) provides direct physical content for the non-trivial}
{\it structure of the IVS representing a condensate of off-shell correlated} 
{\it particle-pairs}. The situation is analogous to the structure of the
physical vacuum state in case of the hard-sphere Bose-gas$^{49}$
and superfluidity$^{147}$. It is therefore , plausible that eqn.(339)
 might lead to interesting consequences for $~T~\neq~0~$, as happens 
in the case of the super-fluid and the hard-sphere Bose-gas.\\
\\
\\

{\Large \bf 12.  Summary, Conclusions and Outlook}\\

In summary, a new scheme of approximation in quantum theory, is presented
which is simple, non-perturbative,  self-consistent  and systematically 
improvable.
The scheme is, in principle, applicable to arbitrary  interacting  systems. We
have, however, confined the application of the method to the quartic, sextic
and octic anharmonic oscillators, to the quartic and sextic double well
oscillators and to the $~\lambda \phi^{4}~$ symmetric QFT in the present work.\\

 The essential method of this scheme of approximation  consists of finding a
``mapping''  which maps the ``interacting system'' on to an ``exactly
solvable'' model, while preserving the major effects of interaction through
the self consistency  requirement of equal quantum averages of observables
in the two systems.\\

This approximation method has the advantage over the$~{ na\ddot{i}ve}~$
 perturbation theory (NPT) and the variational approximation by transcending 
the limitations of both: unlike the variational method, it is systematically 
improvable through the development of an improved perturbation theory (IPT)
whereas, in contrast to the case of the NPT, the latter satisfies the necessary
condition of convergence for {\it all} allowed values of the quadratic and 
the 
anharmonic coupling strenths `$~g~$' and `$\lambda$'.
The method reproduces the results obtained by several earlier 
methods$^{155}$
 but transcends the limitations of these methods in 
respect of  {\it wider  applicability, systematic improvement and}
{\it  better  convergence}.\\

A remarkable feature of the scheme  is that it respects the exact
 predictions of super symmetric quantum mechanics (SUSYQM) to a good degree of
accuracy in case of the sextic anharmonic oscillator and the sextic double
well potential, when these form a set of ``partner potentials''. In particular,
the property of ``iso-spectrality'', ``positivity''of energy levels and the
predictions for the ``exact'' ground state wave function are reproduced with~
~good~ accuracy {\it even in the  lowest  order  of approximation}.\\

We have also investigated the stability~ properties ~and ~the~
structure ~of~ th `effective' vacuum (EV) of the exactly solvable Hamiltonian,
 $~H_{0}~$, which models the fully interacting system in the leading order. In
 particular, it is shown that the free-field(``perturbative'') vacuum is 
unstable for all values of the couple strength in comparision with the EV. 
Moreover, the latter is  endowed  with a rich structure (``dressing'') in 
terms of the free-field  quanta manifested by the increasing number density of
 particles with the  strength of the interaction, in analogy with the case of 
the super fluid Helium and the hard sphere Bose gas. The application of the 
method to quantum  statistics, non-oscillator systems and field theory appear 
to be straight forward.\\

The present approach to $\lambda \phi^{4}$  theory reproduces the main
results of the Gaussian approximation$^{154}$ . This is considered quite 
significant since the two approaches are based upon rather different physical 
assumptions. However, the authors consider the present approach to be more 
general than the Gaussian approximation  since the former provides a dynamical
 explanation  of the latter through the mechanism of altered vacuum-structure 
introduced by the interaction. Besides, the authors go beyond the scope of the 
Gaussian-approach in establishing new results, e.g., the calculation of the 
momentum-distribution of the condensate-structure function `n(k)'. It may be
emphasized that by going beyond the LO of NGAS, the results of the Gaussian
approximation can be systematically improved, order-by-order.\\

As has been demonstrated, application of NGAS in the LO leads to a nontrivial
and stable $~\lambda \phi^{4}~$ theory in the symmetric phase.
It is well-known, however, that lattice investigations$^{156}$ of 
$~\phi^{4}~$-field theory indicate the triviality scenario and miss the 
non-trivial version arrived at here and in ref.[154]. This result can be 
succinctly understood as follows: the lattice regularised version of the theory corresponds  to a finite ultraviolet-cut-off. This means that the bare coupling $~\lambda~$ is small $~\sim 0(1/I_{-1}^{lattice})~$ and negative for the case
 considered in this work. However, the range of the classical field $\sigma$ 
remains unrestricted.  For this reason, there always exist sufficiently large 
values of $~\sigma~$ (for any given lattice-spacing) such that the term 
$\lambda \sigma ^{4}$ occuring in the effective potential (see, Eq.(301) 
dominates over all other terms (Note that integrals, $~I_{n}~$ are all finite 
on the lattice). Therefore, $~V^{lattice}(\sigma)~$ becomes unbounded from 
below hence, unstable since `$\lambda$' is negative !\\

In a continuum theory, however, the ultra-violet cut-off is never actually
present: if a cut-off is introduced to regularise the theory, the same has to
be sent to infinity first $~\it{prior}~$ to considering any other limiting
 behaviour,
such as $~|\sigma |\rightarrow \infty~ $. This crucial difference in the order
 of taking limits: (UV-Cut-off) $\Lambda \rightarrow \infty $ and $|\sigma|
\rightarrow \infty$, makes all the difference in the physical content of the
theory in the two approaches and explains why, for any finite lattice-spacing,
it will not be possible to discover the stable and non-trivial version of the
theory presented in this work and in ref.[154]. For a detail discussion on this
 important point, see ref.[154].\\

The resulting momentum distribution of the vacuum condensate structure  
function `$n(k)$' deserves special mention as it displays the non-standard 
feature of an appreciable spread in $~|k|~$ about the orgin scaled by the 
renormalised mass of the physical quanta. It is reasonable to expect that this
 condensate-structure of the physical vacuum persists to finite temperature 
manifesting  in observable consequences in the thermodynamic properties of the 
associated system. This would, therefore, constitute a test of the basic 
underlying  assumptions  of the  approximation.\\

Finally, it may be worthwhile to compare and  comment on related work in the 
recent literarure. As has been remarked earlier, analogous ansatz for the 
ground state and the field-operators derived by Boguliobov transformation, 
has been used in ref.[38] for the case of the anharmonic oscillator. The 
important conclusion that emerges from this study is that a convergent
 and accurate perturbation theory  for the energy levels results when the 
theory is developed about the trial  vacuum-state. In contrast, the 
perturbation theory is badly divergent$^{12-14,79}$ if developed about the
 non-interacting (perturbative) vacuum.\\

The relation of the present variational approximation (which is equivalent
 in the LO to the Gaussian approximation) to the one-loop approximation 
method$^{157}$ has been discussed in detail in ref.[154].It has been shown
 that the one-loop results are contained, as a special case, in the present 
results (Eqs.(305-306)) in the limit of small bare-coupling, $~\lambda 
\rightarrow 0_{+}~$ when $~0(\lambda ^{2})~$ terms are neglected. This means
 that the one-loop results remain essentially perturbative in nature, even
 though one resums an infinity of ordinary Feynmann diagrams at the one-loop
order. Besides, the one-loop effective potential has a rather restricted 
domain in both $~\lambda~$ and $~\sigma~$ beyond which it shows pathological
 behaviour. In contrast, the effective potential based upon the present  
approximation, even in the LO, has a considerably larger range of validity and
is genuinely non-perturbative in nature.\\

In a spirit  similar to the present approach, massless $~\lambda\phi^
{4}~$ theory has been variationally investigated in ref.[158]. It has 
been  shown $^{158}$ in this work that the preferred vacuum is also described
 by a condensate structure albeit with a different momentum  distribution 
function of the condensate particle density. In lower (1+1) dimension, the GEP
 has been derived$^{55}$ for the $~\lambda\phi^{4}~$ theory employing similar 
ansatz  for the trial vacuum state. However, the underlying condensate 
structure and  renormalisation have not been investigated in these works.\\

As remarked earlier, possible applications of the results derived
here are envisaged in diverse area of current interest including critical
phenomena (involving a scalar field as the order-parameter$^{122}$),
inflationary cosmology$^{153}$, finite temperature field  theory$^{159}$ ,
exploration of the vacuum structure$^{160}$  of pure gluonic-QCD and  Higgs
sector of the standard model ( by extending$^{161}$  the analysis to  the 
spontaneously broken phase, which corresponds to the case of negative bare-
mass $~m^{2}<0~$).\\

The present work can be extended in different directions to include : finite 
temperature  field theory, quantum-statistics, application to non-oscillator
systems, super-symmetric theories and quantum field theories involving fermions
and gauge-fields etc.\\ 
\\
\\


\end{document}